%% file: preprint.tex
\documentclass[numbers]{assets/matterlab}

\input{assets/shared-preamble.tex}
\usepackage{cinzel}
\usepackage{quantikz}
\usepackage{adjustbox}
\usepackage{wrapfig}
\usepackage{booktabs}
\usepackage{tabularx}
\usepackage{xcolor}

\usepackage[most]{tcolorbox}
\usepackage{braket}
\tcbset{textmarker/.style={%
    enhanced,
    frame hidden, 
    parbox=false,
    boxrule=0mm,
    boxsep=0mm,
    arc=3mm,          
    outer arc=3mm,    
    left=3mm,right=3mm,
    top=7pt,bottom=7pt,
    toptitle=1mm,bottomtitle=1mm,
    left skip=0.2cm
}}

\tcbset{textmarker2/.style={%
    enhanced,
    frame hidden, 
    parbox=false,
    boxrule=0mm,
    boxsep=0mm,
    arc=3mm,          
    outer arc=3mm,    
    left=3mm,right=3mm,
    top=7pt,bottom=7pt,
    toptitle=1mm,bottomtitle=1mm,
    left skip=0.8cm
}}

\newtcolorbox{hintBox}{textmarker,
    colback=blue!10!white}

\newtcolorbox{hintBox2}{textmarker2,
    colback=blue!5!white}

\newtcolorbox{importantBox}{textmarker,
    colback=red!10!white}

\newtcolorbox{noteBox}{textmarker,
    colback=matterbg}

\newcommand{\prompt}[1]{\begin{noteBox} \underline{\textbf{Prompt}} #1 \end{noteBox}}

\newcommand{\user}[1]{\begin{noteBox} \underline{\textbf{User:}} Yes #1 \end{noteBox}}

\newcommand{\usern}[1]{\begin{noteBox} \underline{\textbf{User:}} No #1 \end{noteBox}}

\newcommand{\response}[1]{\begin{hintBox} \underline{\textbf{Response}} #1 \end{hintBox}}

\newcommand{\quantumScientist}[1]{\begin{hintBox} \underline{\textbf{Quantum scientist agent}} #1 \end{hintBox}}

\newcommand{\cudaexpert}[1]{\begin{hintBox2} \underline{\textbf{Cudaq expert agent}} #1 \end{hintBox2}}

\newcommand{\tequilaexpert}[1]{\begin{hintBox2} \underline{\textbf{Qiskit expert agent}} #1 \end{hintBox2}}

\newcommand{\qutipexpert}[1]{\begin{hintBox2} \underline{\textbf{Qutip expert agent}} #1 \end{hintBox2}}

\newcommand{\pennylaneexpert}[1]{\begin{hintBox2} \underline{\textbf{Pennylane expert agent}} #1 \end{hintBox2}}

\newcommand{\tenpyexpert}[1]{\begin{hintBox2} \underline{\textbf{Tenpy expert agent}} #1 \end{hintBox2}}

\title{{\cinzel El Agente Cuántico}: \\ \hfill Automating quantum simulations}

\author[1]{Ignacio Gustin}
\author[2,6]{Luis Mantilla Calder\'on}
\author[1]{Juan B. P\'erez-S\'anchez}
\author[2,6]{Chris Crebolder}
\author[8]{J\'er\^ome F. Gonthier}
\author[1,2]{Mohammad Ghazi Vakili}
\author[2,5,6]{Yuma Nakamura}
\author[1,6]{Karthik Panicker}
\author[2,6]{Manav Ramprasad}
\author[2,6]{Zijian Zhang}
\author[2,6]{Yunheng Zou}
\author[1,2,5,*]{Varinia Bernales}
\author[1,2,3,4,5,6,7,8,*]{Al\'an Aspuru-Guzik}

\affiliation[1]{\addressCHEM}
\affiliation[2]{\addressCS}
\affiliation[3]{\addressMSE}
\affiliation[4]{\addressCHEMENG}
\affiliation[5]{\addressAC}
\affiliation[6]{\addressVECTOR}
\affiliation[7]{\addressCIFAR}
\affiliation[8]{\addressNVIDIA}

\abstract{
\input{includes/include-abstract}
}

\date{\today}
\correspondence{\begin{tabular}[t]{@{}l l@{}}
&Al\'an Aspuru-Guzik  \email{alan@aspuru.com}\\[-7pt]
& and Varinia Bernales  \email{varinia@bernales.org}
\end{tabular}}

\begin{document}

\maketitle

\tableofcontents

\input{includes/include-body}


\section*{Acknowledgments}
\input{includes/include-acknowledgement}

\clearpage


{
\small
\bibliography{references}
\bibliographystyle{unsrtnat}
}


\clearpage

\setcounter{page}{1}
\renewcommand{\thepage}{S\arabic{page}}

\setcounter{section}{0}
\renewcommand{\thesection}{S\arabic{section}}

\setcounter{figure}{0}
\renewcommand{\thefigure}{S\arabic{figure}}

\setcounter{table}{0}
\renewcommand{\thetable}{S\arabic{table}}

\include{includes/include-appendix}


\end{document}

%% file: assets/shared-preamble.tex
\usepackage{microtype}
\usepackage{graphicx}
\usepackage{booktabs}
\usepackage{float}
\usepackage{xurl}
\usepackage{hyperref}
\usepackage{titlesec}

\usepackage{amsmath}
\usepackage{amssymb}
\usepackage{mathtools}
\usepackage{amsthm}
\usepackage{bm}

\newcommand{\urlhref}[2]{\href{#1}{\texttt{#2}}}

\usepackage[noabbrev,nameinlink]{cleveref}

\input{assets/math_commands}

\usepackage{orcidlink}
\usepackage{comment}
\usepackage[version=4]{mhchem}
\usepackage{longtable}
\usepackage{array}
\renewcommand{\arraystretch}{1.7}
\usepackage{multirow}           
\usepackage{amsfonts}           
\usepackage{nicefrac}
\usepackage{duckuments}         
\usepackage{thmtools,thm-restate}
\usepackage{enumitem}
\usepackage{xcolor,colortbl}
\usepackage{tikz}
\usepackage{tikz-cd}
\usepackage{caption}
\usepackage{subcaption}
\usepackage{listings}
\usepackage{gensymb}
\usepackage{textcomp} 
\usepackage[inkscapelatex=false]{svg}

\newcommand{\addressCHEM}{Department of Chemistry, University of Toronto,  80 St. George St., Toronto, ON M5S 3H6, Canada}
\newcommand{\addressAC}{Acceleration Consortium, 700 University Ave., Toronto, ON M7A 2S4, Canada}
\newcommand{\addressCS}{Department of Computer Science, University of Toronto, 40 St George St., Toronto, ON M5S 2E4, Canada}
\newcommand{\addressVECTOR}{Vector Institute for Artificial Intelligence, W1140-108 College St., Schwartz Reisman Innovation Campus, Toronto, ON M5G 0C6, Canada}
\newcommand{\addressMSE}{Department of Materials Science \& Engineering, University of Toronto, 184 College St., Toronto, ON M5S 3E4, Canada}
\newcommand{\addressCHEMENG}{Department of Chemical Engineering \& Applied Chemistry, University of Toronto, 200 College St., Toronto, ON M5S 3E5, Canada}
\newcommand{\addressCIFAR}{Canadian Institute for Advanced Research (CIFAR), 661 University Ave., Toronto,
ON M5G 1M1, Canada}
\newcommand{\addressNVIDIA}{NVIDIA Corporation, Santa Clara, CA, USA}

\newcommand{\acknowAC}{This research is part of the University of Toronto’s Acceleration Consortium, which receives funding from the CFREF-2022-00042 Canada First Research Excellence Fund. }
\newcommand{\acknowGEN}[1]{This research was enabled in part by support provided by #1 and the Digital Research Alliance of Canada (\urlhref{www.alliancecan.ca}{alliancecan.ca})}
\newcommand{\acknowSciNet}{Computations were performed on the Niagara supercomputer at the SciNet HPC Consortium. SciNet is funded by: the Canada Foundation for Innovation; the Government of Ontario; Ontario Research Fund - Research Excellence; and the University of Toronto. }

\providecommand{\doi}[1]{\href{https://doi.org/#1}{doi:#1}}


%% file: assets/math_commands.tex
\usepackage{amsmath,amsfonts,bm}

\def\eqref#1{equation~\ref{#1}}

\def\1{\bm{1}}

\DeclareMathAlphabet{\mathsfit}{\encodingdefault}{\sfdefault}{m}{sl}
\SetMathAlphabet{\mathsfit}{bold}{\encodingdefault}{\sfdefault}{bx}{n}



%% file: includes/include-abstract.tex
Quantum simulation is central to understanding and designing quantum systems across physics and chemistry. However, its practical use is often limited by the exponential growth of Hilbert space and by the increasing complexity of modern quantum-simulation software. Here we introduce{\cinzel El Agente Cuántico}, a multi-agent AI system that automates quantum-simulation workflows by translating natural-language scientific intent into executed and validated computations across heterogeneous quantum-software frameworks. By reasoning directly over library documentation and APIs, our agentic system dynamically assembles end-to-end simulations spanning state preparation, closed- and open-system dynamics, tensor-network methods, quantum control, quantum error correction, and quantum resource estimation. The developed system unifies traditionally distinct simulation paradigms behind a single natural-language interface. Beyond reducing technical barriers, this approach opens a path toward scalable, adaptive, and increasingly autonomous quantum simulation, enabling faster exploration of physical models, rapid hypothesis testing, and closer integration between theory, simulation, and emerging quantum hardware.

%% file: includes/include-body.tex
\break
\section{Introduction}
\begin{wrapfigure}{r}{0.5\textwidth}
    \vspace{-28pt}
    \centering
    \includegraphics[
        width=\linewidth,
    ]{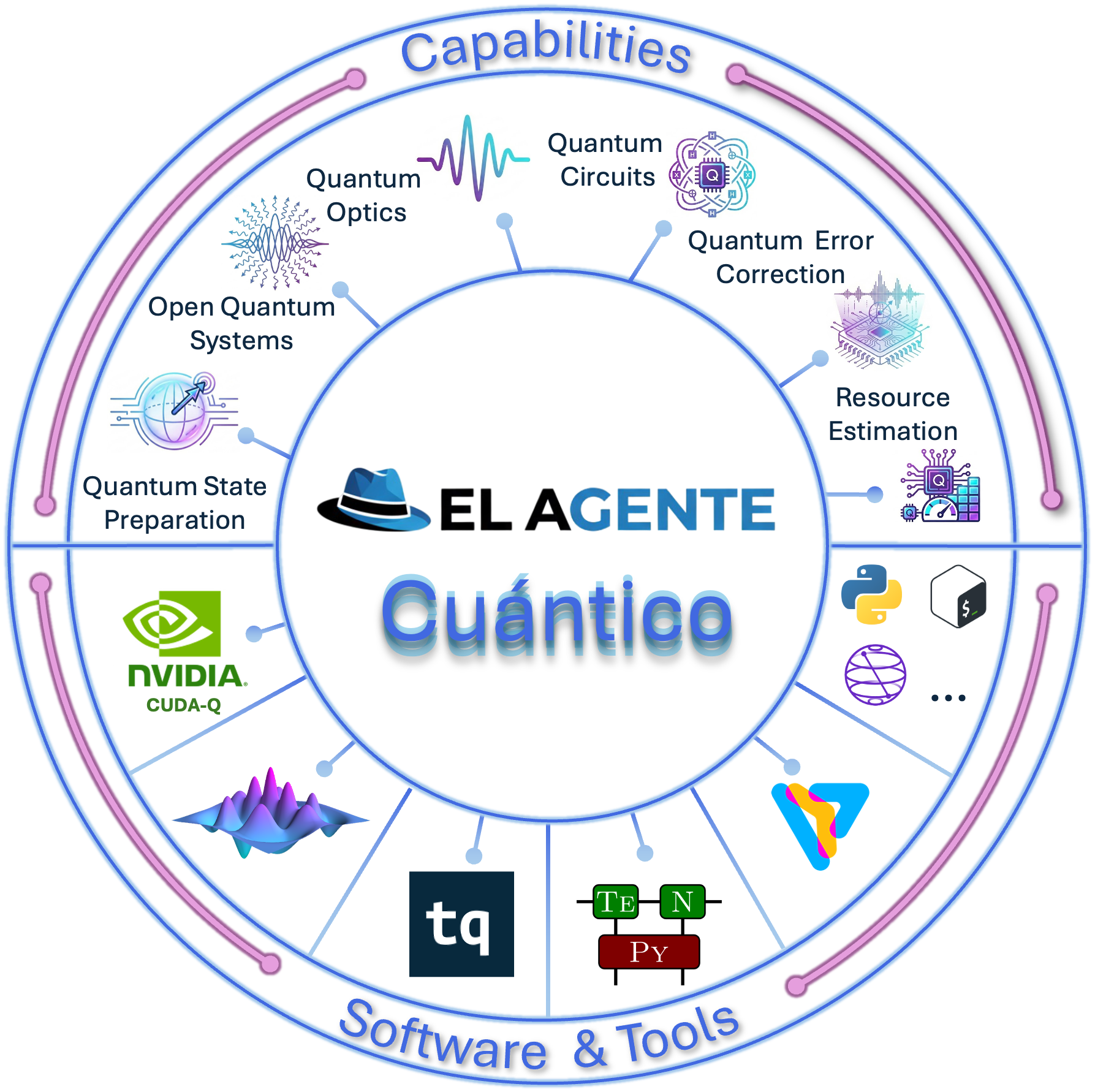}
    \caption{Schematic overview of {\cinzel El Agente Cuántico}, detailing its capabilities (upper half) and the integrated high-performance software stack and tools (bottom half).}
    \label{fig:ElAgente_art}
    \vspace{-12pt}
\end{wrapfigure}
The simulation of quantum systems is central to modern science, providing a systematic way to predict, interpret, and design phenomena governed by quantum mechanics. Across quantum chemistry~\cite{levine2009quantum,cao2019quantum}, condensed-matter physics~\cite{marder2010condensed,anderson2018basic,chaikin1995principles}, quantum optics~\cite{scully1997quantum,gerry2023introductory}, high-energy~\cite{longair2011high,perkins2000introduction}, and materials science~\cite{chawla2012composite, callister2000fundamentals}, a wide range of experimentally relevant observables ultimately require solving, exactly or approximately, the dynamics and/or stationary properties of an underlying quantum system Hamiltonian. In this sense, quantum simulation serves as a computational substrate that connects Hamiltonian models to measurable outcomes, enables systematic validation and refinement of theoretical approximations, and provides access to regimes otherwise inaccessible to direct analysis or experiments.

Despite its importance, quantum simulation remains difficult to access broadly~\cite{georgescu2014quantum}. A core obstacle is that the problem size grows exponentially with system size, making realistic calculations infeasible without strong approximations, careful numerical checks, and substantial computing resources~\cite{troyer2005computational,aspuru2005simulated}. Furthermore, even when powerful techniques are available, such as tensor network methods in suitable many-body regimes~\cite{schollwock2011density, eisert2010colloquium, orus2014practical}, efficient open system solvers~\cite{de2017dynamics,beck2000multiconfiguration,Chen2025}, and variational approaches~\cite{peruzzo2014variational, mcclean2016theory,cerezo2021variational, mantilla2025mbqml,bharti2022noisy,cao2019quantum}, implementing them requires specialized expertise. However, the users, in general, are not all-around specialists in the theory or software ecosystem, and even experts rarely know how to employ right off-the-bat every part of an end-to-end toolchain. As a result, researchers must invest substantial effort in learning new packages, assembling workflows, diagnosing failures, maintaining compatibility across environments, and adapting to different tool interfaces~\cite {heim2020quantum,fingerhuth2018open,larose2019overview}. This overhead often exceeds the time spent on the scientific question itself, and it slows the path from a concrete idea to a working simulation whose results can be trusted.

Large language models (LLMs) offer a potential path to lowering the entry barrier to performing quantum simulations~\cite{ramos2025review,wu2024survey}. Through large-scale pretraining and post-training methods such as reinforcement learning, modern models acquire broad technical knowledge and can apply it through in-context learning~\cite{brown2020language,ouyang2022training,christiano2017deep}. This makes it possible to express simulation intent in natural language and have the system map that intent to concrete implementation choices, including algorithms, software libraries, parameter settings, and correctness checks~\cite{chen2021evaluating,yao2022react,madaan2023self}. Reasoning models have further improved multi-step planning, code synthesis, and iterative self-correction~\cite{wei2022chain,chen2021evaluating,madaan2023self}. Representative examples include OpenAI o1 \cite{openai2024}, DeepSeek R1~\cite{deepseek-ai2025}, Gemini 2.5~\cite{comanici2025}, and Nemotron 4~\cite{nvidia2024}, together with detailed system and model cards that document training procedures, capabilities, and limitations for the GPT-5 series~\cite{openai_gpt5_card}, Gemini models~\cite{deepmind_gemini_card}, and the Claude family~\cite{anthropic_claude}. Consequently, LLMs are increasingly able to support workflows in which the model does not merely describe what to do but helps execute the steps that turn a scientific question into a working simulation.

In parallel, a rapidly expanding ecosystem of scientific agent systems has emerged~\cite{wei2025ai}. General-purpose agents now demonstrate end-to-end research loops that integrate reasoning with tool use, encompassing literature survey, hypothesis generation, verification, implementation, and reporting.\cite{huang2025biomni} NovelSeek~\cite{team2025novelseek} is an early example that explores end-to-end hypothesis generation and validation. PaperQA~\citep{lala2023paperqa} advances literature search and retrieval by enabling question answering over full-text scientific papers. The Kosmos agent~\citep{mitchener2025kosmos} spans literature search, ideation, implementation, and report generation across domains such as biology and materials science. The Virtual Lab~\citep{swanson2025virtuallab} demonstrates collaborative AI agents that design and experimentally validate scientific discoveries. The AI Scientist~\citep{lu2024ai,yamada2025ai} executes independent research workflows in machine learning tasks.

Alongside these general-purpose systems, domain-specific agents are gaining traction in settings that demand specialized tools and deep expertise. In scientific research orchestration, ORGANA~\cite{darvish2024organa} is a domain-specific agent that structures and coordinates complex research workflows by operating over formalized scientific artifacts, objectives, and constraints, rather than acting as an unconstrained general-purpose assistant. In chemistry, early systems such as ChemCrow~\cite{bran2023chemcrow} and Coscientist~\cite{boiko2023autonomous} demonstrated the potential of tool-using agents, while more recent work, including ChemAgent~\cite{tang2025chemagent}, extends this paradigm through multi-agent coordination for literature reasoning and complex laboratory operations. In quantum chemistry and electronic structure, El Agente Q~\cite{elagenteOG} targets ab initio simulation workflows by translating high-level scientific objectives into executable computational pipelines. In physics, SciExplorer~\cite{nagele2025agentic} uses a single LLM agent with access to general-purpose coding and analysis tools to infer equations of motion or Hamiltonians of initially unknown systems from simulated data.
In mathematics, Ax Prover~\cite{breen2025ax} is a multi-agent system for automated theorem proving that can operate autonomously or in collaboration with human experts. In astronomy, StarWhisper Telescope \cite{wang2025starwhisper} automates observational planning and data processing. In scholarly communication and biological research, DORA \cite{naumov2025dora} is a domain-specific agent for academic writing, producing publication-ready manuscripts from structured scientific sources. 

In quantum computing and hardware, AI for quantum computing and agent-based approaches to quantum-physics research have been gaining traction.\cite{alexeev2025artificial,arlt2025towards} For example, \textit{k}-agents~\cite{cao2025automating} provide a framework for autonomously calibrating superconducting qubit experimental devices, QCR-LLM~\cite{flores2025quantum} integrates quantum algorithms for combinatorial problems in the reasoning process of LLMs, and PhIDO~\cite{sharma2025towards} addresses automated design of integrated photonic circuits. Furthermore, Saggio et al.~\cite{saggio2021experimental} report an experimental reinforcement-learning demonstration on a programmable integrated nanophotonic platform, Elliott et al.~\cite{elliott2022quantum} introduce quantum adaptive agents with efficient long-term memory for learning in partially observed/non-Markovian environments, Thompson et al.~\cite{thompson2025energetic} analyze energetic advantages for quantum agents executing complex online strategies and Yun et al.~\cite{yun2022quantum} propose multi-agent reinforcement learning using variational quantum circuits. Quantum-agentic platforms that define quantum agents and outline hybrid architectures have also been proposed~\cite{sultanow2025quantum}. Lemma~\cite{lemma_website}, a commercial tool developed by Axiomatic uses an agentic system to create mathematical models (digital twins) of physics publications. Similarly, Kipu Quantum outlines a roadmap toward market-ready QC and QC+AI products built around platform-based solutions and iterative customer feedback.\cite{kipu_linkedin_roadmap}

Motivated by these developments, and leveraging our previous work and infrastructure on El Agente Q\cite{elagenteOG},
we introduce {\cinzel El Agente Cu\'antico}, a multi-agent system designed to automate quantum simulation workflows by translating natural language prompts into executed and validated computations over a high-performance quantum software stack (see Fig. \ref{fig:ElAgente_art}). The system is grounded in the surrounding software ecosystem through direct access to library documentation and usage examples, which it leverages to recover the relevant APIs and implementation details needed to generate and run code end-to-end. More broadly, this reasoning guided deep research into documentation, which offers an efficient mechanism for software adaptation, enables the agent to rapidly align with new libraries, evolving interfaces, and domain-specific conventions.

The remainder of the paper is organized as follows. Section \ref{sec:architecture} presents the architecture and overall design of the multi-agent system. Section \ref{sec:simulations} tackles representative examples across a broad range of quantum simulation tasks. Section~\ref{sec:benchmark} evaluates the robustness and reproducibility of {\cinzel El Agente Cu{'a}ntico} through systematic benchmarks based on repeated and independent executions of representative problems. Section \ref{sec:beyond} addresses problems that extend beyond quantum simulations. Section \ref{sec:discussion} discusses the broader implications and limitations of the proposed approach, and outlines a strategic roadmap for {\cinzel El Agente Cu\'antico}. Finally, in Section \ref{sec:conclusions}, we present the concluding remarks.

\section{Architecture of {\cinzel El Agente Cu\'antico}} \label{sec:architecture}

\begin{figure*}[ht]
    \centering
    \includegraphics[width=0.8\textwidth]{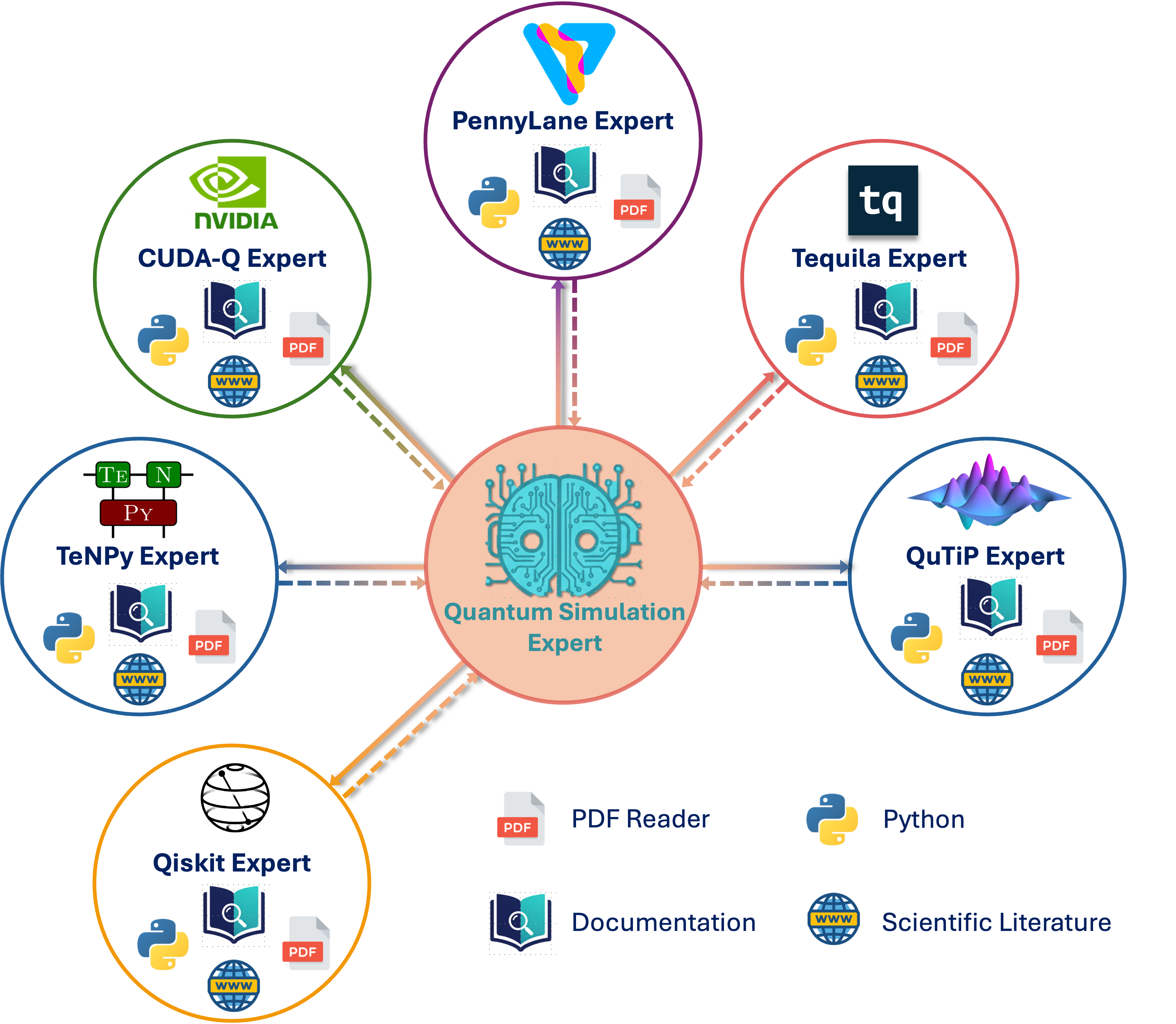} 
    \caption{Multiagentic architecture of {\cinzel El Agente Cu\'antico}. The central agent orchestrates different experts capable of designing and executing code using a specific quantum simulation package.}
    \label{fig:network}
\end{figure*}

{\cinzel El Agente Cu\'antico} uses the cognitive architecture of El Agente \cite{elagenteOG}, designed to enhance the agent’s reasoning, execution, and long-term operation. As a novelty, we deliberately employ a minimalist agent architecture designed to exploit the intrinsic reasoning and execution capabilities of state-of-the-art large language models, rather than constraining behavior through extensive human-crafted tools, prompts, or rigid workflows. The system consists of a set of high-capacity LLM nodes, one per software, connected through a central orchestrator that solely coordinates data exchange, inter-software execution, and user interfacing (see Fig. \ref{fig:network}).

Our agent's architecture incorporates expert agents on quantum software frameworks and platforms, including the general-purpose \texttt{CUDA-Q}\cite{cuda-q_2025_15407754} programming platform for high-performance quantum simulation workflows, \texttt{PennyLane} for quantum and hybrid programming~\cite{bergholm2018pennylane}, \texttt{Qiskit} for quantum circuit development and execution~\cite{javadiabhari2024quantumcomputingqiskit}, \texttt{QuTiP} for general open and closed quantum systems simulation~\cite{QuTiP2025}, \texttt{TeNPy} for tensor-network simulations~\cite{tenpy2024}, and \texttt{Tequila} for flexible quantum algorithm design~\cite{Kottmann_Tequila_A_platform_2020}.

Each agent consists of an LLM that operates with minimal persistent context and without a domain-specific tool or engineered cognition \cite{wang2023voyager,AutoGPT,Yao2023react}. Instead, agents dynamically invoke targeted web search to retrieve software documentation or scientific literature, and Python code for execution. This design contrasts with many other agentic frameworks that rely on large libraries of human-engineered tools, long contextual instructions or reasoning workflows, and task-specific decomposition policies, which can inadvertently limit adaptability and induce brittle behavior when confronted with unanticipated scientific questions. By keeping the agents comparatively ``free” of imposed structure, our approach emphasizes model-internal abstraction and on-demand knowledge acquisition from primary sources, aligning more closely with how human researchers interact with computational tools and documentation. In this sense, our architecture exploits intrinsic LLM capabilities for task execution while being optimized for method discovery in open-ended research settings.

Current quantum-simulation software can be combined, but typically only through workflows that are fixed in advance by developers or users. For example, a researcher may decide ahead of time which program is responsible for generating a circuit, which one runs the
simulation, and which one analyzes the results, and then manually connect these steps. In {\cinzel El Agente Cu\'antico}, the ``quantum simulations expert'' instead makes these decisions at
runtime based on the scientific goal, choosing how to connect available tools on demand.
This allows simulation workflows to adapt to the problem rather than being constrained by
predefined software pipelines.

\section{Experiments for quantum simulations} \label{sec:simulations}

Quantum simulation refers to the use of computational methods to approximate the properties and dynamics of quantum systems starting from an underlying Hamiltonian description. Depending on the physical setting and the observables of interest, this may involve preparing quantum states, simulating real- or imaginary-time evolution, or computing expectation values under unitary or effective non-unitary dynamics.

For closed quantum systems, time evolution is commonly governed by the Schrödinger equation,
\begin{equation}
i\hbar \frac{\partial}{\partial t} |\psi(t)\rangle = H(t)|\psi(0)\rangle.
\label{eq:schrodinger}
\end{equation}
Solving \Cref{eq:schrodinger} requires specifying an initial state $|\psi(0)\rangle$ and a Hamiltonian $H(t)$, which may be time-dependent or time-independent. In practice, the exponential growth of the Hilbert space with system size makes exact solutions intractable beyond small systems. This motivates the use of approximate simulation strategies that balance accuracy, computational cost, and applicability across different physical regimes, and that often require careful choices of algorithms, software tools, and validation procedures.

In this section, we benchmark the ability of {\cinzel El Agente Cu\'antico} to autonomously translate natural-language scientific intent into executed and validated quantum-simulation workflows. Rather than providing a comprehensive review of the underlying techniques, each subsection focuses on a concrete simulation task and evaluates whether the agent can identify appropriate algorithms, software tools, and
validation strategies directly from the prompt. The examples are organized by task type — state preparation, time-independent dynamics, and time-dependent dynamics — and together span a representative cross-section of quantum-simulation methodologies used in current practice.

For each example, we begin with a brief introduction to the relevant scientific context of the problem, followed by a summary of the task posed to the agent, the aspects of El Agente capability showcased by the example, and the agents and software tools involved in the run. We then present the prompt provided to the agent and its corresponding response. For clarity, the agent’s response shown in the blue box corresponds to a lightly edited version of its original output, modified solely for conciseness. Detailed descriptions of the agent available tools, and the complete unedited interaction logs for all experiments are provided in the Supporting Information (Secs. S1 and S2).

\subsection{State preparation}

\subsubsection{Variational quantum eigensolver (VQE)} \label{sec:VQE}
This method provides a practical approach for computing molecular ground states on noisy intermediate-scale quantum hardware \cite{peruzzo2014variational}. The method begins by selecting a parametrized ansatz $|\psi(\theta)\rangle$, where $\theta$ denotes the set of variational parameters that control the quantum circuit used to prepare the state. The algorithm relies on the variational principle,
\begin{equation*}
    E(\theta) =
    \langle \psi(\theta) | H | \psi(\theta) \rangle
    \ge E_0,
\end{equation*}
which states that the expectation value of the Hamiltonian $H$ over any trial state provides an upper bound to the exact ground-state energy $E_0$. Minimizing $E(\theta)$ over $\theta$ therefore approximates the true ground state.

To evaluate the energy on a quantum device, the molecular electronic Hamiltonian is mapped to a qubit Hamiltonian through a fermion-to-qubit encoding such as Jordan--Wigner~\cite{jordan1928paulische,aspuru2005simulated, somma2002simulating} or Bravyi--Kitaev~\cite{seeley2012bravyi}. This yields an operator decomposition of the form
\begin{equation*}
    H = \sum_i h_i P_i,
\end{equation*}
where the coefficients $h_i$ are real scalars determined by the one- and two-electron integrals, and the $P_i$ are tensor products of Pauli operators acting on the qubit register. The corresponding expectation values $\langle P_i \rangle_{\theta}$ are obtained by repeated measurement of the ansatz state on the quantum device.

The VQE algorithm proceeds iteratively: (1) prepare the parametrized state $|\psi(\theta)\rangle$ on the quantum hardware, (2) measure the expectation values $\langle P_i \rangle_{\theta}$ and reconstruct the energy $E(\theta)$, (3) compute a classical update of the parameters $\theta$ using an optimization routine, and (4) repeat until convergence.

For molecular benchmarks, we employ the unitary coupled-cluster singles and doubles (UCCSD) ansatz, which incorporates electron correlation through an anti-Hermitian excitation operator \cite{bartlett2007coupled, romero2018strategies, anand2022quantum},
\begin{equation*}
    |\psi(\theta)\rangle =
    e^{T(\theta) - T^\dagger(\theta)}
    |\phi_{\mathrm{HF}}\rangle,
\end{equation*}
where $|\phi_{\mathrm{HF}}\rangle$ is the Hartree--Fock reference state and
\begin{equation*}
    T(\theta) = 
    \sum_{ai} \theta_{ai}\, a_a^\dagger a_i +
    \frac{1}{4} \sum_{abij} \theta_{abij}\, a_a^\dagger a_b^\dagger a_j a_i
\end{equation*}
contains the single- and double-excitation operators in second quantization. Here, $i,j$ label occupied orbitals, $a,b$ label virtual orbitals, and $a_p^\dagger$/$a_p$ denote fermionic creation and annihilation operators. The unitary form ensures norm preservation and provides a systematically improvable, chemically motivated ansatz.

\noindent\textbf{What we asked.} We asked {\cinzel El Agente Cu\'antico} to read the \texttt{CUDA-Q} manual and compute the VQE ground-state energy of H$_2$ over bond distances from 0.4 to 2.2~\AA, benchmark against the exact energies, and generate/interpret a dissociation-curve plot. 

\textbf{What this showcases.} This example tests end-to-end autonomy for a standard chemistry workflow: selecting an ansatz (UCCSD), executing a parameter sweep, validating against a reference method, and producing publication-quality plots plus a physically meaningful error analysis. 

\textbf{Used agent/tools:} quantum scientist agent, cudaq expert agent, deepsearch, python REPL, submit python jobs and view \texttt{png} file, see supplementary information for detailed description. The answer was verified to be correct.


\prompt{Read the \texttt{CUDA-Q} manual and compute the VQE energy of \ce{H2} over bond distances from 0.4 to 2.2 \AA   \ and compare against the exact answer. For the plot, create a figure of the dissociation curve and explain the physical meaning and where the curve deviates from exact energies.}

\response{

\textbf{Left panel, PES.} Morse-like potential with equilibrium at 0.70 \AA. VQE (coral circles) perfectly overlaps Exact (navy dots) while HF (teal) diverges catastrophically at stretched geometries.

\textbf{Right panel, Errors.}  Log scale ($10^{-6}$ to $10^{4}$ mHa) reveals 5 orders of magnitude separating VQE errors ($\sim10^{-4}$ mHa) from HF errors ($\sim 10^{2}$ mHa), with chemical accuracy threshold clearly in between.

\textbf{Static correlation.} HF fails because a single Slater determinant cannot represent the entangled superposition needed for correct dissociation. VQE's UCCSD ansatz captures this physics exactly.

    \begin{center}
\includegraphics[width=0.9\linewidth]{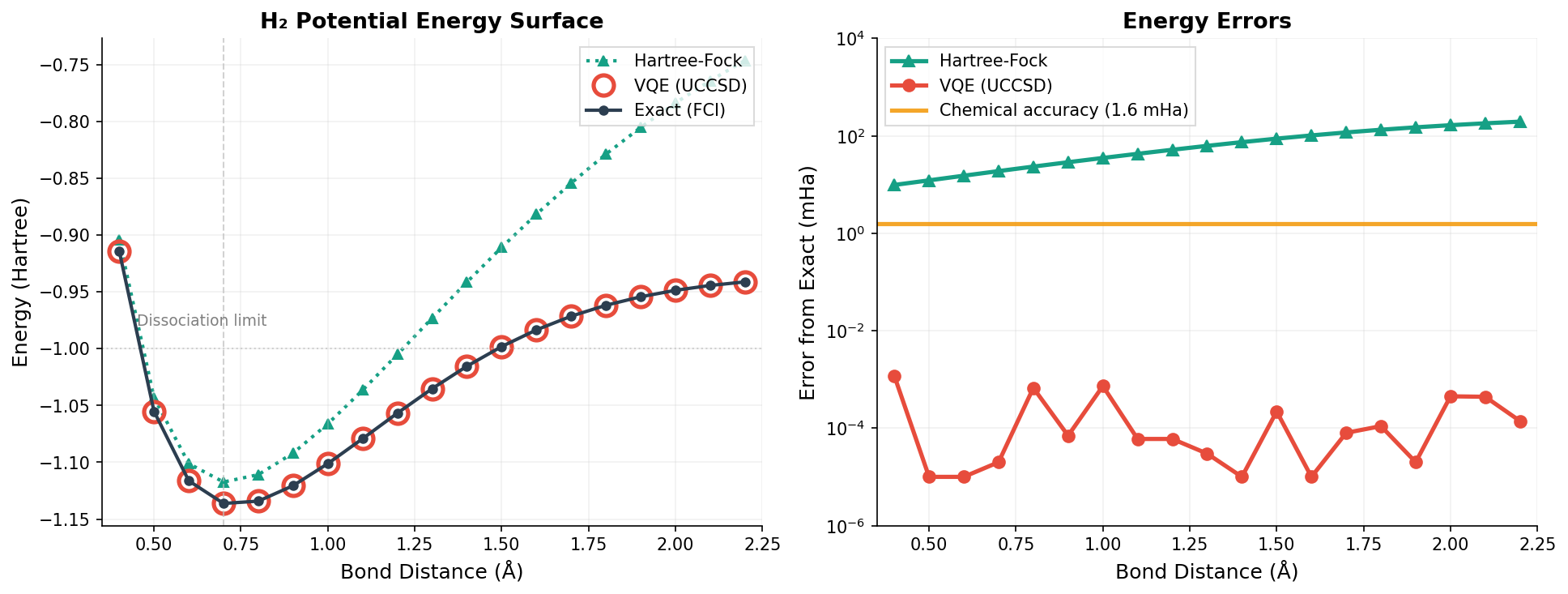}
\captionof{figure}{Plots generated by {\cinzel El Agente Cu\'antico}. On the left, potential energy surface (PES) of \ce{H2} computed via VQE with the UCCSD ansatz, compared to FCI and HF. On the right, absolute energy errors on a log scale.}
    \end{center}
}

\subsubsection{Bell state preparation}\label{sec:Bell}
Bell states are canonical examples of maximally entangled two-qubit states and serve as foundational resources in quantum information science \cite{nielsen_chuang_2010}. The state $(\ket{00}+\ket{11})/\sqrt{2}$, first discussed in the context of quantum nonlocality by Bell \cite{bell1964einstein}, exhibits perfect correlations in both the Z and X measurement bases. Individual outcomes remain random, while joint statistics cannot be explained by any classical model that assumes local realism. Owing to this structure, Bell states support central quantum protocols such as quantum teleportation \cite{bennett1993teleporting}, superdense coding \cite{bennett1992communication}, and a wide range of entanglement-based benchmarking methods \cite{horodecki2009quantum}. They therefore provide a minimal and analytically transparent setting for studying entanglement, coherence, and two-qubit observables using quantum hardware or simulators.

\textbf{What we asked.} We asked {\cinzel El Agente Cu\'antico} to read the \texttt{CUDA-Q} manual and implement a 2-qubit circuit that prepares the Bell state $(|00\rangle+|11\rangle)/\sqrt{2}$, then estimate $\langle Z\otimes Z\rangle$ (computational basis) and $\langle X\otimes X\rangle$ (X basis) from 4096-shot measurements, and plot/analyze the results. 

\textbf{What this showcases.} This example probes whether the agent can translate a minimal entanglement task into correct circuit construction, basis changes, statistics-to-observables postprocessing, and sanity-checking via the known ideal correlations.

\textbf{Used agent/tools:} quantum scientist agent, cudaq expert agent, deepsearch, python REPL, submit python jobs and view \texttt{png} file, see supplementary information for detailed description. The answer was verified to be correct.

\prompt{Read the \texttt{CUDA-Q} manual and create a 2-qubit quantum circuit that prepares a Bell state $(\ket{00}+\ket{11})/\sqrt{2}$. Start in $\ket{00}$. Apply a Hadamard gate on qubit 0 and then a CNOT with control qubit 0 and target qubit 1. Measure both qubits in the computational $\braket{Z}$ basis with 4096 shots and return the measurement counts. From those counts, compute and return the expectation value of $Z \otimes Z$. Then also estimate the expectation value of $X \otimes X$ by measuring in the X basis, again with 4096 shots, and return both the counts and the estimated $\braket{X\otimes X}$. Plot and analyze the results.}

\response{

\textbf{Left panel 1, Z-basis measurements.} The outcomes are nearly equally split between $\ket{00}$ ($51.4\%$) and $\ket{11}$ ($48.6\%$), with no occurrences of $\ket{01}$ or $\ket{10}$. The slight imbalance from a 50/50 distribution is consistent with finite sampling noise from 4096 shots.

\textbf{Left panel 2, X-basis measurements.} The near-$50/50$ split persists in the X basis, confirming strong correlations in multiple measurement bases. Unlike classically correlated states, which appear uncorrelated in X, the Bell state exhibits correlations in both bases, enabling violations of Bell inequalities.

\textbf{Left panel 3, Expectation values.} $\braket{Z\otimes Z}$ = +1 (perfect correlation in Z-basis),
    $\braket{X\otimes X} = +1$ (perfect correlation in X-basis),
$\braket{Y\otimes Y} = -1$ (perfect anti-correlation in Y-basis).
The combination $(+1, +1, -1)$ uniquely identifies $\ket{\Phi^{+}}$ among the four Bell states. 
    \begin{center}
\includegraphics[width=0.9\linewidth]{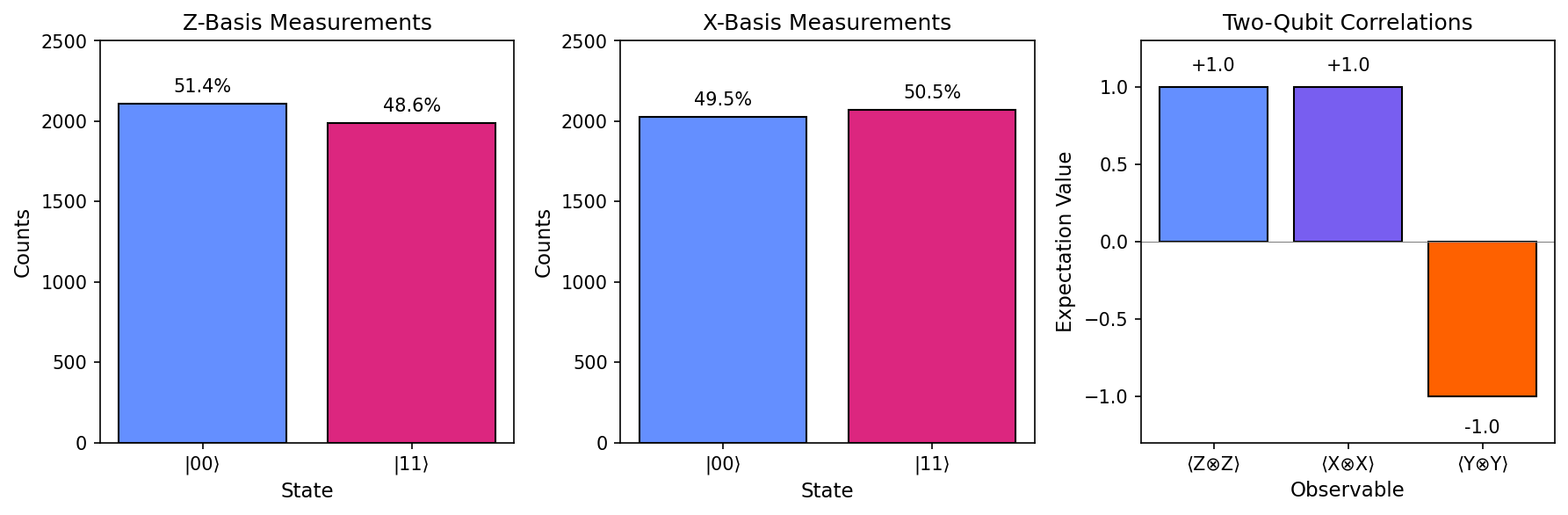}
 \captionof{figure}{Plots generated by {\cinzel El Agente Cu\'antico} characterizing a 2-qubit Bell state prepared on a quantum simulator. The left and center panels show measurement counts in $Z$ and $X$ bases. The right panel shows the qubit correlations $\braket{Z\otimes Z}$, $\braket{X\otimes X}$, and $\braket{Y\otimes Y}$.}
    \end{center}}

\subsubsection{Thermal states via imaginary-time evolution}

Preparing thermal states of interacting many-body systems is a central task in quantum simulation. At inverse temperature $\beta = 1/T$, the equilibrium state is described by the Gibbs density matrix,
\[
\rho(\beta) = \frac{e^{-\beta H}}{Z}, \qquad 
Z = \mathrm{Tr}\!\left[e^{-\beta H}\right],
\]
which encodes the thermodynamic and correlation properties of the system. For lattice models such as the one-dimensional Hubbard Hamiltonian~\cite{essler2005one},
\[
H = -t \sum_{\langle i,j\rangle,\sigma} \left( c^{\dagger}_{i\sigma} c_{j\sigma} + \mathrm{h.c.} \right)
+ U \sum_i n_{i\uparrow} n_{i\downarrow},
\]
imaginary-time evolution~\cite{feiguin2005finite,verstraete2004matrix} provides a practical classical approach to approximating $\rho(\beta)$ by evolving an initial state under $e^{-\beta H}$. The resulting state can be validated by examining thermodynamic observables such as the temperature dependence of the energy and local occupancies.

\textbf{What we asked.} We asked {\cinzel El Agente Cu\'antico} to read the \texttt{TenPy} manual and compute the entanglement entropy for the 1D Bose--Hubbard model using purification-based imaginary-time evolution with a second-order Trotterization. 

\textbf{What this showcases.} This example evaluates whether the agent can move beyond simple state preparation and correctly set up tensor-network finite-temperature machinery (purification + imaginary time), while producing a quantitative observable (entanglement entropy) from the resulting state.

\textbf{Used agent/tools:} quantum scientist agent, tenpy expert agent, deepsearch, python REPL, submit python jobs and view \texttt{png} file, see supplementary information for detailed description. The answer was verified to be correct.

\prompt{ Prepare a Gibbs state of a 1D Hubbard model on a classical computer using imaginary-time evolution. Consider a 10-site Hubbard chain with open boundary conditions, hopping $t = 1$, on-site interaction $U = 4$, and half filling. Target an inverse temperature $1/T = 2$. Explain how to validate that the resulting state is thermal by checking energies and occupancies.
 }

\response{   

\textbf{Panel (a), energy vs 1/T.} As temperature decreases, the Boltzmann distribution favors lower-energy states, so the average energy drops monotonically toward the ground state.

\textbf{Panel (b), site occupancy.} The canonical ensemble at fixed $N=10$ electrons maintains exactly one electron per site on average, as expected for half-filling.

\textbf{Panel (c), double occupancy.} The strong on-site repulsion $U=4$ suppresses double occupancy by 64$\%$. This is the hallmark of Hubbard physics—electrons avoid sharing sites due to Coulomb repulsion.

\textbf{Panel (d), entanglement entropy.} Von Neumann entropy S at each bond cutting the chain. The purification MPS has entanglement from both quantum correlations and thermal mixing. Higher entropy at the center is typical for finite systems with open boundaries.
\begin{center}
\includegraphics[width=0.8\linewidth]{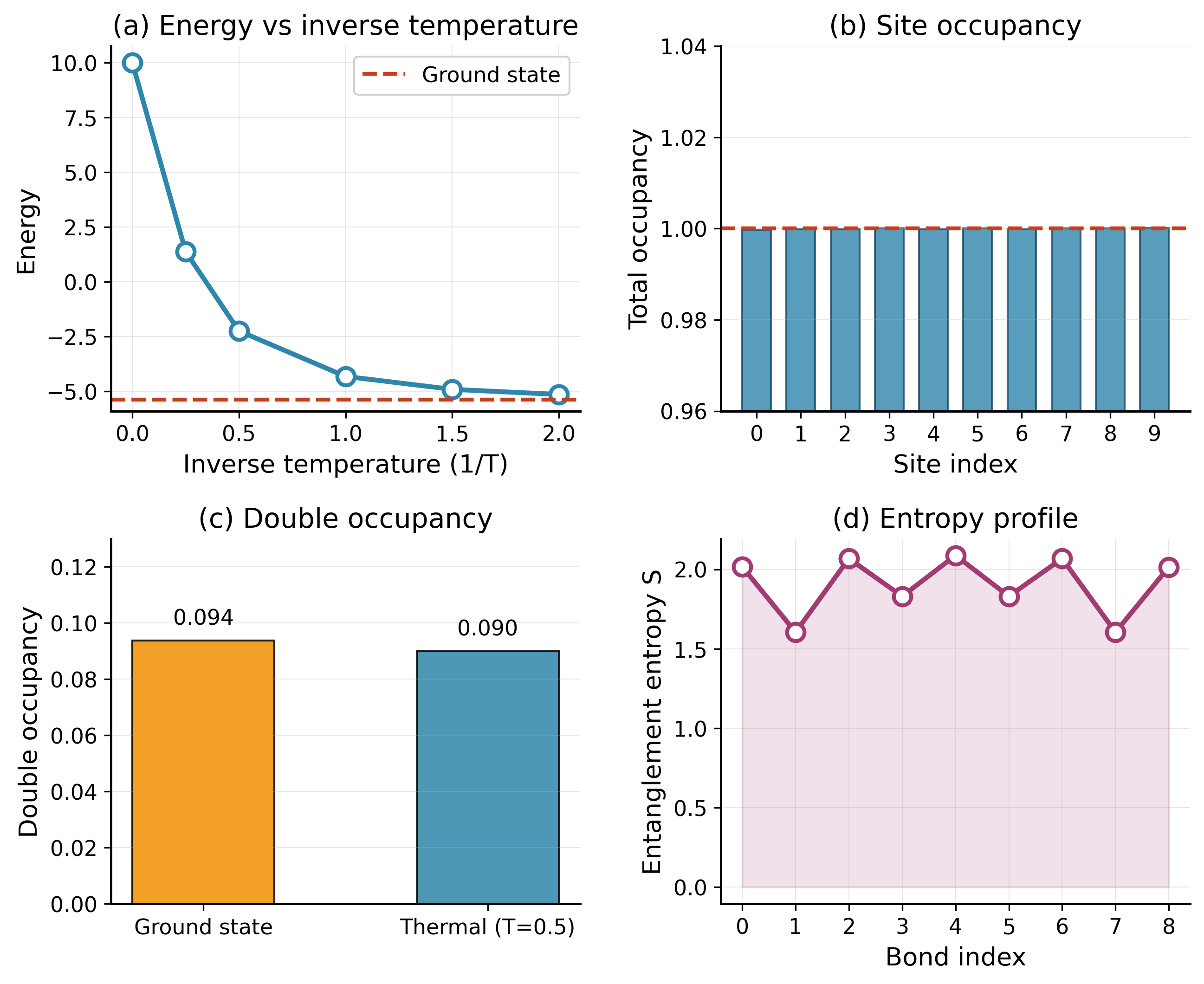}
 \captionof{figure}{Plots generated by {\cinzel El Agente Cu\'antico} preparing and validating a thermal state of a 10-site Hubbard chain using imaginary-time evolution.}
    \end{center}}

\subsection{Time-independent Hamiltonian simulation}

\subsubsection{Trotter decomposition} \label{sec:ising}
When simulating the dynamics of Hamiltonians of the form $H= H_1 + H_2 + \cdots + H_n$, where $[H_i, H_j] \neq 0$ for $i\neq j$ and each $H_i$ is easily diagonalizable, it is common to perform Trotterization \cite{suzuki1990fractal}. This method is a standard technique for simulation using quantum computers and approximates the exponential of $H$ by $r$ repetitions of the individual dynamics of each $H_j$,
\begin{equation*}
    e^{iHt} \approx \left(\prod_{k=0}^n e^{iH_k t/r}\right)^r.
\end{equation*}
There are multiple ways to compute the products of single-term exponentials to obtain a better approximation of $e^{iHt}$. For example, high-order product formulas \cite{childs2021theory} or corrected product formulas \cite{bagherimehrab2025faster} achieve better error scaling than the first-order approximation mentioned before. 

\textbf{What we asked.} We asked {\cinzel El Agente Cu\'antico} to simulate closed-system time evolution for a time-independent Hamiltonian using a Trotterized product-formula approach, including an explicit comparison/validation against a reference evolution where possible. 

\textbf{What this showcases.} This example focuses on method selection for Hamiltonian simulation (product formulas) and tests whether the agent can reason about algorithmic accuracy (step size/Trotter error), implement the workflow in a concrete software stack, and validate dynamics with a controlled benchmark.

\textbf{Used agent/tools:} quantum scientist agent, cudaq expert agent, deepsearch, python REPL, submit python jobs and view \texttt{png} file, see supplementary information for detailed description. The answer was verified to be correct.

\prompt{Read the \texttt{CUDA-Q} manual and implement a simulation of the transverse-field Ising model for a 10-qubit chain with open boundary conditions. Compute the dynamics of the magnetization $\langle Z \rangle$ for two parameter regimes, $J_{\text{coup}} = 10\,J$ with $h = 0.1\,J$, and $J_{\text{coup}} = 0.1\,J$ with $h = 10\,J$. Start the evolution from a Haar-random initial state and simulate the dynamics over a total time of 10 seconds using a second-order product formula with 1000 time steps. Generate a heatmap where the x-axis corresponds to qubit position, the y-axis to time, and the color scale to the $Z$-magnetization, for each of these two phases. Analyze the resulting magnetization dynamics between the two parameter choices.}

\response{

\textbf{Left panel, strong coupling.} Slow, smooth dynamics dominated by $ZZ$ interactions; coherent spatial correlations across the chain; and spatial structure reflecting collective behavior. 

        \textbf{Right panel, strong field.} Fast oscillatory dynamics with horizontal banding; nearly independent qubit precession at frequency $\sim 2h$; rapid oscillation period around $0.06\,\text{s}$; and weak inter-qubit correlations due to small coupling.

The contrast clearly illustrates the quantum phase transition behavior of the Transverse-field Ising model: the ferromagnetic phase (strong coupling) vs. the paramagnetic phase (strong field).
    \begin{center}
\includegraphics[width=0.9\linewidth]{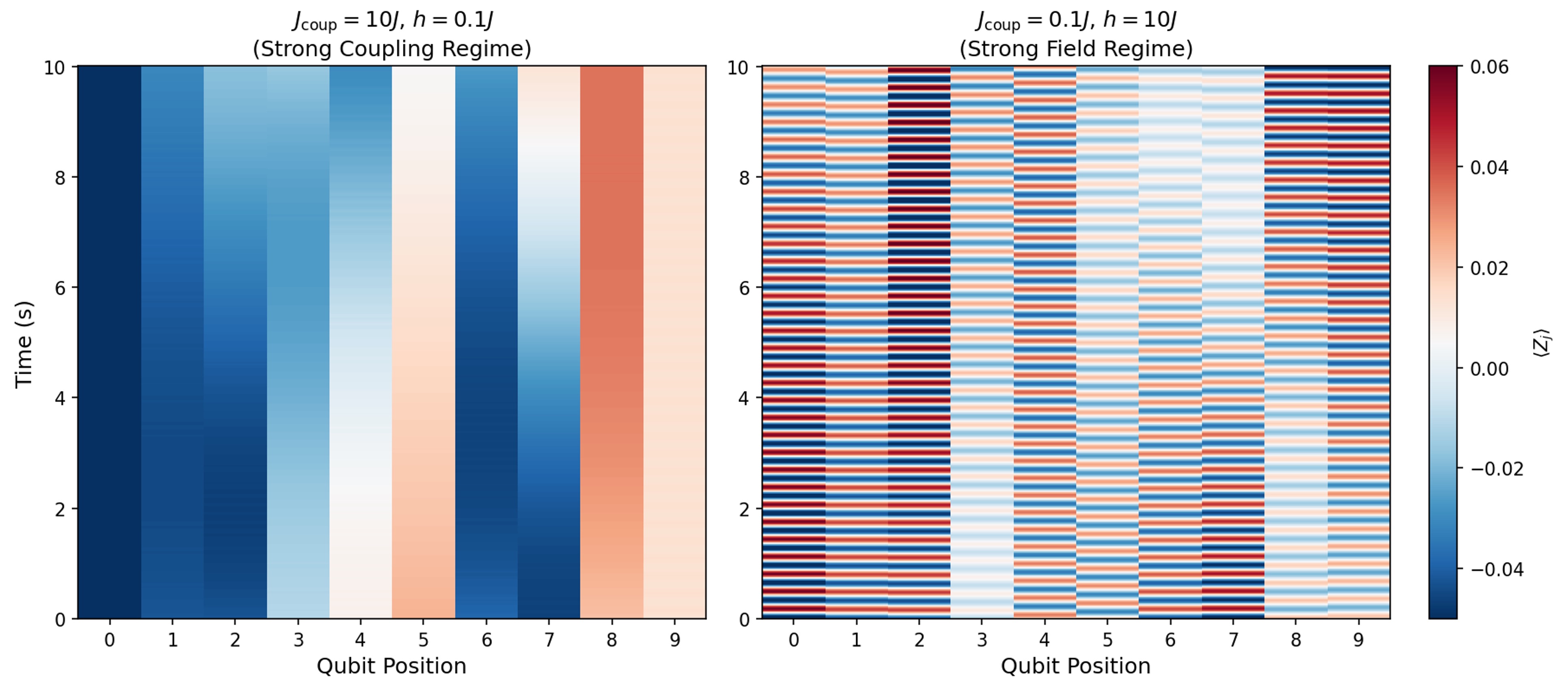}
 \captionof{figure}{Plots generated by {\cinzel El Agente Cu\'antico} time evolving a 10-qubit transverse-field Ising chain using Trotterization in two parameter regimes: strong coupling (left) and strong field (right).}
    \end{center}
}

\subsubsection{Open systems dynamics using the Lindblad approximation} \label{sec:lindblad} 

Realistic quantum systems interact with surrounding degrees of freedom, which leads to decoherence~\cite{gustin2023mapping, joos2013decoherence,gustin2025decoherence,schlosshauer2007decoherence,hu2022tuning} and dissipation~\cite{gustin2025dissipation,weiss2012quantum,gustin2025general,whitfield2010quantum,rebentrost2009environment} that cannot be captured by unitary evolution. In such cases, the dynamics of the reduced density matrix $\rho(t)$ is described by a quantum master equation. Under the assumptions of complete positivity, trace preservation, and Markovianity, the generator of the evolution takes the Gorini–Kossakowski–Sudarshan–Lindblad (GKSL) form \cite{Gorini1976,Lindblad1976,BreuerPetruccione2002}
\begin{equation}
\frac{d\rho}{dt} = - i[H,\rho] + \sum_{k} \mathcal{D}[C_{k}](\rho),
\end{equation}
where the dissipator is defined as
\begin{equation}
\mathcal{D}[C](\rho)
= C \rho C^{\dagger}
- \tfrac{1}{2} \left\{ C^{\dagger} C , \rho \right\}.
\end{equation}
The operators $C_{k}$ represent environmental channels and determine how populations and coherences decay. The GKSL structure ensures that the evolution remains physical and preserves positivity and the trace of the density matrix.

For a single qubit, the Lindblad equation provides a clear description of how coherent Bloch sphere motion competes with irreversible relaxation. A standard example is amplitude damping generated by $C = \sqrt{\gamma}\,\sigma_{-}$, which drives population from the excited state to the ground state at rate $\gamma$ while attenuating coherences. The expectation value
\begin{equation}
\langle Z(t) \rangle = \mathrm{Tr}[\rho(t) Z]
\end{equation}
highlights the contrast between closed- and open-system evolution. Unitary dynamics generates undamped oscillations, whereas in the presence of dissipation, the trajectories decay toward the stationary state.

\texttt{QuTiP} \cite{QuTiP2025} provides efficient solvers for the numerical propagation of Schrödinger and Lindblad equations. The following example illustrates how {\cinzel El Agente Cuántico} uses this functionality to model dissipative single-qubit dynamics.

\textbf{What we asked.} We asked {\cinzel El Agente Cu\'antico} to read the \texttt{QuTiP} manual and compare a single qubit’s dynamics under (i) unitary evolution and (ii) Lindblad evolution, for $H=\sigma_x+\tfrac{1}{2}\sigma_z$, initial $X$ eigenstates $|+\rangle,|-\rangle$, and a decay channel $C=\sqrt{\gamma}\sigma_-$ with $\gamma=0.2$, reporting $\langle Z(t)\rangle$ over $t\in[0,40]$. 

\textbf{What this showcases.} This example tests solver/tool choice (Schr\"odinger vs. master equation), correct handling of collapse operators and expectations, and the agent’s ability to present a clear closed-vs-open comparison plot and interpretation.

\textbf{Used agent/tools:} quantum scientist agent, qutip expert agent, deepsearch, python REPL, submit python jobs and view \texttt{png} file, see supplementary information for detailed description. The answer was verified to be correct.

\prompt{ Read the \texttt{QuTiP} manual and compare the dynamics of a single qubit under unitary evolution and under open system Lindblad evolution using: 

\textbf{Hamiltonian:} $H = \sigma_{x} + \frac{1}{2}\sigma_{z}$,
\textbf{Initial condition:} Use the plus and minus eigenstates of X.
\textbf{Dissipation:} Use a collapse operator $C = \sqrt\gamma  \sigma_{-}$, with $\gamma$ = 0.2. \textbf{Evolution:} Use a time range from $t = [0,40]$ with 400 time steps. \textbf{Output:} Plot the expectation value of Z as a function of time for each initial state and compare closed and open dynamics. }
\response{

\textbf{Panel (a), Unitary Evolution:} Shows persistent Rabi-like oscillations for both $\ket{+}$ (dark blue, solid) and $\ket{-}$  (light blue, dashed) initial states with amplitude $\sim$0.4

\textbf{Panel (b), Lindblad Evolution:} Shows damped oscillations converging to a steady state with a horizontal dashed line at $\braket{Z} \approx$  -0.33

\textbf{Physical interpretation:} The Hamiltonian $H = \sigma_{x} + 0.5\sigma_{z}$ creates a tilted effective magnetic field causing Rabi-like precession. The collapse operator $C = \sqrt{\gamma} \sigma_{-}$ induces spontaneous emission ($\ket{1} \to \ket{0}$ decay). Competition between coherent driving (H) and dissipation ($\gamma$) produces a non-trivial steady state where $\sim67\%$ population is in $\ket{0}$.
Crucially, the steady state is unique. Both initial states converge to the same value, demonstrating loss of memory of initial conditions (a hallmark of open quantum systems).
    \begin{center}
\includegraphics[width=1.0\linewidth]{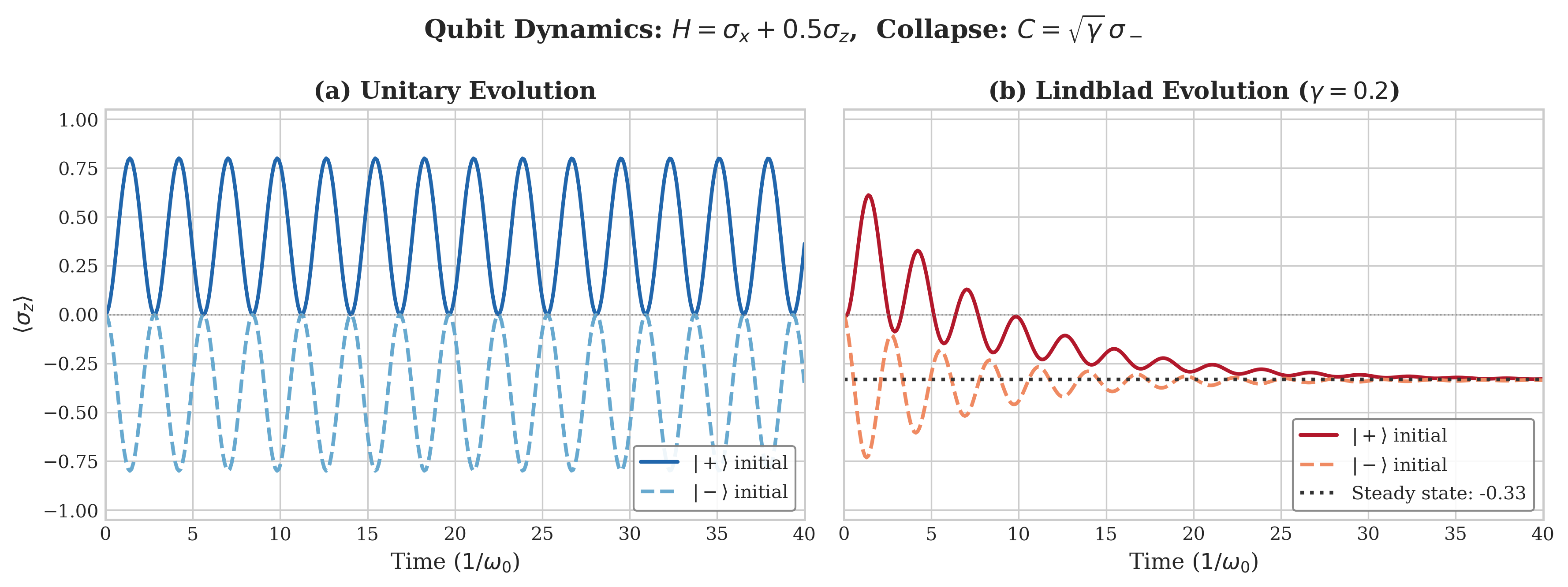}
 \captionof{figure}{Plots generated by {\cinzel El Agente Cu\'antico} comparing the oscillatory behaviour of unitary and Lindblad dynamics on a single qubit.}
\end{center}
}

\subsubsection{Hierarchical equations of motion (HEOM)}\label{sec:HEOM}

The hierarchical equations of motion provide a powerful and widely used framework for modeling open quantum dynamics when the assumptions underlying the Lindblad equation are no longer valid. Many condensed phase systems interact with environments characterized by strong coupling and long-lived correlations, which prevents a Markovian master equation from describing the evolution of the reduced density matrix. In the HEOM\cite{tanimura2020numerically, Ikeda2020} formalism, the system Hamiltonian $H$ evolves together with a hierarchy of auxiliary density operators that encode the full influence of the bath. This approach retains non-Markovian memory and offers an essentially exact treatment of dissipation for environments with structured spectral features.

In practical applications, the bath is described by a spectral density, such as the Drude–Lorentz form, with parameters including the reorganization energy ($\lambda$), the cutoff rate ($\gamma$), and the temperature ($T$). The corresponding bath correlation functions are written as sums of exponentials. Each exponential generates one level of auxiliary operators, whose couplings capture the exchange of coherence and population between the system and the environment. Truncation at a finite hierarchy depth ($L$) together with a chosen number of Matsubara terms ($K$) yields a controllable approximation that converges to the exact dynamics as these parameters are increased. \texttt{QuTiP} provides an efficient implementation of this method, enabling accurate simulation of excitonic dynamics on physically relevant timescales.

\textbf{What we asked.} We asked {\cinzel El Agente Cu\'antico} to use \texttt{QuTiP}’s HEOM implementation to simulate exciton population dynamics in the FMO complex at two temperatures (300~K and 77~K), propagating for 1~ps and plotting all site populations versus time with an analysis of temperature-dependent coherence. 

\textbf{What this showcases.} This example stresses a ``heavier'' open-quantum-systems workflow: selecting a specialized method (HEOM), executing it with realistic parameters, and extracting physically interpretable data from the computed trajectories.

\textbf{Used agent/tools:} quantum scientist agent, qutip expert agent, literature search, deepsearch, python REPL, submit python jobs and view \texttt{png} file, see supplementary information for detailed description. The answer was verified to be correct.

\prompt{ 
Read \texttt{QuTiP's} manual and use the HEOM implementation to simulate exciton dynamics in the Fenna--Matthews--Olson (FMO) complex. Use the following parameters:
\[
\textbf{System Hamiltonian}\cite{adolphs2006proteins,kreisbeck2011high,kreisbeck2014scalable} =
{\scriptsize
\setlength{\arraycolsep}{3pt}
\begin{pmatrix}
240.0 & -87.7 & 5.5 & -5.9 & 6.7 & -13.7 & -9.9 \\
-87.7 & 315.0 & 30.8 & 8.2 & 0.7 & 11.8 & 4.3 \\
5.5 & 30.8 & 0.0 & -53.5 & -2.2 & -9.6 & 6.0 \\
-5.9 & 8.2 & -53.5 & 130.0 & -70.7 & -17.0 & -63.3 \\
6.7 & 0.7 & -2.2 & -70.7 & 285.0 & 81.1 & -1.3 \\
-13.7 & 11.8 & -9.6 & -17.0 & 81.1 & 435.0 & 39.7 \\
-9.9 & 4.3 & 6.0 & -63.3 & -1.3 & 39.7 & 245.0
\end{pmatrix}
}.
\]
\textbf{Bath parameter:} Drude--Lorentz spectral density with reorganization energy $\lambda = 35~\mathrm{cm}^{-1}$, cutoff time $1/\gamma = 50~\mathrm{fs}$, and temperature $T = 300~\mathrm{K}$ and $T = 77~\mathrm{K}$. Hierarchy depth $L = 4$, number of Matsubara terms $K = 2$.
\textbf{Initial condition:} system localized on site 0.
\textbf{Dynamics:} propagate for $1~\mathrm{ps}$ and record site populations vs.\ time.
\textbf{Report:} Plot all site populations as a function of time at the two different temperatures and analyze the results.
}

\response{

\textbf{Left panel, T = 300 K.} The initially excited Site 0 decays smoothly from population 1.0 to about 0.27 with only weak oscillations. Population spreads broadly across the complex: Sites 1–3 each reach roughly 13–22 percent, and even high-energy Site 5 reaches about 4 percent. The overall behavior is predominantly incoherent and diffusive.\par
\textbf{Right panel, T = 77 K.} Site 0 undergoes a rapid drop in the first $\sim100$ fs followed by clear oscillations and partial recovery, retaining about 46 percent population at 1 ps. Oscillatory features appear across several sites, and population remains concentrated on lower-energy sites (mainly 2 and 3), while higher-energy sites (4–6) stay below 2 percent. The dynamics are more coherent and wavelike than at room temperature.

\textbf{Consistency with Literature}
    \begin{itemize}
        \item The energy funnel behavior toward Site 3 (BChl 3) is well-documented
        \item Temperature-dependent rates agree with theoretical predictions
    \end{itemize}
    
    \begin{center}
\includegraphics[width=0.89\linewidth]{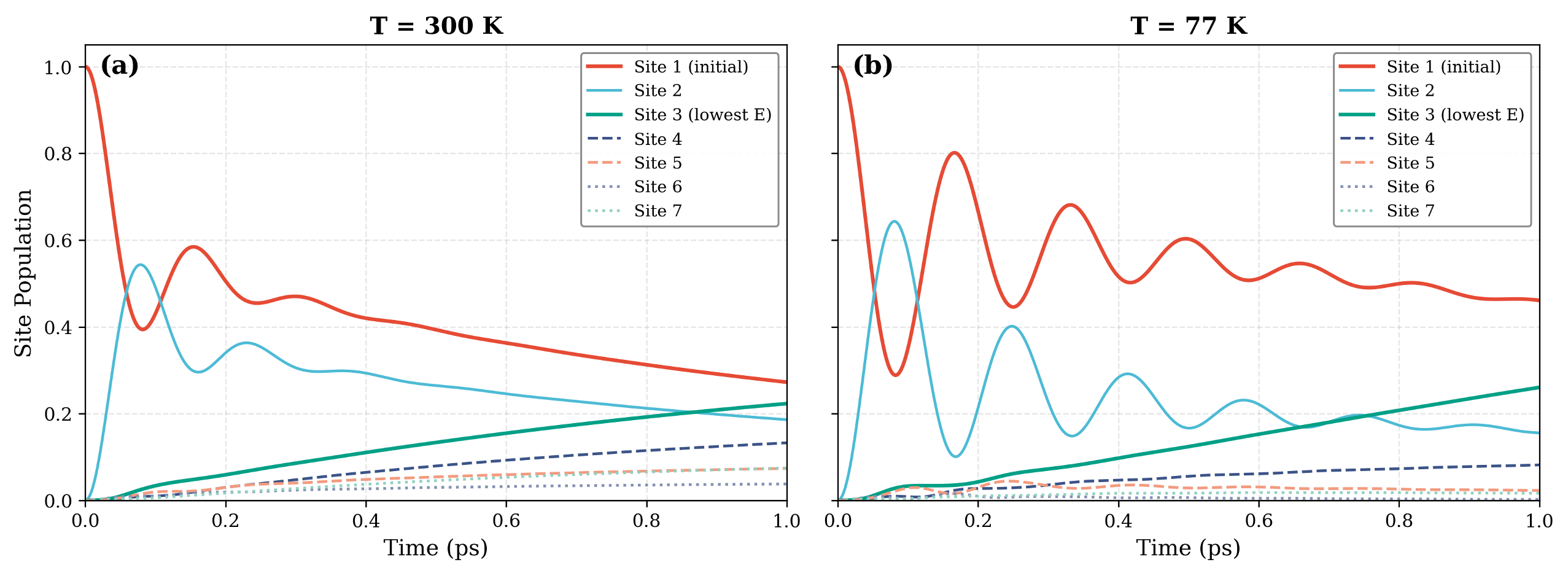}
 \captionof{figure}{Plots generated by {\cinzel El Agente Cu\'antico} simulating exciton dynamics in the FMO complex using HEOM at two temperatures: 300 K (left) and 77 K (right).}
    \end{center} }

\subsection{Time-dependent Hamiltonians}

\subsubsection{Quantum optimal control}
Quantum optimal control provides a principled way to design time-dependent fields that guide a system toward a chosen target state with high fidelity. The Gradient Ascent Pulse Engineering (GRAPE) algorithm \cite{khaneja2005optimal} is widely used because it computes analytic gradients of the fidelity with respect to the controls, thereby enabling efficient optimization over large pulse parametrizations. A Lambda type three-level system offers a simple and instructive platform for this task \cite{vitanov2017stimulated}. The states 
$\ket{1}$, $\ket{2}$, and $\ket{3}$ are connected through two independent control Hamiltonians that couple 
$\ket{1}$ to $\ket{2}$ and $\ket{2}$ to $\ket{3}$. The objective is to maximize population transfer from the initial state $\ket{1}$ to the target state $\ket{3}$ while suppressing occupation of the intermediate state $\ket{2}$. This setting provides a clear benchmark for gradient-based optimization and illustrates how GRAPE identifies high-fidelity solutions that exploit coherent control pathways within the system Hamiltonian.

\textbf{What we asked.} We asked {\cinzel El Agente Cu\'antico} to read  \texttt{QuTiP} manual and solve a GRAPE optimal-control problem for a $\Lambda$-type three-level system, transferring population $|1\rangle\!\to\!|3\rangle$ while minimizing occupation of $|2\rangle$. 

\textbf{What this showcases.} This example shows that the agent can run a full optimization-and-validation loop: mapping a control objective to the correct library routines, handling time-dependent controls, and producing diagnostics that expose both success (high fidelity) and limitations of the chosen cost function.

\textbf{Used agent/tools:} quantum scientist agent, qutip expert agent,  deepsearch, python REPL, submit python jobs and view \texttt{png} file, see supplementary information for detailed description. The answer was verified to be correct.

\prompt{ Read the \texttt{QuTiP} manual and solve a quantum control problem for a $\Lambda$-type three-level system (levels $\ket{1}$, $\ket{2}$, $\ket{3}$). The goal is to transfer population from the ground state $\ket{1}$ to the target state $\ket{3}$ while minimizing the occupation of the intermediate state $\ket{2}$. 

\textbf{Control Hamiltonians:} $H_{c1}=(\ket{1}\bra{2}+\ket{2}\bra{1})$ (Pump coupling). $H_{c2}=\ket{2}\bra{3}+\ket{3}\bra{2})$ (Stokes coupling).
\textbf{The Optimization:} Use the GRAPE algorithm. 
\textbf{Duration:} T=40 (arbitrary time units) with time slots = 3000.
\textbf{Validation:} After optimizing the pulses, run a time-evolution simulation using the generated pulses to verify the dynamics.
\textbf{Output:} A plot showing the Control Pulses (Amplitude vs Time) and a plot showing the populations vs Time. Print the final fidelity.}

\response{

\textbf{Left panel (Control pulses):} Both pulses are oscillatory and overlap in time. 

\textbf{Right panel (Populations):} $P_{1}$ (blue): Smoothly depletes from $1 \to 0$. $P_{2}$ (orange): Peaks at $83 \%$ mid-transfer, return to 0. $P_{3}$ (green): build to $\sim 1.0$ (99.99999$\%$ fidelity)

\textbf{Important Note:} The standard GRAPE algorithm optimizes only for final state fidelity, not for minimizing intermediate state occupation during the transfer. The solution found is mathematically optimal for the endpoint but does NOT minimize $\ket{2}$ occupation during evolution.
    \begin{center}
\includegraphics[width=1.0\linewidth]{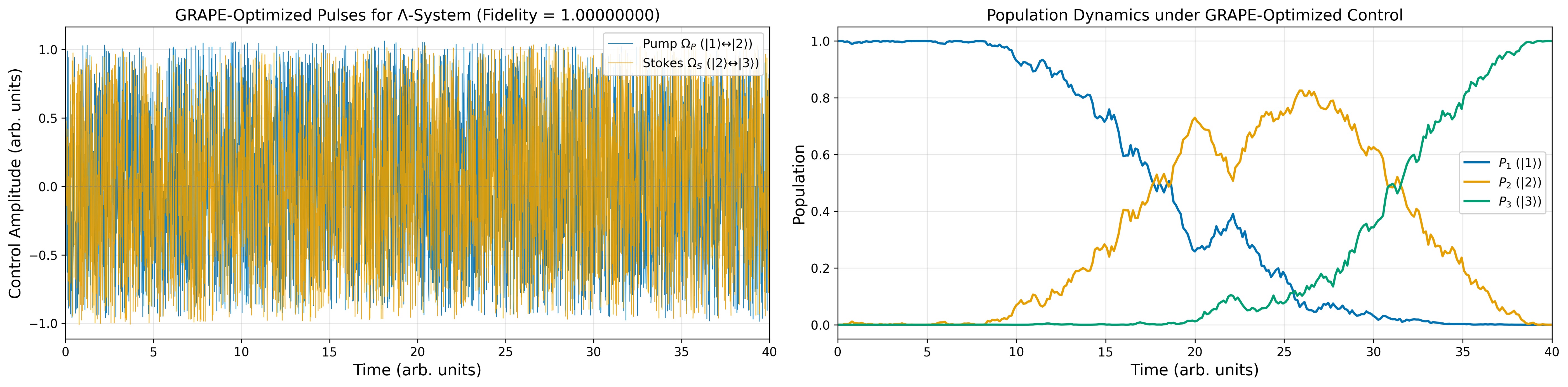}
\captionof{figure}{Plots generated by {\cinzel El Agente Cu\'antico} solving a quantum optimal control problem for a Lambda-type three-level system using the GRAPE algorithm. The left panel shows the control pulses, and the right panel shows the state populations over time.} 
    \end{center} }

\subsubsection{Time-dependent product formulas}

Product formulas can also be used for simulating time-dependent Hamiltonians \cite{wiebe2011simulating}. A common example of these Hamiltonians is Floquet models, such as kicked systems \cite{prosen2002general}, which are periodically updated by instantaneous changes to the Hamiltonian every period $T$. The standard form of such Hamiltonian is given by
\begin{equation*}
    H(t) = H_1 + \sum_{k=-\infty}^\infty \delta(t-kT) H_2.
\end{equation*}
Simulating the dynamics of these systems requires approximating the time-ordered exponential of the Hamiltonian:
\begin{equation*}
    U(t) = \mathcal{T} \exp\left(-i \int_0^t H(t') dt'\right) \approx 
    e^{-i H_1 \tau} \left(e^{-i H_2} e^{-i H_1 T}\right)^{n}, \quad t = nT + \tau
\end{equation*}
Here, we show {\cinzel El Agente Cuántico} approximating the Floquet dynamics of a kicked Ising chain using product formulas implemented in \texttt{CUDA-Q}.

\textbf{What we asked.} We asked {\cinzel El Agente Cu\'antico} to read the \texttt{CUDA-Q} manual and simulate Floquet dynamics of a periodically kicked Ising chain (10 qubits, open boundaries) using time-dependent product-formula methods, then visualize and analyze the resulting driven dynamics. 

\textbf{What this showcases.} This example tests whether the agent can correctly translate a time-ordered exponential problem into an implementable Floquet step (drift + kicks), choose an appropriate simulation strategy in \texttt{CUDA-Q}, and extract time-domain signatures of nontrivial driven behavior.

\textbf{Used agent/tools:} quantum scientist agent, cudaq expert agent,  deepsearch, python REPL, submit python jobs and view \texttt{png} file, see supplementary information for detailed description. The answer was verified to be correct.

\prompt{

Read \texttt{CUDA-Q} manual and simulate the Floquet dynamics of a periodically kicked Ising chain with 10 qubits and open boundary conditions. The system has nearest-neighbor $ZZ$ interactions with coupling $J=1$, and is kicked every $T=1$ by a global transverse field in X. Use a strong kick with strength $h = 2J$ so that the effect of the drive is clearly visible in the dynamics. Initialize the system in a Haar-random state and simulate the evolution for $10$ Floquet periods. Plot a heatmap of $\langle Z_i(t)\rangle$, with qubit index on the x-axis, Floquet step (time) on the y-axis, and color representing the local magnetization.
}

\response{

We observe irregular red--blue patches fluctuating around white, with all
magnetization values remaining very small ($\lVert \langle Z \rangle \rVert < 0.09$). 
There is no discernible spatial pattern or temporal ordering.

\textbf{Physical interpretation.} 
\begin{itemize}
    \item \textbf{Thermalization:} The near-zero magnetization across the entire space--time grid indicates that the system has effectively thermalized to an infinite-temperature state, as expected for chaotic Floquet dynamics.
    \item \textbf{Scrambling:} The rapid, sign-changing fluctuations reflect strong information scrambling. The large kicks ($h = 2J$) drive efficient mixing of quantum states and suppress any coherent structure.
\end{itemize}

The featureless, noise-like fluctuations around zero confirm that the kicked Ising chain 
is deep in the chaotic regime, acting as a quantum thermalizer rather than 
supporting stable or long-lived coherent dynamics.

    \begin{center}
\includegraphics[width=0.6\linewidth]{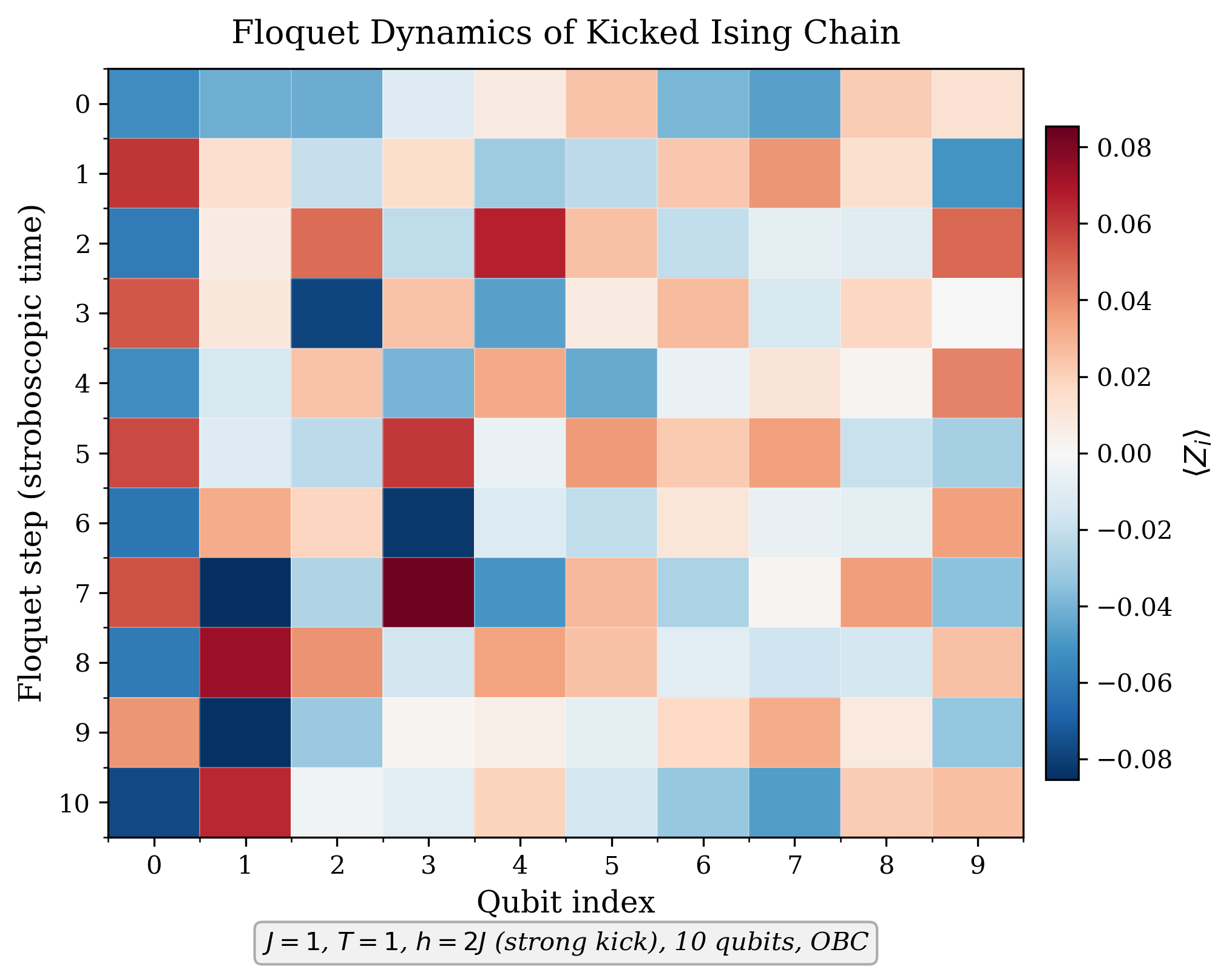}
\captionof{figure}{Plots generated by {\cinzel El Agente Cu\'antico} simulating the Floquet dynamics of a kicked Ising chain using time-dependent product formulas. The heatmap shows the local magnetization $\langle Z_i(t)\rangle$ over space and time.} 
    \end{center}
}

\subsubsection{Time-evolving block decimation (TEBD)}

Many interesting phenomena in quantum dynamics arise from time-dependent Hamiltonians. An example of such is Floquet time crystals (FTC) \cite{else2020discrete,wilczek2012quantum}. These systems are characterized by having a $T$-periodic Hamiltonian $H(t+T)=H(t)$, but there exists an order parameter $O$ with a longer than $T$ period, $\langle O(t+nT)\rangle = \langle O(t)\rangle$ with $n>1$. When the lifetime of these oscillations is finite but long compared to the driving period, the system is called a prethermal Floquet time crystal \cite{else2017prethermal}. A common method to use for simulating the dynamics of one-dimensional quantum many-body systems is time-evolving block decimation (TEBD) \cite{vidal2003efficient, zeng2017prethermal}. Below, we show {\cinzel El Agente Cuántico} simulating the dynamics of a prethermal Floquet time crystal using TEBD.

\textbf{What we asked.} We asked {\cinzel El Agente Cu\'antico} to simulate nonequilibrium dynamics of a 1D many-body system using TEBD (tensor-network time evolution), and to extract observables that diagnose the dynamical phase of interest.

\textbf{What this showcases.} This example probes whether the agent can select and configure a tensor-network method, using \texttt{TeNPy}, appropriate for larger 1D systems, manage time stepping and truncation controls, and connect the computed observables to a clear physical conclusion (rather than only producing raw plots).

\textbf{Used agent/tools:} quantum scientist agent, tenpy expert agent, literature search,  deepsearch, python REPL, submit python jobs and view \texttt{png} file, see supplementary information for detailed description. The answer was verified to be correct.

\prompt{

Simulate Floquet dynamics with TD-DMRG for a 1D disordered spin-1/2 chain. Use a periodically driven Hamiltonian with period $T=\tau_1+\tau_2$. During $0<t<\tau_1$, evolve under a strong global $x$-field plus interactions and disorder:
$H = \sum_i \Omega_x S_i^x + \Delta_i S_i^z + J(S_i^xS_{i+1}^x + S_i^yS_{i+1}^y - S_i^zS_{i+1}^z)$.
During $\tau_1<t<T$, switch to a global $y$-field,
$H = \sum_i \Omega_y(1-\varepsilon) S_i^y + \Delta_i S_i^z + H_{i,i+1}$.
Choose $\Omega_x\tau_1 \approx \pi$ so each period approximately flips the $x$-polarization. Use the following parameters: system size $L=10$ spins, maximum bond dimension $\chi_{\max}=100$, time step $\delta t=0.03$, evolve for $N_{\text{periods}}=50$ Floquet periods, average over $N_{\text{disorder}}=5$ disorder realizations, and scan $\varepsilon \in \{0, 0.15, 0.3\}$. Keep the Hamiltonian parameters as before: $J=1$, $\Omega_x = \Omega_y = 4J$, $\tau_1 = \pi/(2\Omega_x) \approx 0.393$, $\tau_2 = 0.5$, and disorder strength $W=2J$. 

Initialize the system in a product state polarized along $+x$. Evolve stroboscopically for many periods using TD-DMRG (TEBD or TDVP), keeping truncation errors under control. Use disorder $\Delta_i\in(-W,W)$ and average over several realizations. After each Floquet period, measure the bulk $x$-magnetization and plot it versus the number of periods. Vary $\varepsilon$ to identify regimes with long-lived period-doubling (subharmonic $2T$ oscillations) versus rapid decay due to heating. Track entanglement growth and discuss how it limits the reachable simulation time.
}

\response{  

\textbf{Panel (a) - Magnetization:} Stroboscopic $\langle S^{x}\rangle$ vs Floquet period with clear period-doubling oscillations. The alternating pattern demonstrates discrete-time crystal behavior, with the decay depending on the imperfection $\varepsilon$.

\textbf{Panel (b) - Entanglement:} Slow logarithmic growth of entanglement entropy $S$, confirming MBL protection ($S \approx 2.0 << 3.47$ thermal limit).

Three curves for $\varepsilon \in {0, 0.15, 0.3}$ with shaded error bands ($\pm 1 \sigma$ from disorder averaging).

This simulation provides direct numerical evidence for the MBL-protected discrete time crystal phase: a driven quantum system that spontaneously oscillates at twice the driving period, protected from thermalization by many-body localization.

\begin{center}
    \includegraphics[width=0.9\linewidth]{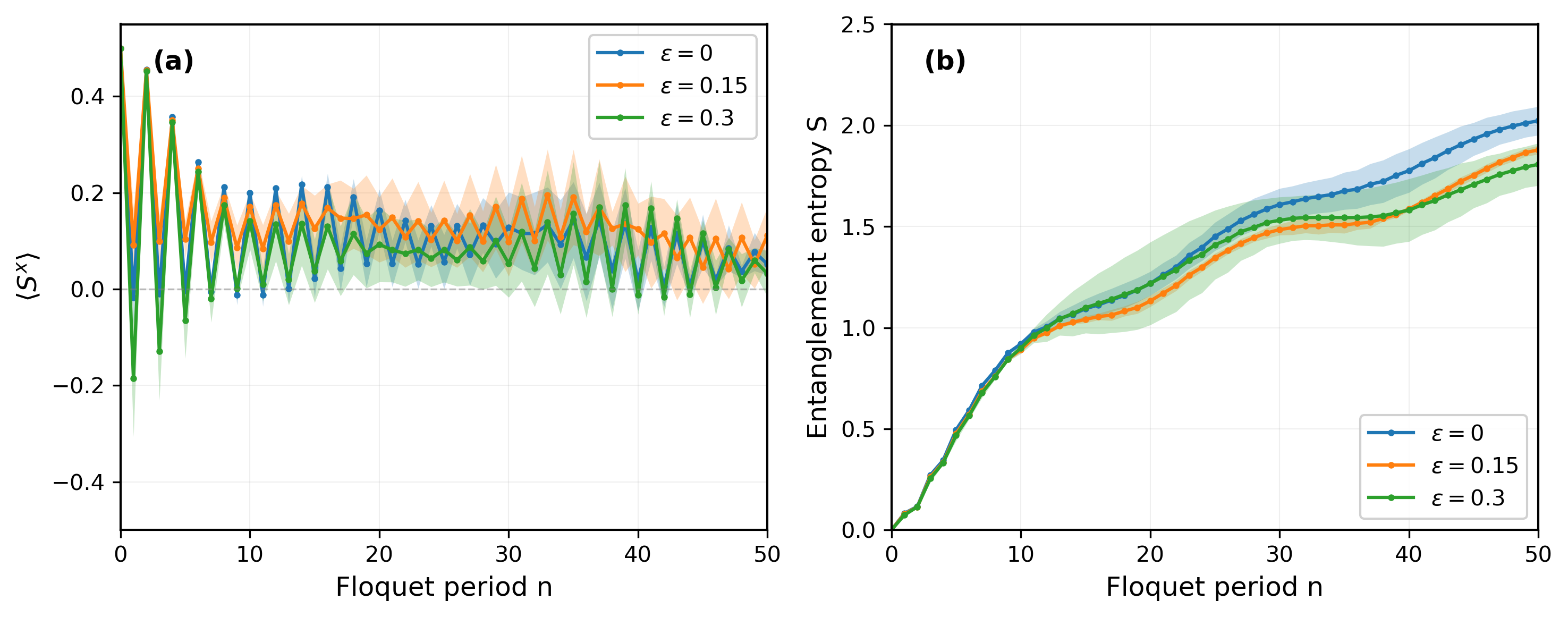}
    \captionof{figure}{Plots generated by {\cinzel El Agente Cu\'antico} simulating a prethermal Floquet time crystal using TD-DMRG (TEBD). Panel (a) shows the stroboscopic magnetization exhibiting period-doubling oscillations, and panel (b) shows the entanglement entropy growth over time.}
    \end{center}}

\section{Reproducibility benchmark}\label{sec:benchmark}

To evaluate the robustness and reproducibility of {\cinzel El Agente Cu{\'a}ntico}, we perform a systematic benchmark based on repeated, independent executions of representative quantum-simulation tasks. The aim of this benchmark is to assess consistency under identical prompts, rather than to optimize peak performance. We consider some of the problems introduced in the previous section, including variational quantum eigensolver (VQE) (Sec.~\ref{sec:VQE}), Bell-state preparation (Sec.~\ref{sec:Bell}), transverse-field Ising dynamics (Sec.~\ref{sec:ising}), open-system Lindblad dynamics (Sec.~\ref{sec:lindblad}), and hierarchical equations of motion (HEOM) (Sec.~\ref{sec:HEOM}). For the VQE benchmark, we deliberately use the \texttt{Tequila} framework instead of \texttt{CUDA-Q} in order to test the agent’s performance across a different software stack.

Each task is executed ten times using the same prompt, with all runs fully independent and no shared memory or persistent context between executions. Each run is evaluated by a human expert using a structured rubric that assesses three aspects, namely correctness of the implementation, quality and validation of the results, and clarity of the final report, with a maximum score of 100 points per run. Partial credit is assigned where appropriate. Five evaluators, consisting of two postdoctoral researchers and three PhD students, participate in the assessment, with a single evaluator assigned consistently to each benchmark task across all repetitions. The full evaluation rubric and detailed point assignments for each benchmark task are provided in the Supporting Information (Sec.~\ref{sec:eval_rubrics}).

Table~\ref{tab:benchmark_summary} summarizes the benchmark results. Across all tasks, the agent attains consistently high scores over repeated executions, demonstrating robust and reproducible performance across a diverse set of quantum-simulation problems and software frameworks. While the specific code generated by the agent may differ between runs, this variability reflects the existence of multiple valid implementations for the same physical task and does not affect the correctness of the results. Perfect reproducibility is observed for the VQE and HEOM benchmarks, where all repetitions yield fully correct implementations and analyses.

The deviations from the maximum score can be attributed to isolated and well-defined issues that do not compromise the validity of the overall workflows. In the Bell-state benchmark, Run 10 produced correct data and executable code but exhibited minor visualization problems, including a missing axis label and overlapping annotations for which we deducted 20 points. In the Lindblad benchmark, Run 5 computed $\langle \sigma_y(t) \rangle$ instead of the requested $\langle \sigma_z(t) \rangle$, for which a substantial score reduction was applied even though the dynamical simulation and physical interpretation were otherwise correct. For the transverse-field Ising benchmark, some variability across runs is expected because the dynamics originate from random initial states. Nonetheless, in Run 4 and Run 6 the initial states were random but not strictly Haar distributed, which motivated a small score reduction despite consistent physics and numerical evolution across repetitions.

\begin{table}[t]
\centering
\scriptsize
\caption{Reproducibility benchmark for {\cinzel El Agente Cu{\'a}ntico}.
Each benchmark task was executed independently ten times using identical prompts.
Scores correspond to expert evaluation using a structured rubric with a maximum
score of 100 per run. The detailed grading criteria and evaluation rubric are provided
in the Sec.~\ref{sec:eval_rubrics} of the Supporting Information.}
\label{tab:benchmark_summary}
\resizebox{\textwidth}{!}{
\begin{tabular}{l c c c c c c c c c c c c }
\hline
\textbf{Task} & \textbf{Software} &
\textbf{R1} & \textbf{R2} & \textbf{R3} & \textbf{R4} & \textbf{R5} &
\textbf{R6} & \textbf{R7} & \textbf{R8} & \textbf{R9} & \textbf{R10} &
\textbf{Average Score}  \\
\hline
VQE  & \texttt{Tequila}
& 100 & 100 & 100 & 100 & 100 & 100 & 100 & 100 & 100 & 100 & 100  \\
Bell state  & \texttt{CUDA-Q}
& 100 & 100 & 100 & 100 & 100 & 100 & 100 & 100 & 100 & 80 & 98  \\
Ising  & \texttt{CUDA-Q}
& 100 & 100 & 100 & 90 & 100 & 90 & 100 & 100 & 100 & 100 & 98  \\
Lindblad  & \texttt{QuTiP}
& 100 & 100 & 100 & 100 & 70 & 100 & 100 & 100 & 100 & 100 & 97   \\
HEOM & \texttt{QuTiP}
& 100 & 100 & 100 & 100 & 100 & 100 & 100 & 100 & 100 & 100 & 100  \\
\hline
\end{tabular}
}
\end{table}

\section{Beyond quantum simulation}\label{sec:beyond}
Solving the Schrödinger equation, or an approximation of it, is often a step in a larger problem. In this section, we demonstrate the capabilities of {\cinzel El Agente Cuántico} in solving problems beyond quantum simulation. This includes, for example, constructing phase diagrams of quantum systems, combining multiple toolkits to study quantum systems, estimating qubit requirements for quantum simulation and quantum error correction.
\subsection{Phase diagrams of quantum systems}

Many-body quantum systems exhibit rich phase diagrams that can be explored through quantum simulation. Often, an order parameter is used to characterize the system's different phases. An interesting example is found in the study of computational phases of quantum matter \cite{doherty2009identifying, miyake2010quantum}. For example, it was shown that for the one-dimensional transverse field cluster Hamiltonian defined by,
\begin{equation*}
    H = -\cos \theta \sum_{i=1}^{N-1} Z_i X_{i+1} Z_{i+2} - \sin \theta \sum_{i=1}^{N} X_i,
\end{equation*}
the string-order parameter $\langle ZXIX \cdots IXZ \rangle$ can be used to identify if the ground state can be used for measurement-based quantum computation (MBQC) \cite{raussendorf2023measurement}. We show {\cinzel El Agente Cuántico} using \texttt{tenpy} \cite{tenpy2024} to compute the phase diagram of this model and estimate the computational phase transition with a simple but complete prompt below.

\textbf{What we asked.} We asked {\cinzel El Agente Cu\'antico} to use DMRG to compute the ground state of the 1D transverse-field cluster Hamiltonian (ring geometry), evaluate the nonlocal string-order parameter $ZXIX\cdots IXZ$ for system sizes $N=11,21,41$, plot the order parameter versus coupling, and estimate the critical point. 

\textbf{What this showcases.} This example tests autonomy on an end-to-end many-body study: selecting a tensor-network ground-state algorithm, implementing a nonlocal diagnostic observable, performing finite-size comparisons, and extracting a phase-transition estimate from computed data.

\textbf{Used agent/tools:} quantum scientist agent, tenpy expert agent, literature search,  deepsearch, python REPL, submit python jobs and view \texttt{png} file, see supplementary information for detailed description. The answer was verified to be correct.

\prompt{Use DMRG to prepare the ground state of the one-dimensional transverse field cluster Hamiltonian over a ring of $N$ qubits (interpolate with cos and sin between these cluster Hamiltonian and the magnetic field). Compute the string-order parameter $ZXIX\cdots IXZ$ for multiple strengths of the cluster Hamiltonian over 11, 21, and 41 qubits. Plot the string-order parameter as a function of the cluster Hamiltonian strength for the different system sizes and estimate the critical point.}

\response{

I implemented DMRG simulations (tolerance $10^{-5}$) for the transverse field cluster Hamiltonian. 

\textbf{Physical Interpretation:} $\bullet$ SPT Phase ($\theta < \theta_c$): String order $O \approx 1$, indicating symmetry-protected topological order $\bullet$ Trivial Phase ($\theta > \theta_c$): String order $\theta \rightarrow 0$, no topological order $\bullet$ Critical Point: $\theta_c / \pi \approx 0.236$ (extrapolated), close to theoretical $\pi/4 = 0.25$ $\bullet$ The transition sharpens with increasing system size, consistent with a quantum phase transition.

    \begin{center}
    \includegraphics[width=1.0\linewidth]{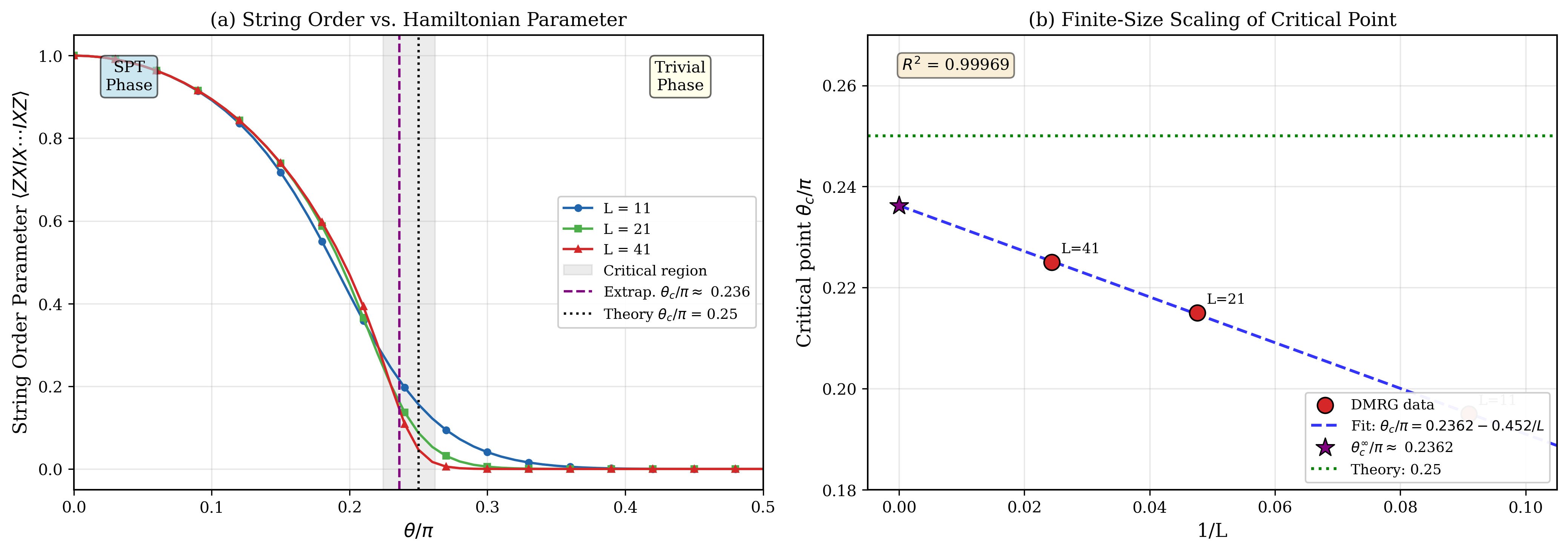}
    \captionof{figure}{Plots generated by {\cinzel El Agente Cu\'antico} simulating the computational phase diagram of the transverse field cluster Hamiltonian using DMRG.}
    \end{center}
}


\subsection{Bell-state correlations under depolarizing noise}
Bell states provide a minimal setting to probe how entanglement degrades under realistic noise channels.
Whereas the previous example considered an ideal implementation using \texttt{CUDA-Q}, here we test whether {\cinzel El Agente Cuántico} can use the \texttt{Qiskit} framework to prepare the same Bell state and evaluate how depolarizing noise affects multi-basis quantum correlations.

\textbf{What we asked.} We asked {\cinzel El Agente Cu\'antico} to simulate Bell-state measurements in the presence of depolarizing noise and analyze how noise degrades correlations, including a comparison between measurement bases. 

\textbf{What this showcases.} This example highlights cross-tool robustness and ``glue'' reasoning: the agent must find a backend that actually supports noisy simulation in the installed environment, run the experiment, and interpret basis-dependent degradation as an accumulated-gate-error effect.

\textbf{Used agent/tools:} quantum scientist agent, qiskit expert agent, literature search,  deepsearch, python REPL, submit python jobs and view \texttt{png} file, see supplementary information for detailed description. The answer was verified to be correct.

\prompt{ Read the \texttt{Qiskit} manual and prepare the Bell state $(|00\rangle + |11\rangle)/\sqrt{2}$. Start in $|00\rangle$, apply a Hadamard gate on qubit 0, and then a CNOT with control qubit 0 and target qubit 1. Add a depolarizing noise channel with probability $p$ to both qubits after the CNOT gate. Simulate the circuit for $p \in \{0, 0.05, 0.1, 0.2, 0.3\}$. For each value of $p$, run 4096 shots in the $Z$ basis, return the measurement counts, and compute $\langle Z \otimes Z \rangle$. Then insert Hadamard gates on both qubits to measure in the $X$ basis, run 4096 shots, return the measurement counts, and compute $\langle X \otimes X \rangle$. Plot both correlators as functions of $p$ and comment on how the noise degrades the Bell-state correlations.}

\response{

\textbf{Physical interpretations.}  The measurement count plots visually demonstrate that $\bullet$ ideal Bell state produces only correlated outcomes ($\ket{00}$, $\ket{11}$) $\bullet$ Depolarizing noise introduces anti-correlated outcomes ($\ket{01}$, $\ket{10}$) $\bullet$ High noise limit: Distribution approaches uniform (maximally mixed state) $\bullet$ X-basis degradation is 2-3x faster due to additional Hadamard gates also experiencing noise. 

    \begin{center}
\includegraphics[width=0.8\linewidth]{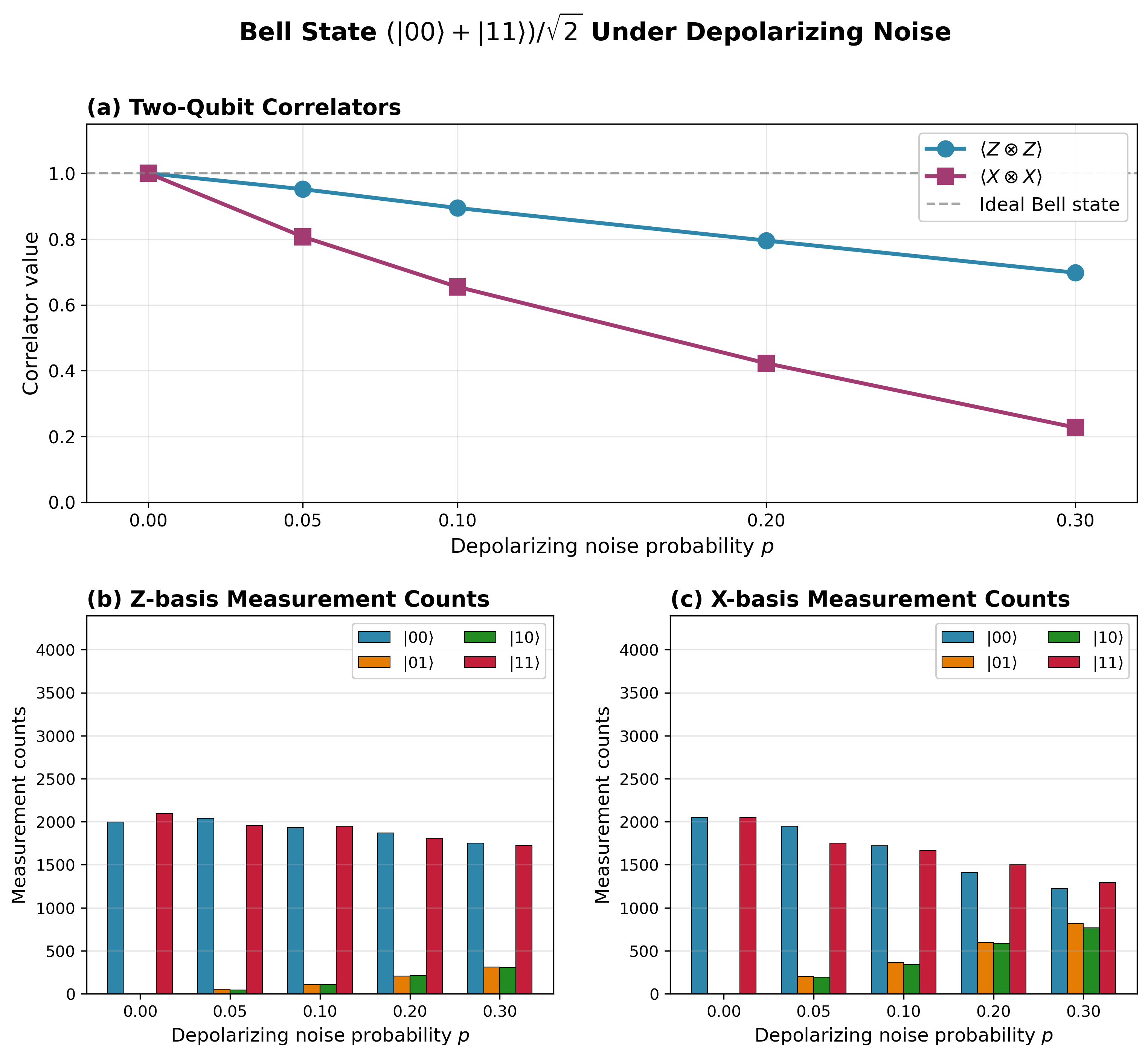}
\captionof{figure}{Plots generated by {\cinzel El Agente Cu\'antico} simulating Bell-states under depolarizing noise. The top plot shows 2-qubit correlations as a function of noise strength. The bottom plots show the tomographic measurement counts for different noise levels in both Z and X bases.}
    \end{center}}

\subsection{Estimating qubit requirements for quantum simulation}

Quantum computation is an emerging technology for simulating quantum systems. One of the most promising algorithms for estimating ground-state energies is quantum phase estimation (QPE) \cite{kitaev1995quantum, nielsen_chuang_2010}, relevant for quantum chemistry applications \cite{aspuru2005simulated}. Below, we show the general structure of QPE for estimating the ground-state energy of a Hamiltonian $H$ using a unitary operator $U = e^{-iHt}$ and an input state $|\psi\rangle$ with non-zero overlap with the ground state $|E_0\rangle$:

\textbf{What we asked.} We asked {\cinzel El Agente Cu\'antico} to read the PennyLane manual, build the second-quantized electronic Hamiltonian of water (STO-3G), map it to qubits (Jordan--Wigner), construct a logical (non-fault-tolerant) QPE circuit targeting chemical accuracy, and use PennyLane’s resource-estimation tools to estimate logical qubits and non-Clifford gate counts from Hamiltonian-simulation segments. 

\textbf{What this showcases.} This example tests whether the agent can move from ``simulate a system'' to ``estimate feasibility'': assembling a chemistry Hamiltonian, selecting a canonical algorithm (QPE), and reporting meaningful logical-level space/time costs with clear scaling interpretation.

\textbf{Used agent/tools:} quantum scientist agent, pennylane expert agent, literature search,  deepsearch, python REPL, submit python jobs and view \texttt{png} file, see supplementary information for detailed description. The answer was verified to be correct.

\begin{center}
\adjustbox{width=.7\textwidth}{%
\begin{quantikz}
    \lstick[wires=4]{$n\ \text{qubits}$}  \midstick{$\ket{0}$}  & \gate{{\rm H}} & \ctrl{4} & \qw &\qw& \qw &\qw    & \gate[wires=4]{{\rm QFT}^\dagger} & \meter{}  \\
                                     \midstick{$\ket{0}$}  &  \gate{{\rm H}}  & \qw           &\qw & \ctrl{3} & \qw &\qw &  \qw & \meter{}  \\ 
                                      \midstick{$\vdots$} &  \setwiretype{n} &&  & & \midstick{$\vdots$}  & &&\midstick{$\vdots$} &   \\ 
                                     \midstick{$\ket{0}$}  & \gate{{\rm H}} & \qw  & \qw    & \qw &\qw        & \ctrl{1} &  \qw & \meter{}  \\ 
    \lstick{$\ket{\psi}$} &  \qwbundle[s]{m}    &\gate{U^{2^{0}}} &        &\gate{U^{2^{1}}}    &\ \ldots\,          &\gate{U^{2^{n-1}}}  &\qw         & \rstick{$\lvert E_0\rangle$}
\end{quantikz}
    }
\end{center}

However, implementing QPE requires significant quantum resources, commonly measured in terms of logical qubits and non-Clifford gates (e.g., $T$ gates), which measure the space and gate complexity, respectively. Below, we show {\cinzel El Agente Cuántico} estimating these resources for simulating the water molecule in the STO-3G basis using \texttt{PennyLane} \cite{bergholm2018pennylane}.

\prompt{ Read the \texttt{PennyLane} manual and construct the second-quantized electronic Hamiltonian of water in the STO-3G basis. Use the Jordan-Wigner transformation to map the Hamiltonian to a qubit representation. Then build a logical (non–fault-tolerant) quantum phase estimation (QPE) circuit that would estimate the ground-state energy of water within chemical accuracy, and use \texttt{PennyLane}'s resource-estimation tools to approximate: (1) the number of logical qubits required and (2) the number of non-Clifford (e.g., $T$) gates arising from the Hamiltonian-simulation segments of the QPE routine. Report and discuss these logical-level resource estimates.
}

\response{

\textbf{(a) Left panel, qubit breakdown.} 14 system (4.8$\%$) vs 276 ancilla (95.2$\%$) = 290 total.

\textbf{(b) Right panel, required Toffoli gates.} Toffoli gates vs error (log-scale), showing $\mathcal{O}(1/\epsilon)$ scaling with chemical accuracy marked in green. 

For water QPE at chemical accuracy, you need 290 logical qubits and $\sim$112M Toffoli gates.
    \begin{center}
    \includegraphics[width=0.9\linewidth]{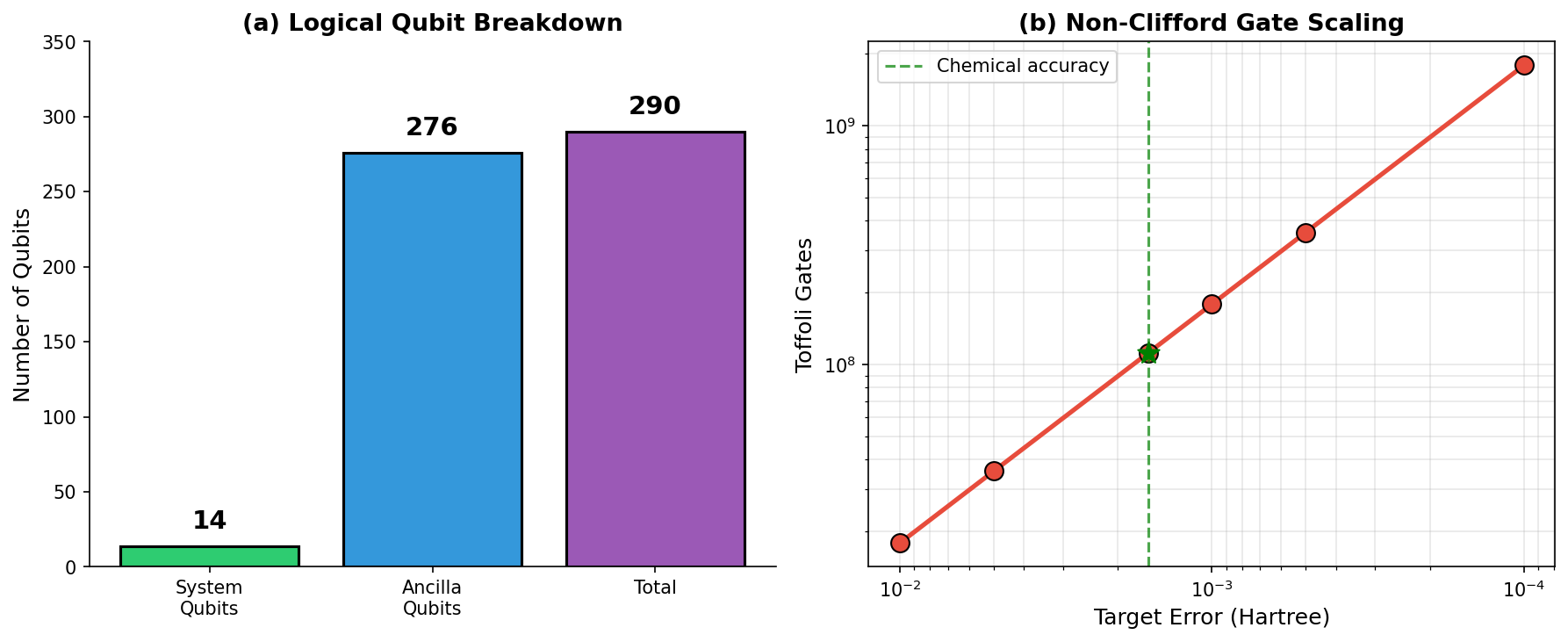}
    \captionof{figure}{Plots generated by {\cinzel El Agente Cu\'antico} estimating space-time resources for QPE of the water molecule. (Left) Qubit breakdown. (Right) Required Toffoli gates for various target errors.}
    \end{center}
}

\subsection{Quantum error correction}

Physical qubits are highly susceptible to noise arising from imperfect control, measurement errors, and decoherence, which can rapidly destroy stored information. Quantum error correction (QEC) overcomes this limitation by encoding logical quantum information into correlated states of multiple physical qubits, enabling errors to be detected and corrected without directly measuring the encoded state. Since its introduction in the mid-1990s, QEC has become the central component of fault-tolerant quantum computing, establishing that arbitrarily long quantum computations are possible provided physical error rates remain below a threshold value~\cite{shor1995scheme,steane1996error,knill1998resilient,devitt2013quantum}.

Among the many QEC schemes developed to date, surface codes have emerged as one of the leading candidates for large-scale implementations due to their high threshold value. These codes arrange qubits on a two-dimensional lattice and rely only on local measurements, making them well-suited to current hardware architectures. This family of codes, when using an appropriate decoder such as minimum-weight perfect matching (MWPM)~\cite{Higgott2025sparseblossom} or Union-Find (UF)~\cite{delfosse2021almost}, enables logical error rates to be exponentially reduced by increasing the code's distance~\cite{kitaev2003fault,fowler2012surface}. Recent experimental progress has begun to demonstrate logical memories whose performance improves with system size, showing promise toward practical fault-tolerant quantum devices~\cite{google2023suppressing,bluvstein2024logical, google2025quantum, aghaee2025scaling,zhao2022realization}.

\textbf{What we asked.} We asked {\cinzel El Agente Cu\'antico} to implement a surface-code logical $Z$-basis memory experiment using Stim , for distances $d=3,5,7$, rounds $R=d$, and $3\times10^6$ shots per distance, with two-qubit depolarizing noise ($p=0.003$) applied on every CNOT, then compute the logical observable $Z_L$ from final data-qubit parities. 

\textbf{What this showcases.} This example evaluates whether the agent can execute a modern QEC workflow end-to-end: selecting a decoder-ready simulation backend, running large-shot Monte Carlo experiments, computing logical observables, and connecting results to the expected distance-scaling behavior of logical error rates.

\textbf{Used agent/tools:} quantum scientist agent, cudaq expert agent,  deepsearch, python REPL, submit python jobs and view \texttt{png} file, see supplementary information for detailed description. The answer was verified to be correct.

\prompt{ Implement and run a surface-code logical memory experiment in the Z basis using \texttt{CUDA-Q} QEC with the \texttt{Stim} backend on CPU. Use the following parameters:

\textbf{Distances:} d = 3, 5, 7. \textbf{Rounds:} R = d stabilizer-measurement rounds. \textbf{Shots:} N = 3000000 per distance \textbf{State:} prepare logical $\ket{0_{L}}$ (Z-basis memory). \textbf{Noise:} apply two-qubit depolarizing noise with probability p = 0.003 on every CX (CNOT) gate only. \textbf{Sampling:} for each shot, sample \texttt{Stim} detector events and final data-qubit Z measurements. \textbf{Logical observable:} compute measured $Z_{L}$ as the parity of the final data-qubit Z outcomes along the chosen logical $Z_{L}$ string. \textbf{Decoding:} build the \texttt{Stim}-generated detector error model (DEM) and decode using \texttt{PyMatching} (MWPM). \textbf{Plots:} report a single plot of d vs logic error gate for the data with and without the decoder, and add theoretical estimation.
}

\response{ The figure shows the logical error rate versus code distance for a Z-basis surface-code memory, comparing raw measurements (red), MWPM-decoded results (blue), and a threshold-model prediction (green). Without decoding, the logical error rate increases with distance because larger codes require more physical qubits, CNOT gates, and measurement rounds, leading to accumulated errors that scale approximately as \(p_{\mathrm{raw}} \sim d^{3}\). In contrast, decoding produces a rapidly decreasing logical error rate with distance, consistent with correction of up to \(\lfloor (d-1)/2 \rfloor\) errors and the expected scaling \(p_{L} \sim (p/p_{\mathrm{th}})^{(d+1)/2}\). Quantitatively, the decoded logical error drops from \(0.193\%\) at \(d=3\) natural-language request can become an end-to-end pipeline that spans model definition, execution, and analysis across circuit simulators, open-system solvers, tensor network methods, and resource-estimation suppression as distance increases.

    \begin{center}
    \includegraphics[width=0.7\linewidth]{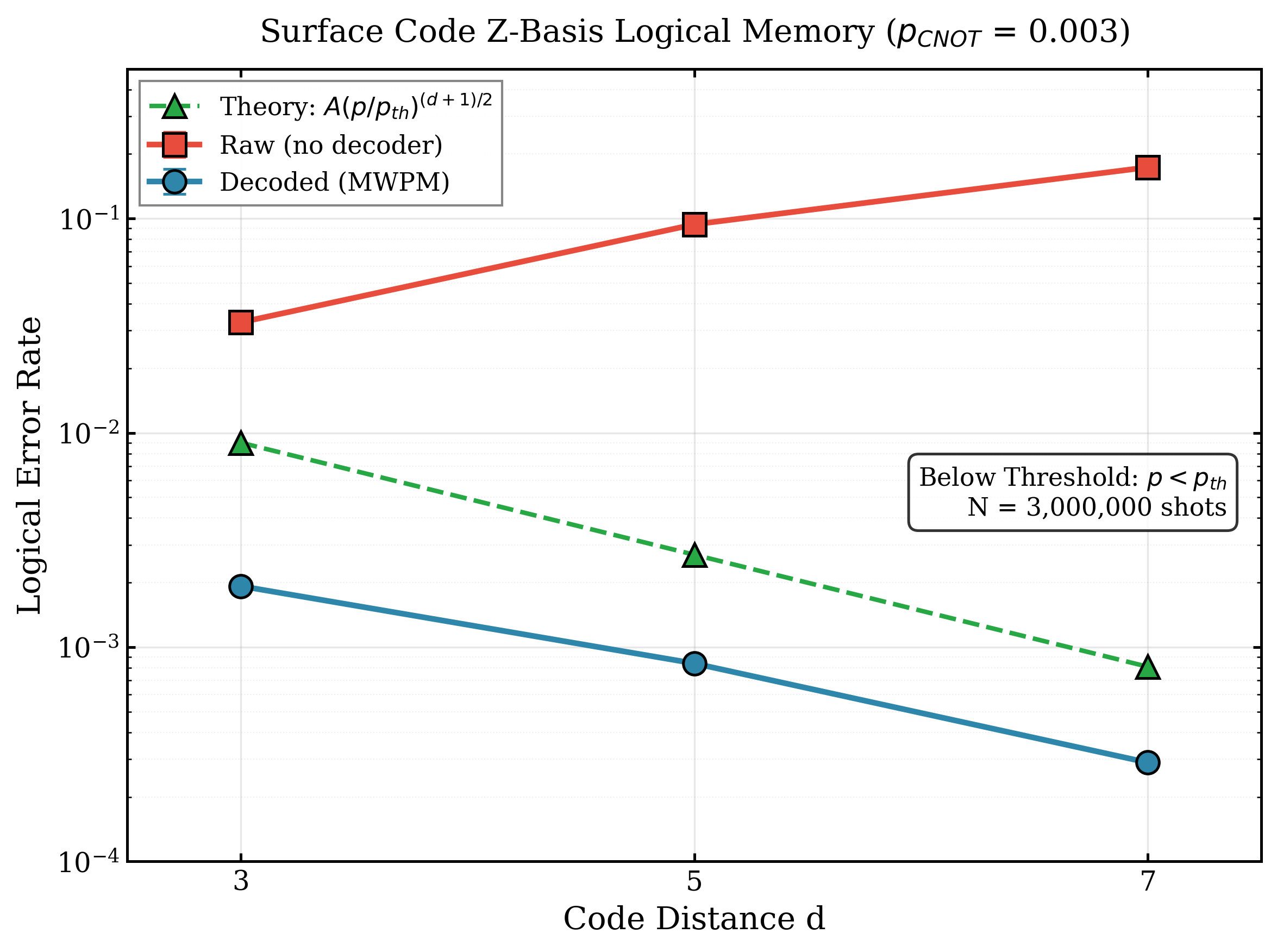}
    \captionof{figure}{Plot generated by {\cinzel El Agente Cu\'antico} showing the exponential suppression of logical error rates in a surface-code quantum memory as a function of code distance, both with and without decoding.}
    \end{center}
}

\section{Discussion}\label{sec:discussion}

In this work, we showed how {\cinzel El Agente Cu\'antico} can simulate quantum-mechanical systems using a wide range of tools while keeping the workflow close to how researchers frame questions. A single natural-language request can become an end-to-end pipeline that spans model definition, execution, and analysis across circuit simulators, open-system solvers, tensor network methods, and resource-estimation utilities. This positions the agent in a useful role between scientific intent and the practical steps required by different quantum software stacks.

A central observation emerging from these experiments is that the reliability and efficiency of autonomous execution depend strongly on the clarity and internal coherence of available software documentation. When reference materials present well-defined usage patterns, consistent terminology, and sufficiently explicit examples, the agent can translate physical objectives into appropriate computational abstractions with little iteration, allowing its reasoning effort to remain focused on scientific modeling decisions rather than on reconstructing intended interfaces. When documentation instead contains ambiguous guidance, underspecified examples, or overlapping conventions, the agent must infer correct usage through iterative interaction with the software, relying on runtime feedback to resolve uncertainty. Although this adaptive process often succeeds, it introduces additional overhead and variability in execution paths. 

These observations highlight an important systemic relationship between autonomous scientific agents and the broader software ecosystems in which they operate. Clear, well-maintained documentation and consistent interface conventions amplify the agent’s ability to reason effectively across tools and domains. In this sense, {\cinzel El Agente Cuántico} not only benefits from mature software infrastructure but also actively exercises it, revealing how scientific knowledge, software design, and executable practice interact in an integrated workflow. As agent-based systems become more prevalent, this interaction may encourage documentation practices that are simultaneously optimized for human users and machine-mediated reasoning.

Furthermore, the results demonstrate the capacity of our agentic framework to unify simulation methodologies that are traditionally treated as distinct. Circuit-based algorithms, open quantum system dynamics, tensor network techniques, and fault-tolerant resource estimation are all accessed through a common natural language interface, despite their differing mathematical formalisms and implementation paradigms. This unification lowers the conceptual overhead required to move between methods and enables researchers to explore complementary approaches to the same physical problem within a single coherent workflow.

Taken together, these results suggest a shift in how quantum-simulation expertise can be expressed and deployed. Rather than requiring deep familiarity with every library, syntax, and backend, researchers can articulate goals at the level of physical models and observables. At the same time {\cinzel El Agente Cuántico} translates those goals into executable implementations across heterogeneous simulation tools. In this role, the agent functions as an enabling layer for quantum simulation rather than a replacement for human expertise, complementing scientific intuition by reducing technical friction, integrating diverse methods, and supporting iterative exploration. This reframing has the potential to accelerate exploratory research, support rapid hypothesis testing, and make advanced simulation techniques more accessible across disciplinary boundaries. Building on this foundation, the roadmap presented in the following subsection outlines how these capabilities can be extended toward richer forms of autonomy, coordination, and scientific discovery.


\subsection{ {\cinzel El Agente Cu\'antico} roadmap}\label{sec:roadmap}


In Figure \ref{fig:timeline} we summarize a staged roadmap for the evolution of {\cinzel El Agente Cuántico} from task-level automation toward fully autonomous scientific discovery. The current work establishes Stage 0 (Automated Quantum Simulation), in which natural-language prompts are translated into quantum-simulation workflows exploiting the state of the art software stacks~\cite{QuTiP2025,cuda-q_2025_15407754,tenpy2024,Kottmann_Tequila_A_platform_2020,bergholm2018pennylane,gidney2021stim,javadiabhari2024quantumcomputingqiskit}. This capability builds directly on recent advances in reasoning models and scientific agents that map high-level scientific intent to concrete computational actions~\cite{lu2024ai, yamada2025ai,mitchener2025kosmos,wang2025starwhisper, elagenteOG}.

\begin{center}
\includegraphics[width=1.0\linewidth]{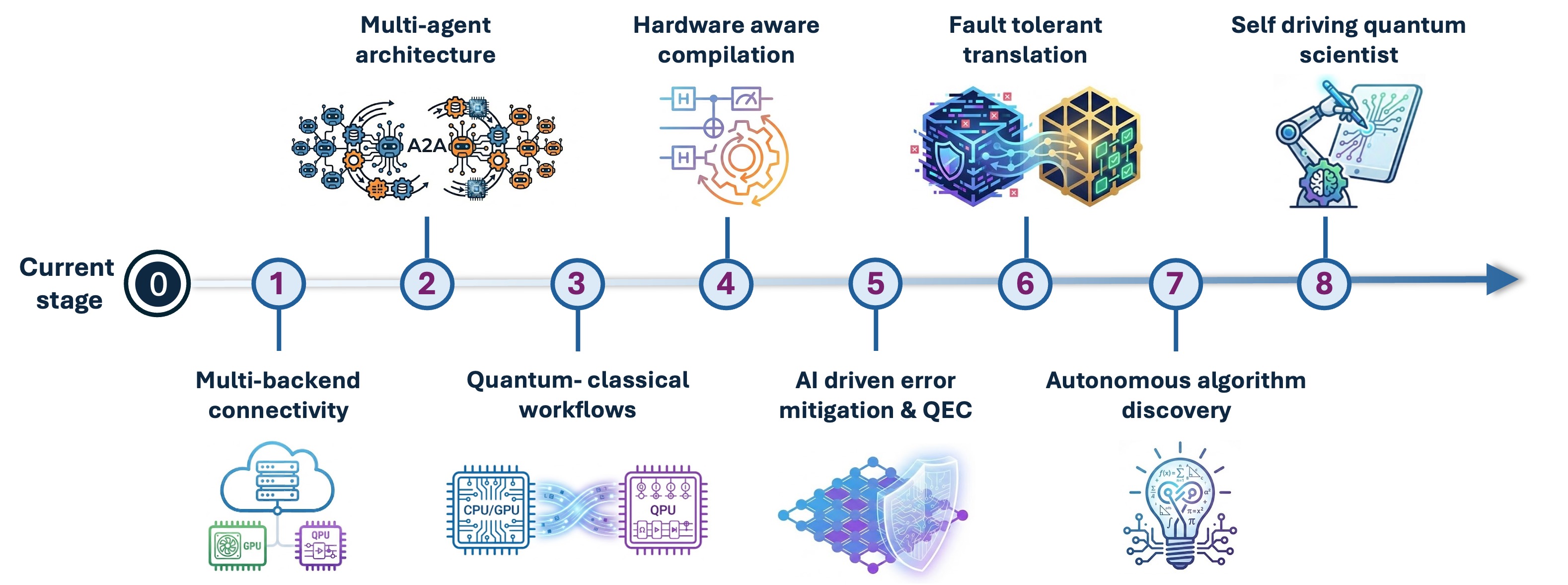} 
\captionof{figure}{Future stages required to develop {\cinzel El Agente Cuántico} from an automated quantum simulation tool toward a self-driving quantum scientist capable of closed-loop hypothesis generation, experimental execution, and result interpretation.}
\label{fig:timeline}
\end{center}

Subsequent development will introduce multi-backend connectivity (Stage 1), enabling seamless execution across heterogeneous computing resources, including cloud servers, GPUs, and emerging QPU platforms, with backend selection guided by the problem structure and hardware availability. As an immediate next step toward this goal, we will release a cloud-enabled alpha version of {\cinzel El Agente Cuántico}, providing web-based access to its core automated quantum-simulation capabilities via \urlhref{https://elagente.ca}{elagente.ca}.
This capability will be complemented by a multi-agent scientific architecture (Stage 2), in which specialized domain agents such as quantum chemistry, materials, and drug-discovery agents developed within the Matter Lab\footnote{These papers will be published in parallel.} ecosystem collaborate through agent-to-agent communication to decompose tasks, cross-validate results, and assemble end-to-end scientific workflows, in line with broader trends in multi-agent scientific AI~\cite{cao2025automating, Arlt2025AIMandel,elagenteOG,sharma2025towards}.

At the workflow level, hybrid quantum-classical integration (Stage 3) will establish closed-loop pipelines that combine classical solvers, tensor network methods, and quantum circuits. This approach follows the dominant strategy for near-term quantum simulation and variational algorithms~\cite{peruzzo2014variational, Murali2019NoiseAdaptive,childs2021theory,aspuru2005simulated}. As execution moves closer to hardware, hardware-aware compilation and scheduling (Stage 4) will account for device-specific constraints such as qubit connectivity, native gate sets, noise properties, and HPC or QPU scheduling policies. In doing so, the agent’s reasoning layer will be directly connected to modern quantum compilation, hardware software co-design, and performance-aware execution frameworks~\cite{Murali2019NoiseAdaptive, Sivarajah2020TKet,  bagherimehrab2025faster, wiebe2011simulating}.

As circuit depth and system size increase, AI-driven error mitigation and quantum error correction (Stage 5) will support the automated selection, implementation, and evaluation of mitigation techniques and error correcting codes. This will rely on fast stabilizer simulation, decoder optimization, and learning based control methods~\cite{gidney2021stim, bayraktar2023cuquantum, cao2025automating}. At the logical level, a fault-tolerant translation layer (Stage 6) will raise agent outputs from physical circuits to logical, fault-tolerant representations. This layer will also provide resource estimates and runtime analysis, connecting high-level algorithm descriptions with fault-tolerant quantum architectures.

Beyond execution, autonomous algorithm discovery (Stage 7) will enable the agent to generate, compile, and evaluate candidate quantum algorithms, classical simulation strategies, hardware-aware fault-tolerant schemes, and control protocols using automated simulation and benchmarking loops. Instead of relying on fixed algorithm designs, the agent will explore families of circuit constructions and iteratively refine them based on accuracy, resource requirements, and noise sensitivity. Recent work by Quantinuum and Hiverge~\cite{algorithm_discovery_quantinuum} demonstrates this approach feasible in quantum chemistry, where AI-driven workflows search variational algorithm spaces starting from simple templates and improve them through simulation-guided optimization, highlighting the potential of automated discovery within near-term hardware constraints.

Finally, the roadmap culminates in a self-driving quantum scientist (Stage 8), in which hypothesis generation, computational modeling, experimental execution, and result interpretation are integrated into a closed, autonomous feedback loop. At this stage, {\cinzel El Agente Cuántico} will  autonomously formulate and refine scientific hypotheses, select and adapt computational and experimental strategies, and iteratively update its internal models based on newly generated data. This vision directly builds on recent advances in agentic scientific systems and self-driving laboratories, where hierarchical planning, adaptive decision-making, and continuous validation enable accelerated, reproducible, and scalable discovery across computational and experimental domains~\cite{tom2024self, bai2022platform, vogler2024autonomous,bai2024dynamic,rauschen2024universal,caramelli2018networking}.

\section{Conclusions}\label{sec:conclusions}

In this work, we presented {\cinzel El Agente Cuántico}, a multi-agent AI system that translates natural-language descriptions of quantum-simulation tasks into executed and analyzed workflows across a heterogeneous quantum-software stack. A central advantage of the approach is that the agent grounds its reasoning in direct searches of library manuals and documentation, allowing it to recover current APIs, implementation details, and best practices at runtime. This design enables scalable adaptation to multiple libraries and evolving interfaces without task-specific engineering, and we demonstrated this capability across state preparation, closed- and open-system dynamics, tensor-network simulations, quantum control, quantum error correction, and quantum resource estimation.

Furthermore, this work shows how agentic AI systems can change how quantum simulations are done in the community, placing the emphasis on physical reasoning rather than on implementation details. Through {\cinzel El Agente Cuántico}, domain expertise is articulated directly in terms of physical models, assumptions, and observables, rather than being scattered across framework-specific code. By reducing the effort required to assemble and maintain complex workflows, agentic systems enable researchers to engage more deeply with scientific questions and their interpretation, rather than with computational overhead.

Looking forward, the roadmap presented in this work outlines a steady evolution from automated quantum-simulation workflows toward an autonomous quantum scientist. Initial advances will focus on expanding computational reach and on improving the integration of classical and quantum devices to address larger, more realistic problems. As these capabilities mature, greater autonomy in execution, decision-making, and interpretation will enable tighter integration among simulation, experimentation, and analysis. Ultimately, this progression aims to support scalable and reproducible discovery by allowing scientific intent to be translated directly into adaptive, self-directed research workflows. In this sense, the goal is not to replace scientific judgment, but to amplify it by enabling more systematic, scalable, and reproducible exploration of complex scientific problems.\cite{krenn2022scientific}

\section*{Data and code availability}
All the data required to evaluate the presented conclusions are available via \url{https://doi.org/10.5683/SP3/UAKARI}

%% file: includes/include-acknowledgement.tex
The authors would like to acknowledge valuable discussions with Jiaru Bai, Naixu Guo, Mohsen Bagherimehrab, Shuxiang Cao, and Ignacio Franco. We gratefully acknowledge the longstanding contributions of the Matter Lab’s current and past group members (\urlhref{https://www.matter.toronto.edu}{matter.toronto.edu}), and in particular from El Agente team. I.G. and J.B.P. acknowledge funding of this project by the National Sciences and Engineering Research Council of Canada (NSERC) Alliance Grant \#ALLRP587593-23. L.M.C. is supported by the
Novo Nordisk Foundation, Grant number NNF22SA0081175, NNF Quantum Computing Programme. 
K.P. and M.R. acknowledge the generous support of
the Canada 150 Research Chairs Program through A.A.-G.
Y.N. gratefully acknowledge support from the NSERC CREATE for Accelerated Discovery (AccelD) training program hosted by the Acceleration Consortium (Grant \#596133-2025) and was partially supported through a collaborative partnership with Moderna Inc.
Z.Z. and Y.Z. acknowledge support from NSERC - IRCPJ 547644.
\acknowAC
\acknowGEN{\emph{SciNet}}
\acknowSciNet 
A.A.-G. thanks Anders G. Fr{\o}seth, for his generous support. A.A.-G. and V.B. also acknowledge the generous support of Natural Resources Canada and the Canada 150 Research Chairs program. This research is part of the University of Toronto’s Acceleration Consortium, which receives funding from the Canada First Research Excellence Fund (CFREF) via CFREF-2022-00042. This work was supported by the AI2050 program of Schmidt Sciences. This work was supported by the Defense Advanced Research Projects Agency (DARPA) under Agreement No. HR0011262E022.

%% file: includes/include-appendix.tex
\section*{Supporting information El Agente Cu{\'a}ntico: Automating quantum simulations}
\label{app:related}

\section{Agents and tools}

\begin{table}[ht!]
\centering
\caption{Summary of the current agents (A) and tools (T) comprised in {\cinzel El Agente cu\'antico}.}
\label{tab:agents}

\renewcommand{\arraystretch}{1.35}
\setlength{\tabcolsep}{6pt}

\begin{tabularx}{\linewidth}
{l >{\raggedright\arraybackslash}X >{\centering\arraybackslash}X}
\toprule
\textbf{Label} 
& \textbf{Description} 
& \textbf{LLM model} \\
\midrule

\rowcolor{gray!10}
\textbf{Quantum scientist (A)}
&
A quantum-physics agent with access to the literature and software documentation that can delegate tasks to specialist experts. Responsible for quantum simulation tasks and high-level planning.
&
claude-opus-4-5~\cite{anthropic_claude}\\

\textbf{Cudaq expert (A)}
& A specialized quantum computing and algorithms agent that generates CUDA-QX simulation code.
& claude-opus-4-5~\cite{anthropic_claude} \\

\rowcolor{gray!10}
\textbf{Pennylane expert (A)}
& A specialized quantum computing agent that generates PennyLane simulation code.
& claude-opus-4-5~\cite{anthropic_claude} \\

\textbf{Qiskit expert (A)}
& A specialized quantum computing agent that generates Qiskit simulation code.
& claude-opus-4-5~\cite{anthropic_claude} \\

\rowcolor{gray!10}
\textbf{Qutip expert (A)}
& A specialized quantum physics agent that generates QuTiP 5.2 simulation code.
& claude-opus-4-5~\cite{anthropic_claude} \\

\textbf{Tenpy expert (A)}
& A specialized physicist with expertise in TeNPy and tensor-network methods.
& claude-opus-4-5~\cite{anthropic_claude} \\

\rowcolor{gray!10}
\textbf{Tequila expert (A)}
& A specialized quantum algorithms agent that generates Tequila simulation code.
& claude-opus-4-5~\cite{anthropic_claude} \\

\textbf{Literature search (T)}
& Search the scientific literature for information.
& gpt-5.2~\cite{openai_gpt5_card}\\

\rowcolor{gray!10}
\textbf{PDF reader (T)}
& Answers questions by searching and extracting information from a provided PDF document.
& gpt-5-mini \\

\textbf{Deepsearch (T)}
& Search the expert manuals (e.g., \texttt{CUDA-Q}, \texttt{QuTiP}, \texttt{PennyLane}, etc) for API documentation.
& gpt-5.2~\cite{openai_gpt5_card} \\

\rowcolor{gray!10}
\textbf{Python REPL (T)}
& Accesses a Python terminal to run Python code and interacts with the local
machine as needed.
& N/A \\

\textbf{Submit python jobs (T)}
&  Executes Python code on the HPC via SLURM with full workspace synchronization.
& N/A \\

\rowcolor{gray!10}
\textbf{View \texttt{png} file (T)}
& Allows visualization of \texttt{.png} files.
& N/A \\

\textbf{Ask for clarification  (T)}
& Ask for clarifications from user.
& N/A \\
\bottomrule
\end{tabularx}
\end{table}

\break
\section{Log sessions}
Here we outline the log sessions corresponding to all problems presented in the main text. For clarity, we omit comments related to file paths and figure handling, and we do not display the generated code. Nevertheless, all code and figures are publicly available via \url{https://doi.org/10.5683/SP3/UAKARI}.

\subsection{Variational quantum eigensolver}

\prompt{

Read the CUDA manual and compute the VQE energy of \ce{H2} over bond distances from 0.4 to 2.2 \AA   \ and compare against the exact answer. For the plot, create a figure of the dissociation curve and explain the physical meaning and where the curve deviates from exact energies.}

\quantumScientist{

\textbf{Action:}
I'll help you compute the VQE energy of H2 across bond distances and compare with exact results. Let me first search the CUDA-Q manual to understand the proper approach for molecular VQE calculations.

\textbf{Tool used:} cudaq deepsearch \\
\underline{query:} VQE variational quantum eigensolver molecular hydrogen H2 energy calculation. 

\underline{query:} CUDA-Q molecular Hamiltonian chemistry create molecular hamiltonian spin op exact diagonalization.
}

\quantumScientist{

\textbf{Action:}
I now have sufficient information from the CUDA-Q manual to propose a high-level plan. Based on my search, here's what I found:

\textbf{Proposed High Level Plan:} 
\begin{itemize}
    \item Step 1: Generate Data (VQE + Exact Energies)
Run in a single script that loops over bond distances from 0.4 to 2.2 \AA:
    \begin{itemize}
        \item For each bond distance, create the $H_{2}$ molecular Hamiltonian with STO-3G basis using cudaq.chemistry.create\_molecular\_hamiltonian().
        \item Compute VQE energy using an appropriate variational ansatz (e.g., UCCSD-inspired or hardware-efficient). Store all results (bond distance, VQE energy, exact energy)
        \item Compute exact ground-state energy via matrix diagonalization of the Hamiltonian
Store all results (bond distance, VQE energy, exact energy)
    \end{itemize}
    \item Step 2: Post-Analysis and Visualization. 
    \begin{itemize}
        \item Generate publication-quality dissociation curve plot comparing VQE vs. exact energies.
        \item Compute and plot the energy error (VQE-exact) across bond distances.
        \item Physical Context (Important for Interpretation):
        \begin{itemize}
            \item Near equilibrium ($\sim$0.74 \AA): Single-reference character dominates.
            \item Stretched bonds ($>1.5$ \AA): Strong correlation and multi-reference character emerge as the bond breaks.
            \item Expected VQE challenge: At large distances, $H_{2}$ dissociates to 2 open-shell H atoms. A simple single-determinant ansatz may struggle, while exact diagonalization captures all correlations. A simple single-determinant ansatz may struggle, while exact diagonalization captures all correlations
        \end{itemize}  
    \end{itemize}
\end{itemize}
\textbf{Shall I proceed with this 2-step plan? (yes / no / modify)}  If you approve, I will delegate Step 1 to the CUDA-Q expert to generate the computational code.
}

\user{}

\quantumScientist{

\textbf{Action:} Proceeding with Step 1. I will delegate to the CUDA-Q expert to generate the VQE code for the $H_{2}$ dissociation curve.

\textbf{Tool used:} cudaq expert \\
\underline{message2agent:} \textit{Omitting for brevity}
} 

\cudaexpert{

\textbf{Action:}
Let me search for more specific CUDA-Q VQE documentation:

\textbf{Tool used:} cudaq deepsearch \\
\underline{query:} cudaq VQE optimize observe kernel hardware efficient ansatz

\underline{query:} cudaq.chemistry molecular hamiltonian pyscf openfermion spin\_op to\_matrix
}

\cudaexpert{

\textbf{Action:}
Now I have good documentation. Let me create a minimal test first to verify the API works, then expand to the full dissociation curve:

\textbf{Tool used:} submit\_python\_slurm \\
\underline{code:} \textit{Omitting for brevity}
}

\cudaexpert{

\textbf{Task Completed:}
The VQE calculation completed successfully! The results show excellent agreement between VQE (UCCSD) and exact energies. Now let me delegate the visualization analysis and physical interpretation to the quantum scientist.
}

\quantumScientist{

\textbf{Action:} I'll analyze the $H_{2}$ VQE dissociation curve results. Let me start by reading the data file and displaying the generated plot, then provide a comprehensive analysis

\textbf{Tool used:} repl\_tool

\textbf{code:} \textit{Omitting for brevity}

}

\quantumScientist{

\textbf{Action:} Task Complete: $H_2$ VQE Dissociation Curve Analysis.

\textbf{Summary:} I successfully computed the VQE energy of $H_2$ across bond distances from 0.4 to 2.2 \AA using CUDA-Q on CPU (qpp-cpu target) and compared against exact (FCI) energies.

\textbf{Report Highlights:} 
\begin{itemize}
    \item \textbf{LEFT PANEL (PES):}
    \begin{itemize}
        \item  Morse\-like curve: Repulsive wall $\to$ equilibrium (0.7 \AA) $\to$  dissociation limit (-1.0 Ha).
        \item VQE $\approx$ Exact: Coral circles surround navy dots perfectly - sub- $\mu$Ha agreement.
        \item HF failure: Diverges at stretched geometries (19$\to$195 mHa error)
    \end{itemize}

    \item \textbf{RIGHT PANEL (Errors, Log Scale):}
    \begin{itemize}
        \item  HF errors: $\sim$$10^2$ mHa (ABOVE chemical accuracy) - fails at all points.
        \item VQE errors: $\sim$$10^{-3}$ mHa (FAR BELOW chemical accuracy) — succeeds at all points.
        \item 5 orders of magnitude separate VQE from HF.
    \end{itemize}
    \item \textbf{Physical Interpretation:}
    \begin{itemize}
        \item Static correlation problem: HF uses single determinant, cannot represent entangled dissociation state.
        \item HF predicts 50$\%$ $H^{+}H^{-}$ ionic character at dissociation — physically wrong.
        \item VQE/UCCSD captures the correct superposition via parameterized quantum circuit.
    \end{itemize}
\end{itemize}

}

\subsection{Bell states preparation}

\prompt{

Read the \texttt{CUDA-Q} manual and create a 2-qubit quantum circuit that prepares a Bell state $(\ket{00}+\ket{11})/\sqrt{2}$. Start in $\ket{00}$. Apply a Hadamard gate on qubit 0 and then a CNOT with control qubit 0 and target qubit 1. Measure both qubits in the computational $\braket{Z}$ basis with 4096 shots and return the measurement counts. From those counts, compute and return the expectation value of $Z \otimes Z$. Then also estimate the expectation value of $X \otimes X$ by measuring in the X basis, again with 4096 shots, and return both the counts and the estimated $\braket{X\otimes X}$. Plot and analyze the results.}

\quantumScientist{

\textbf{Action:}
I'll help you create a CUDA-Q simulation for Bell state preparation and measurement. Let me first search the CUDA-Q manual to ensure we use the correct syntax for this task.

\textbf{Tool used:} cudaq deepsearch \\
\underline{query:} Bell state preparation Hadamard CNOT gate measurement expectation value sample. 

\underline{query:} cudaq sample shots\_count kernel x.h cx measure X basis rotation.
}

\quantumScientist{

\textbf{Action:}
I now have sufficient information from the CUDA-Q manual.

\textbf{Proposed High Level Plan:} 

\textbf{Step 1:} Create CUDA-Q code that: (a) Prepares Bell state, (b) Measures in Z-basis (4096 shots) $\to$ compute $\langle Z \otimes Z \rangle$, (c) Measures in X-basis (4096 shots) $\to$ compute $\langle X \otimes X \rangle$, (d) Generates publication-quality plots of measurement counts and expectation values with error bars.

This is a single-step workflow since all computations are independent once the code is written, and the plotting can be done immediately after obtaining results within the same script.

\textbf{Outputs to deliver:}
\begin{itemize}
    \item Measurement counts for Z-basis (dictionary)
    \item Measurement counts for X-basis (dictionary)
    \item Computed $\langle Z \otimes Z \rangle$ and $\langle X \otimes X \rangle$
    \item Publication-quality figure showing histogram of counts and expectation values
\end{itemize}
\textbf{Shall I proceed with this plan?} (yes / no / modify)
}

\user{}

\quantumScientist{

\textbf{Action:} Proceeding with the plan. I'll delegate this task to the CUDA-Q expert.

\textbf{Tool used:} cudaq expert \\
\underline{message2agent:} \textit{Omitting for brevity}
}

\cudaexpert{

\textbf{Action:}
I'll first search the CUDA-Q manual to understand the correct syntax for measurements and sampling, then create the script.

\textbf{Tool used:} cudaq deepsearch \\
\underline{query:} cudaq.sample mz mx measurement kernel example

\underline{query:} cudaq qvector qubit Bell state cx CNOT Hadamard h gate example
}

\cudaexpert{

\textbf{Action:}
Now I have the correct syntax. Let me first debug with a minimal version to ensure the measurement extraction works correctly:

\textbf{Tool used:} submit\_python\_slurm \\
\underline{code:} \textit{Omitting for brevity}
}

\cudaexpert{

\textbf{Action:}
I see the issue - the result object conversion and X-basis measurement need different handling. Let me fix this:

\textbf{Tool used:} submit\_python\_slurm \\
\underline{code:} \textit{Omitting for brevity}
}

\cudaexpert{

\textbf{Task Completed:}
The CUDA-Q simulation completed successfully! I created a complete CUDA-Q Python script for Bell state preparation and measurement.
}

\quantumScientist{

\textbf{Action:} Excellent! The CUDA-Q expert completed the simulation. Let me view the generated plot to verify the quality:

\textbf{Tool used:} repl\_tool

\textbf{code:} \textit{Omitting for brevity}

}

\quantumScientist{

\textbf{Action:} Task Complete: Bell State Experiment.

\textbf{Summary:} 
\begin{itemize}
    \item 2-qubit system initialized in $\ket{00}$.
    \item Hadamard gate on qubit 0.
    \item CNOT with control=0, target=1.
    \item Prepared Bell state: $\ket{\Phi^{+}}$ = $(\ket{00} + \ket{11})/\sqrt{2}$
\end{itemize}

\textbf{Bell State $\ket{\Phi^{+}}$ Measurement Analysis}
\begin{itemize}
    \item \textbf{Panel 1: Z-Basis Measurements.} The near-equal split between $\ket{00}$ (51.4\%) and $\ket{11}$ (48.6\%) reveals the essence of quantum entanglement. When we measure qubit 0 and find it in state $\ket{0}$ , qubit 1 is always found in $\ket{0}$; likewise, measuring $\ket{1}$ on one qubit guarantees $\ket{1}$ on the other. The complete $\ket{01}$ and $\ket{10}$ outcomes demonstrates perfect correlation—the qubits share a definite relationship even though neither has a predetermined individual value. The small deviation from exactly 50/50 is statistical noise from finite sampling (4096 shots).
    \item \textbf{Panel 2: X-Basis Measurements.} The $\sim$50/50 split persists when measuring in the X-basis, confirming that correlations exist in multiple measurement bases simultaneously. This is the hallmark of genuine quantum entanglement. A classically correlated state (like a coin flip that sets both qubits identically) would show correlations in Z but appear random in X. The Bell state maintains perfect correlations in both bases—a feat impossible without quantum superposition. This multi-basis correlation is what enables violations of Bell inequalities.
    \item \textbf{Panel 3: Expectation Values.} 
    \begin{itemize}
        \item $\braket{Z\otimes Z}$ = +1 (perfect correlation in Z-basis),
        \item $\braket{X\otimes X} = +1$ (perfect correlation in X-basis),
        \item $\braket{Y\otimes Y} = -1$ (perfect anti-correlation in Y-basis).
    \end{itemize}
    The combination $(+1, +1, -1)$ uniquely identifies $\ket{\Phi^{+}}$ among the four Bell states. 
\end{itemize}
}

\subsection{Thermal states via imaginary-time evolution}

\prompt{ Prepare a Gibbs state of a 1D Hubbard model on a classical computer using imaginary-time evolution. Consider a 10-site Hubbard chain with open boundary conditions, hopping $t = 1$, on-site interaction $U = 4$, and half filling. Target an inverse temperature $1/T = 2$. Explain how to validate that the resulting state is thermal by checking energies and occupancies.
 }

\quantumScientist{

\textbf{Action:} This is a well-defined finite-temperature quantum many-body simulation task. Let me analyze the problem and propose a plan.

\textbf{Recommended approach:} I recommend using TeNPy with the purification method for imaginary-time evolution. TeNPy has a built-in FermiHubbardModel and excellent support for finite-temperature simulations via Matrix Product Density Operators (MPDO) or purification.

\textbf{Proposed High-Level Plan:}
\begin{itemize}
    \item \textbf{Step 1: Setup and Reference Computation.} 
    \begin{itemize}
        \item Define the 1D Hubbard Hamiltonian in TeNPy with specified parameters.
    \end{itemize}
    \item  \textbf{Step 2: Imaginary-Time Evolution and Validation.}
    \begin{itemize}
        \item Prepare initial maximally mixed (infinite-temperature) purified state.
        \item Perform imaginary-time TEBD evolution from $\beta=0$ to $\beta=2$.
        \item Measure thermal observables:
        \begin{itemize}
            \item Total energy $\langle H \rangle$.
            \item Site occupancies $\langle n_{i} \uparrow \rangle$, $\langle n_{i} \downarrow \rangle$ 
            \item Double occupancy $\langle n_{i} \uparrow n_{i} \downarrow\rangle$
            \item Von Neumann entropy S.

        \end{itemize}

    \end{itemize}
\end{itemize}
\textbf{Would you like me to proceed with this 2-step plan?} 
}

\user{}

\quantumScientist{

\textbf{Action:} Proceeding with the plan. I'll delegate this to the TeNPy expert who specializes in tensor network simulations of strongly correlated systems.

\textbf{Tool used:} tenpy expert \\
\underline{message2agent:} \textit{Omitting for brevity}
}

\tenpyexpert{

\textbf{Action:}
I need to search the TeNPy manual first to ensure I use the correct API for the Hubbard model and purification-based imaginary-time evolution.

\textbf{Tool used:} tenpy deepsearch \\
\underline{query:} FermiHubbardModel purification imaginary time evolution thermal state TEBD

\underline{query:} PurificationMPS from\_infiniteT charge\_sector canonical\_ensemble particle number conservation expectation\_value

}

\tenpyexpert{

\textbf{Action:}
Now I have all the information needed. Let me create a minimal working version first to test the approach:

\textbf{Tool used:} submit\_python\_slurm \\
\underline{code:} \textit{Omitting for brevity}
}

\tenpyexpert{

\textbf{Action:}
The exact diagonalization failed due to size issues. Let me fix this by using sparse diagonalization and properly setting max\_size:

\textbf{Tool used:} submit\_python\_slurm \\
\underline{code:} \textit{Omitting for brevity}
}

\tenpyexpert{

\textbf{Action:}
The exact diagonalization failed due to size issues. Let me fix this by using sparse diagonalization and properly setting max\_size:

\textbf{Tool used:} submit\_python\_slurm \\
\underline{code:} \textit{Omitting for brevity}
}

\tenpyexpert{

\textbf{Action:}
I see several issues:
\begin{itemize}
    \item The charge sector extraction from ED is incorrect (found only 1 state).
    \item The from\_infiniteT\_canonical API is different.
    \item The TEBD is not evolving (bond dim stays at 1).
\end{itemize}
Let me search for the correct API and fix these issues:

\textbf{Tool used:} tenpy deepsearch \\
\underline{query:} PurificationTEBD run\_imaginary imaginary time evolution bond dimension not growing

\underline{query:} PurificationTEBD options parameters TEBD order dt trunc\_params initialization
}

\tenpyexpert{

\textbf{Action:}
Now I understand the issues. Let me fix the code with proper TEBD options and correct ED charge sector extraction:

\textbf{Tool used:} submit\_python\_slurm \\
\underline{code:} \textit{Omitting for brevity}
}

\tenpyexpert{

\textbf{Action:}
The simulation completed successfully. Let me now delegate the visualization and analysis to the quantum scientist.

\textbf{message2agent:} \textit{Omitting for brevity}
}

\quantumScientist{

\textbf{Action:} I'll analyze and visualize the Gibbs state preparation results for the 1D Hubbard model. Let me start by loading the data and exploring its structure.

\textbf{Tool used:} repl\_tool

\textbf{code:} \textit{Omitting for brevity}}

\quantumScientist{

\textbf{Action:} Analysis Completed. I've created all requested visualizations and comprehensive physical interpretations for the Gibbs state preparation of the 10-site Hubbard chain at $\beta=2$.

\textbf{Physical Insights} 
\begin{itemize}
    \item \textbf{Panel (a), energy vs 1/T.} As temperature decreases, the Boltzmann distribution favors lower-energy states, so the average energy drops monotonically toward the ground state.
    \item \textbf{Panel (b), site occupancy.} The canonical ensemble at fixed $N=10$ electrons maintains exactly one electron per site on average, as expected for half-filling.
    \item \textbf{Panel (c), Double occupancy.} The strong on-site repulsion $U=4$ suppresses double occupancy by 64$\%$. This is the hallmark of Hubbard physics—electrons avoid sharing sites due to Coulomb repulsion.
    \item \textbf{Panel (d), entanglement entropy.} Von Neumann entropy S at each bond cutting the chain. The purification MPS has entanglement from both quantum correlations and thermal mixing. Higher entropy at the center is typical for finite systems with open boundaries.
\end{itemize}
}

\subsection{Trotter decomposition}

\prompt{

Read the \texttt{CUDA-Q} manual and implement a simulation of the transverse-field Ising model for a 10-qubit chain with open boundary conditions. Compute the dynamics of the magnetization $\langle Z \rangle$ for two parameter regimes, $J_{\text{coup}} = 10\,J$ with $h = 0.1\,J$, and $J_{\text{coup}} = 0.1\,J$ with $h = 10\,J$. Start the evolution from a Haar-random initial state and simulate the dynamics over a total time of 10 seconds using a second-order product formula with 1000 time steps. Generate a heatmap where the x-axis corresponds to qubit position, the y-axis to time, and the color scale to the $Z$-magnetization, for each of these two phases. Analyze the resulting magnetization dynamics between the two parameter choices.}

\quantumScientist{

\textbf{Action:}
I'll help you simulate the transverse-field Ising model dynamics using CUDA-Q. This is an interesting task that will compare magnetization dynamics in two distinct quantum phases - the ferromagnetic phase (strong coupling) and paramagnetic phase (strong transverse field).
Let me first search the CUDA-Q manual to understand the best approach for Hamiltonian simulation with product formulas.

\textbf{Tool used:} cudaq deepsearch \\
\underline{query:} Hamiltonian simulation product formula Trotter time evolution dynamics

\underline{query:} state vector simulation expectation value measurement pauli operators spin operators.

\underline{query:} cudaq.State from\_data numpy array initial state qvector initialization.
}

\quantumScientist{

\textbf{Action:}
Based on my search of the CUDA-Q documentation, I have gathered enough information to propose a high-level plan. Let me outline the key considerations and the proposed workflow.

\textbf{Proposed Workflow (2 Steps)}

\begin{itemize}
    \item \textbf{Step 1: Implement and Run Simulations} (Both regimes can run in parallel)
    \begin{itemize}
        \item Implement Trotter-based time evolution for the 10-qubit TFIM with open boundary conditions
        \item Generate a Haar-random initial state (using NumPy/SciPy unitary sampling, then converted to initial circuit or direct state manipulation)
        \item Simulate dynamics for both parameter regimes:
        \begin{itemize}
            \item Regime 1: $J_{\text{coup}} = 10$, $h = 0.1$ (ferromagnetic)
            \item Regime 2: $J_{\text{coup}} = 0.1$, $h = 10$ (paramagnetic)
        \end{itemize}
        \item Store $\langle Z_{i}(t)\rangle$ for all 10 qubits at all 1000 time steps
    \end{itemize}

    \item \textbf{Step 2: Visualization and Analysis} (Depends on Step 1 output)
    \begin{itemize}
        \item Generate publication-quality heatmaps for both regimes:
        \begin{itemize}
            \item x-axis: qubit position (0–9)
            \item y-axis: time (0–10)
            \item color: Z-magnetization
        \end{itemize}
        \item Perform comparative analysis of the magnetization dynamics between the two phases
        \item Interpret results in terms of the known physics of TFIM phase behavior
    \end{itemize}
\end{itemize}
\textbf{Should I proceed with this plan?} (yes / no / modify)
}

\user{}

\quantumScientist{

\textbf{Action:} This is a quantum dynamics simulation task that should be handled by the CUDA-Q expert. Let me delegate this to them with all the specific requirements.

\textbf{Tool used:} cudaq expert.

\textbf{message2agent:} \textit{Omitting for brevity}}

\cudaexpert{

\textbf{Action:} I'll first search the CUDA-Q manual to understand the proper way to implement quantum dynamics simulations with Trotter decomposition.

\textbf{Tool used:} cudaq deepsearch

\underline{query:} quantum dynamics simulation Trotter decomposition product formula time evolution

\underline{query:} exp\_pauli cudaq second order Trotter Suzuki decomposition time step

\underline{query:} cudaq observe expectation value spin operator get\_state random initial state

}

\cudaexpert{

\textbf{Action:} Now I have all the information needed. Let me create a complete CUDA-Q simulation for the transverse field Ising model dynamics using second-order Trotter decomposition.

\textbf{Tool used:} submit\_python\_slurm

\textbf{code:} \textit{Omitting for brevity}

}

\cudaexpert{

\textbf{Action:} I see the issue - the exp\_pauli function needs to be called differently within kernels. Let me search for the correct syntax

\textbf{Tool used:} cudaq deepsearch

\underline{query:} exp\_pauli kernel syntax pauli\_word cudaq.kernel decorator

}

\cudaexpert{

\textbf{Action:} I see the issue - inside a @cudaq.kernel, you call exp\_pauli directly without the cudaq. prefix. Let me fix the code:

\textbf{Tool used:} submit\_python\_slurm

\textbf{code:} \textit{Omitting for brevity}

}

\cudaexpert{

\textbf{Action:} The simulation completed successfully. Let me now delegate to the postprocessing expert to analyze the results and provide enhanced visualizations.

\textbf{message2agent:} \textit{Omitting for brevity}

}

\quantumScientist{

\textbf{Action:} I'll analyze the CUDA-Q simulation results for the transverse field Ising model dynamics. Let me load the data and create comprehensive visualizations

\textbf{Tool used:} repl\_tool

\textbf{code:} \textit{Omitting for brevity}

}

\quantumScientist{

\textbf{Action:} Simulation Successfully Executed.
The CUDA-Q simulation of the transverse field Ising model dynamics has been completed for both parameter regimes using second-order Trotter-Suzuki decomposition.

\textbf{Physical Insights} 

\textbf{Left panel, strong coupling  (J\_coup=10J, h=0.1J).} 
\begin{itemize}
    \item Slow dynamics with characteristic relaxation time $\sim$2s
    \item Collective behavior dominates with correlations propagating through the chain
    \item The system shows slow thermalization toward $\langle Z \rangle \approx 0$ 
\end{itemize}

\textbf{Right Panel - Strong Field Regime (J\_coup=0.1J, h=10J):} 
\begin{itemize}
    \item Fast oscillatory dynamics (Rabi-like precession) with period $\sim$0.06s
    \item Each qubit precesses nearly independently due to weak coupling
    \item Clear striped pattern from rapid local oscillations
    \item Short correlation length between neighboring qubits
\end{itemize}       

The contrast clearly illustrates the quantum phase transition behavior of the Transverse-field Ising model: the ferromagnetic phase (strong coupling) vs. the paramagnetic phase (strong field).

}

\subsection{Lindbladian dynamics}

\prompt{

Read the \texttt{QuTiP} manual and compare the dynamics of a single qubit under unitary evolution and under open system Lindblad evolution using: 

\textbf{Hamiltonian:} $H = \sigma_{x} + \frac{1}{2}\sigma_{z}$,
\textbf{Initial condition:} Use the plus and minus eigenstates of X.
\textbf{Dissipation:} Use a collapse operator $C = \sqrt\gamma  \sigma_{-}$, with $\gamma$ = 0.2. \textbf{Evolution:} Use a time range from $t = [0,40]$ with 400 time steps. \textbf{Output:} Plot the expectation value of Z as a function of time for each initial state and compare closed and open dynamics. }

\quantumScientist{

\textbf{Action:}
I'll help you set up this qubit dynamics comparison using QuTiP. Let me first search the QuTiP manual to ensure we use the correct syntax for both solvers.

\textbf{Tool used:} qutip deepsearch \\
\underline{query:} sesolve mesolve Lindblad master equation qubit evolution collapse operator expectation value

}

\quantumScientist{

\textbf{Action:}
I have the necessary information from the QuTiP manual. This is a well-defined task that can be executed efficiently. Let me propose a high-level plan:

\textbf{Proposed Plan}

\textbf{Step 1: Single execution step} — Generate QuTiP code that performs all simulations and creates the comparison plot.

The code will:
\begin{itemize}
    \item Define Hamiltonian: $H = \sigma_x + \frac{1}{2}\sigma_z$
    \item Prepare initial states: $\ket{+} = \frac{1}{\sqrt{2}}(\ket{0} + \ket{1})$ and $\ket{-}= \frac{1}{\sqrt{2}}(\ket{0} - \ket{1})$
    \item Define collapse operator: $C = \sqrt{0.2},\sigma_-$
    \item Run 4 simulations in parallel (conceptually):
    \begin{itemize}
        \item sesolve (unitary) for $\ket{+}$
        \item sesolve (unitary) for $\ket{-}$
        \item mesolve (Lindblad) for $\ket{+}$
        \item mesolve (Lindblad) for $\ket{-}$
    \end{itemize}
    \item Plot $\langle\sigma_z\rangle(t)$ for all 4 cases with clear labels distinguishing closed vs. open dynamics
\end{itemize}
\textbf{Should I proceed with this plan?} (yes / no / modify)
}

\user{}

\quantumScientist{

\textbf{Action:} I'll delegate this task to the QuTiP expert now.

\textbf{Tool used:} qutip expert.

\textbf{message2agent:} \textit{Omitting for brevity}}

\qutipexpert{

\textbf{Action:} II'll search the QuTiP 5.2 manual first to ensure I use the correct API for sesolve and mesolve.

\textbf{Tool used:} qutip deepsearch

\underline{query:} sesolve mesolve Lindblad dynamics expectation values collapse operators QuTiP 5.2

}

\qutipexpert{

\textbf{Action:} Now I have the correct QuTiP 5.2 API information. Let me generate the code and execute it:

\textbf{Tool used:} submit\_python\_slurm

\textbf{code:} \textit{Omitting for brevity}
}

\qutipexpert{

\textbf{Action:} Task Completed Successfully. 

\textbf{message2agent:}  I generated and executed QuTiP 5.2 code comparing unitary vs Lindblad dynamics for a single qubit.
}

\quantumScientist{

\textbf{Action:} Task Complete. Here's the summary:

\textbf{Figure Description:}
\begin{itemize}
    \item \textbf{Panel (a), Unitary Evolution:} Shows persistent Rabi-like oscillations for both $\ket{+}$ (dark blue, solid) and $\ket{-}$  (light blue, dashed) initial states with amplitude $\sim$0.4.
    \item \textbf{Panel (b), Lindblad Evolution:} Shows damped oscillations converging to a steady state with a horizontal dashed line at $\braket{Z} \approx$  -0.33
\end{itemize}

\textbf{Physical interpretation:} 
\begin{itemize}
    \item The Hamiltonian $H = \sigma_{x} + 0.5\sigma_{z}$ creates a tilted effective magnetic field causing Rabi-like precession.
    \item The collapse operator $C = \sqrt{\gamma} \sigma_{-}$ induces spontaneous emission ($\ket{1} \to \ket{0}$ decay).
    \item Competition between coherent driving (H) and dissipation ($\gamma$) produces a non-trivial steady state where $\sim67\%$ population is in $\ket{0}$.
    \item Crucially, the steady state is unique.
    \item Both initial states converge to the same value, demonstrating loss of memory of initial conditions (a hallmark of open quantum systems).
\end{itemize}  
}

\subsection{Hierarchical equations of motion}

\prompt{ 
Read \texttt{QuTiP's} manual and use the HEOM implementation to simulate exciton dynamics in the Fenna--Matthews--Olson (FMO) complex. Use the following parameters:
\[
\textbf{System Hamiltonian}\cite{adolphs2006proteins,kreisbeck2011high,kreisbeck2014scalable} =
{\scriptsize
\setlength{\arraycolsep}{3pt}
\begin{pmatrix}
240.0 & -87.7 & 5.5 & -5.9 & 6.7 & -13.7 & -9.9 \\
-87.7 & 315.0 & 30.8 & 8.2 & 0.7 & 11.8 & 4.3 \\
5.5 & 30.8 & 0.0 & -53.5 & -2.2 & -9.6 & 6.0 \\
-5.9 & 8.2 & -53.5 & 130.0 & -70.7 & -17.0 & -63.3 \\
6.7 & 0.7 & -2.2 & -70.7 & 285.0 & 81.1 & -1.3 \\
-13.7 & 11.8 & -9.6 & -17.0 & 81.1 & 435.0 & 39.7 \\
-9.9 & 4.3 & 6.0 & -63.3 & -1.3 & 39.7 & 245.0
\end{pmatrix}
}.
\]
\textbf{Bath parameter:} Drude--Lorentz spectral density with reorganization energy $\lambda = 35~\mathrm{cm}^{-1}$, cutoff time $1/\gamma = 50~\mathrm{fs}$, and temperature $T = 300~\mathrm{K}$ and $T = 77~\mathrm{K}$. Hierarchy depth $L = 4$, number of Matsubara terms $K = 2$.
\textbf{Initial condition:} system localized on site 0.
\textbf{Dynamics:} propagate for $1~\mathrm{ps}$ and record site populations vs.\ time.
\textbf{Report:} Plot all site populations as a function of time at the two different temperatures and analyze the results.
}

\quantumScientist{

\textbf{Action:}
I'll help you simulate the exciton dynamics in the FMO complex using QuTiP's HEOM implementation. This is a classic problem in quantum biology for understanding energy transfer in photosynthetic systems.
Let me first search the QuTiP manual to understand the HEOM implementation details and ensure we use the correct API.

\textbf{Tool used:} qutip deepsearch \\
\underline{query:} HEOM hierarchical equations of motion Drude Lorentz spectral density bath solver.

}

\quantumScientist{

\textbf{Action:}
I found detailed information about QuTiP's HEOM implementation. Now let me formulate a high-level plan for this FMO simulation.

\textbf{High-Level Plan: FMO Exciton Dynamics with HEOM}\\
Key Considerations I've Identified:
\begin{itemize}
    \item \textbf{Multi-bath setup:} The FMO complex has 7 chromophores, each coupled to its \textbf{own independent bath} (7 baths total). This is crucial for correct physics.
    \item \textbf{Unit consistency:} The Hamiltonian is in cm$^{-1}$, bath parameters need consistent units:
    \begin{itemize}
        \item $\gamma$ = 1/(50 fs) needs conversion to cm$^{-1}$ (using $\hbar$ = 5.309 cm$^{-1}$$\cdot$ps)
        \item Temperature needs conversion via $k_{B}T$ in cm$^{-1}$ ($k_B$ = 0.695 cm$^{-1}$/K)
    \end{itemize}
    \item \textbf{Coupling operators:} Each site i has coupling operator $Q_i = \ket{i}\bra{i}$ (local site projector)
    \item \textbf{Computational cost:} 7 baths x K=2 Matsubara terms x L=4 depth can generate a large hierarchy - this is manageable but not trivial.
\end{itemize}
\textbf{Proposed Workflow (2 Steps):}
\begin{itemize}
    \item \textbf{Step 1: Code Generation \& Simulation}
    \begin{itemize}
        \item Generate QuTiP HEOM simulation code for the 7-site FMO complex
        \item Include both T = 300 K and T = 77 K simulations
        \item Set up proper unit conversions (cm$^{-1}$ throughout)
        \item Use DrudeLorentzBath with proper multi-bath configuration
        \item Run dynamics for 1 ps with initial condition on site 0
        \item Save population data for all 7 sites vs. time
    \end{itemize}
    \item \textbf{Step 2: Post-Processing \& Analysis}
    \begin{itemize}
        \item Generate publication-quality plots of site populations vs. time
        \item Create two panels (or two figures): one for 300 K, one for 77 K
        \item Analyze and compare coherence timescales and relaxation behavior at both temperatures
        \item Identify key physical observations (coherent oscillations, thermalization timescales)
    \end{itemize}
\end{itemize}
\textbf{Do you want me to proceed with this plan?} Also, would you like me to perform a literature search on FMO dynamics benchmarks to validate our results against published simulations?
}
\user{}

\quantumScientist{

\textbf{Action:}
Proceeding with \textbf{Step 1:} Delegating to the QuTiP expert to generate the HEOM simulation code for FMO dynamics.

\textbf{Tool used:} qutip expert \\

\textbf{message2agent: } \textit{Omitting for brevity}

}

\qutipexpert{

\textbf{Action:}
I'll help you generate QuTiP 5.2 code for simulating exciton dynamics in the FMO complex using the HEOM solver. Let me first search the QuTiP manual to ensure I use the correct API for HEOM in version 5.2.

\textbf{Tool used:} qutip deepsearch 

\underline{query:} HEOM solver DrudeLorentzBath multi-bath exciton dynamics 
}

\qutipexpert{

\textbf{Action:}
I have the information needed about the QuTiP 5.2 HEOM API. Now let me generate and execute the FMO simulation code.

\textbf{Tool used:} qsubmit\_python\_slurm 

\textbf{code: } \textit{Omitting for brevity}

}

\qutipexpert{

\textbf{Action:}
The HEOM simulations completed successfully for both temperatures. The results show interesting physics:
\begin{itemize}
    \item T = 300 K: Slower, more thermally-driven dynamics with population spreading across sites
    \item T = 77 K: More coherent dynamics with population staying concentrated on fewer sites
\end{itemize}
Now let me delegate the visualization and analysis to the quantum scientist 

\textbf{message2agent:} \textit{Omitting for brevity} 
}

\quantumScientist{

\textbf{Action:} I'll create publication-quality visualizations for the FMO complex exciton dynamics. This is a classic quantum biology problem - the FMO complex is a pigment-protein complex that facilitates energy transfer in green sulfur bacteria.
Let me start by reading the data and creating the visualizations:

\textbf{Tool used:} repl\_tool

\textbf{code:} \textit{Omitting for brevity} 
}

\quantumScientist{

\textbf{Action:} I see the data contains complex number strings. Let me parse them properly and check the actual format:

\textbf{Tool used:} repl\_tool

\textbf{code:} \textit{Omitting for brevity} 
}

\quantumScientist{

\textbf{Action:} Now I understand the format - the data contains complex numbers. Let me parse them correctly and extract real population values::

\textbf{Tool used:} repl\_tool

\textbf{code:} \textit{Omitting for brevity} 
}

\quantumScientist{

\textbf{Action:} Data parsed successfully. Now let me create the publication-quality visualizations:

\textbf{Tool used:} repl\_tool

\textbf{code:} \textit{Omitting for brevity} 
}

\quantumScientist{

\textbf{Action:} Simulation Completed Successfully.
The HEOM simulation of the Fenna-Matthews-Olson complex has been completed using QuTiP 5.2. Here are the key results:

\begin{itemize}
    \item \textbf{Left Panel (T = 300 K, Room Temperature):}
    \begin{itemize}
        \item The initially excited Site 0 decays smoothly from population 1.0 to about 0.27
with only weak oscillations.
        \item Population spreads broadly across the complex: Sites 1–3 each reach
roughly 13–22 percent, and even high-energy Site 5 reaches about 4 percent.
        \item The overall behavior is
predominantly incoherent and diffusive.
    \end{itemize}
    \item \textbf{Right Panel (T = 77 K, Cryogenic):}
    \begin{itemize}
        \item Site 0 undergoes a rapid drop in the first $\sim$ 100 fs followed by clear oscillations and partial recovery, retaining about 46 percent population at 1 ps.
        \item Oscillatory features appear across several sites, and population remains concentrated on lower-energy sites (mainly 2 and 3), while higher-energym sites (4–6) stay below 2 percent.
        \item The dynamics are more coherent and wavelike than at room temperature.
    \end{itemize}
    \item \textbf{Consistency with Literature}
    \begin{itemize}
        \item The energy funnel behavior toward Site 3 (BChl 3) is well-documented
        \item Temperature-dependent rates agree with theoretical predictions
    \end{itemize}
\end{itemize}

}

\subsection{Quantum optimal control}

\prompt{

Read the \texttt{QuTiP} manual and solve a quantum control problem for a Lambda-type three-level system (levels $\ket{1}$, $\ket{2}$, $\ket{3}$). The goal is to transfer population from the ground state $\ket{1}$ to the target state $\ket{3}$ while minimizing the occupation of the intermediate state $\ket{2}$. 

\textbf{Control Hamiltonians:} $H_{c1}=(\ket{1}\bra{2}+\ket{2}\bra{1})$ (Pump coupling). $H_{c2}=\ket{2}\bra{3}+\ket{3}\bra{2})$ (Stokes coupling).
\textbf{The Optimization:} Use the GRAPE algorithm. 
\textbf{Duration:} T=40 (arbitrary time units) with time slots = 3000.
\textbf{Validation:} After optimizing the pulses, run a time-evolution simulation using the generated pulses to verify the dynamics.
\textbf{Output:} A plot showing the Control Pulses (Amplitude vs Time) and a plot showing the populations vs Time. Print the final fidelity.}

\quantumScientist{

\textbf{Action:}
I'll help you with this quantum control problem using QuTiP's GRAPE algorithm for a Lambda-type three-level system. This is a classic STIRAP-like problem where we want to achieve population transfer while avoiding the intermediate state.
Let me first search the QuTiP manual to understand the GRAPE implementation details in QuTiP 5.2

\textbf{Tool used:} qutip deepsearch \\
\underline{query:} GRAPE algorithm quantum control optimization pulse

\underline{query:} optimize\_pulse state transfer fidelity three level system qutip-qtrl

}

\quantumScientist{

\textbf{Action:}
Excellent! I now have comprehensive information about QuTiP's GRAPE implementation. This is a well-defined Lambda-system quantum control problem (similar to STIRAP). Let me propose a high-level plan.

\textbf{High-Level Plan}\\
\textbf{Objective:} Optimize control pulses for population transfer $\ket{1} \to \ket {3}$ in a Lambda-type three-level system using GRAPE, while minimizing intermediate state $\ket{2}$ occupation.

\textbf{Step 1: GRAPE Optimization and Validation (single step - all tasks use same input)}
\begin{itemize}
    \item \textbf{Task 1a:} Set up the Lambda system Hamiltonians and run GRAPE optimization
    \begin{itemize}
        \item Drift Hamiltonian: $H_{0} = 0$ (no detunings specified)
        \item Control Hamiltonians: $H_{c1}$ (pump), $H_(c2)$ (Stokes)
        \item Initial state: $\ket{1}$, Target state: $\ket{3}$
        \item Duration T=40, time slots=3000
    \end{itemize}
    \item \textbf{Task 1b:} After optimization, run time evolution with mesolve using the optimized pulses
    \item \textbf{Task 1c:} Generate publication-quality plots:
    \begin{itemize}
        \item Plot 1: Control pulse amplitudes vs time
        \item Plot 2: Populations of $\ket{1}$, $\ket{2}$, $\ket{3}$ vs time   (print fidelity)
    \end{itemize}
\end{itemize}
\textbf{Should I proceed with this plan? (yes / no / modify)} }

\user{}

\quantumScientist{

\textbf{Action:}
Proceeding with the plan. I'll delegate this task to the QuTiP expert with all the necessary specifications.

\textbf{Tool used:} qutip expert

\textbf{message2agent:} \textit{Omitting for brevity} 
}

\qutipexpert{

\textbf{Action:}
I'll help you implement a GRAPE-based quantum control for a Lambda-type three-level system. First, let me search the QuTiP 5.2 documentation to verify the correct API for GRAPE optimization.

\textbf{Tool used:} qutip deepsearch

\underline{query:} GRAPE optimize\_pulse quantum control qutip-qtrl pulseoptim \\
\underline{query:}  GRAPE state transfer fid\_type UNIT PSU fidelity optimize\_pulse initial target state 
}

\qutipexpert{

\textbf{Action:}
Excellent! Now I have all the information I need. Let me start with a minimal working version to debug the GRAPE optimization, then expand to the full implementation.

\textbf{Tool used:}  submit\_python\_slurm

\textbf{code:} \textit{Omitting for brevity}
}

\qutipexpert{

\textbf{Action:}
Good progress! The imports work. Let me fix the numpy import order and run a proper minimal test:

\textbf{Tool used:}  submit\_python\_slurm

\textbf{code:} \textit{Omitting for brevity}
}

\qutipexpert{

\textbf{Action:}
The minimal test works. Now let me implement the full solution with 3000 time slots, proper plotting, and mesolve validation:

\textbf{Tool used:}  submit\_python\_slurm

\textbf{code:} \textit{Omitting for brevity}
}

\qutipexpert{

\textbf{Action:}
The coefficient function format has changed in QuTiP 5.2. Let me search for the correct syntax:

\textbf{Tool used:} qutip deepsearch

\underline{query:} GRAPE optimize\_pulse quantum control qutip-qtrl pulseoptim \\
\underline{query:}  mesolve time-dependent Hamiltonian coefficient function QobjEvo QuTiP 5
}

\qutipexpert{

\textbf{Action:}
I see the issue. QuTiP 5.2 has a different coefficient signature. Let me fix the code using array coefficients, which are better suited for GRAPE piecewise-constant pulses:

\textbf{Tool used:}  submit\_python\_slurm

\textbf{code:} \textit{Omitting for brevity}
}

\qutipexpert{

\textbf{Action:}
I see the issue - the optimization is terminating abnormally and the pulses aren't working properly. The "LIN" initial guess may be causing problems. Let me try with a random initial guess and add better debugging:

\textbf{Tool used:}  submit\_python\_slurm

\textbf{code:} \textit{Omitting for brevity}
}

\qutipexpert{

\textbf{Action:}
I found the issue - there's a discrepancy between GRAPE's reported fidelity and actual evolution. Let me investigate and fix this:

\textbf{Tool used:}  submit\_python\_slurm

\textbf{code:} \textit{Omitting for brevity}
}

\qutipexpert{

\textbf{Action:}
Let me fix the Qobj handling:

\textbf{Tool used:}  submit\_python\_slurm

\textbf{code:} \textit{Omitting for brevity}
}

\qutipexpert{

\textbf{Action:}
Now I understand - GRAPE's internal evolution is correct (fidelity=1.0) but there's a mismatch with manual reconstruction. Let me extract the dynamics properly and create the final working solution:

\textbf{Tool used:}  submit\_python\_slurm

\textbf{code:} \textit{Omitting for brevity}
}

\qutipexpert{

\textbf{Action:}
The optimization achieved perfect fidelity! But there's an issue with the population dynamics extraction. The fwd\_evo contains propagators, not states. Let me fix this properly:

\textbf{Tool used:}  submit\_python\_slurm

\textbf{code:} \textit{Omitting for brevity}
}

\qutipexpert{

\textbf{Action:}
The population extraction is still showing unphysical values. Let me fix this properly by using sesolve with correct coefficient handling for QuTiP 5.2:

\textbf{Tool used:}  submit\_python\_slurm

\textbf{code:} \textit{Omitting for brevity}
}

\qutipexpert{

\textbf{Action:}
I need to handle the QuTiP 5.2 return types correctly. Let me fix this:

\textbf{Tool used:}  submit\_python\_slurm

\textbf{code:} \textit{Omitting for brevity}
}

\qutipexpert{

\textbf{Action:}
There's still a discrepancy between GRAPE's fidelity and my manual propagation. Let me investigate GRAPE's dynamics generator convention

\textbf{Tool used:}  submit\_python\_slurm

\textbf{code:} \textit{Omitting for brevity}
}

\qutipexpert{

\textbf{Action:}
Now I understand the issue. GRAPE uses a linearized propagator internally (for gradient computation), not the full matrix exponential. The optimized pulses ARE correct, but I need to use proper step-by-step unitary evolution matching GRAPE's time discretization exactly:

\textbf{Tool used:}  submit\_python\_slurm

\textbf{code:} \textit{Omitting for brevity}
}

\qutipexpert{

\textbf{Action:}
The optimization now works perfectly. Both GRAPE and true unitary evolution show fidelity = 1.0. Let me save the complete final code and organize the files:

\textbf{Tool used:}  submit\_python\_slurm

\textbf{code:} \textit{Omitting for brevity}
}

\qutipexpert{

\textbf{Action:}
I see there's a fundamental discrepancy - GRAPE uses linearized dynamics internally. Let me search for the proper way to do GRAPE with exact unitary propagators in QuTiP:

\textbf{Tool used:} qutip deepsearch

\underline{query:} qutip-qoc GRAPE optimize exact unitary propagator prop\_type DEF dynamics generator matrix exponential
}

\qutipexpert{

\textbf{Action:}
I found the issue! I need to explicitly specify dyn\_type='UNIT' to get proper unitary dynamics. Let me fix this:

\textbf{Tool used:}  submit\_python\_slurm

\textbf{code:} \textit{Omitting for brevity}
}

\qutipexpert{

\textbf{Action:}
The optimization now works perfectly with dyn\_type='UNIT'. Both GRAPE and true unitary evolution show fidelity = 1.0. Let me save the complete final code and organize the files:

\textbf{Tool used:}  submit\_python\_slurm

\textbf{code:} \textit{Omitting for brevity}
}

\qutipexpert{

\textbf{Action:}
Task Completed Successfully.
The GRAPE quantum control optimization for the Lambda-type three-level system has been successfully implemented and validated.

\textbf{Critical Implementation Detail
}  

The key fix that made GRAPE work correctly was adding dyn\_type='UNIT' to the optimize\_pulse() call. This ensures proper unitary dynamics with exact matrix exponential propagators (exp(-iHt)), rather than linearized approximations.

}

\quantumScientist{

\textbf{Action:}
The GRAPE optimization completed successfully with excellent results. 

\textbf{Physics Observations} \\
The population dynamics plot shows:
\begin{itemize}
    \item \textbf{Left panel (Control pulses):} Both pulses are oscillatory and overlap in time. No STIRAP-like counterintuitive order. Amplitudes vary between $\pm 5$ units.
    \item \textbf{Right panel (Populations):} $P_{1}$ (blue): Smoothly depletes from $1 \to 0$. $P_{2}$ (orange): Peaks at $83 \%$ mid-transfer, return to 0. $P_{3}$ (green): build to $\sim 1.0$ (99.99999$\%$ fidelity)
\end{itemize}
\textbf{Important Note:} The standard GRAPE algorithm optimizes only for final state fidelity, not for minimizing intermediate state occupation during the transfer. The solution found is mathematically optimal for the endpoint but does NOT minimize $\ket{2}$ occupation during evolution.\\
If minimizing $\ket{2}$ population is critical, this would require:
\begin{itemize}
    \item Adding a penalty term to the cost function for intermediate state population
    \item Using STIRAP-like pulse shapes as initial guesses (counter-intuitive ordering: Stokes before Pump)
\end{itemize}
Would you like me to re-run the optimization with a modified cost function that penalizes intermediate state occupation? This would yield a solution closer to the adiabatic STIRAP protocol.
}

\usern{}

\subsection{Time-dependent product formulas}

\prompt{

Read \texttt{CUDA-Q} manual and simulate the Floquet dynamics of a periodically kicked Ising chain with 10 qubits and open boundary conditions. The system has nearest-neighbor $ZZ$ interactions with coupling $J=1$, and is kicked every $T=1$ by a global transverse field in X. Use a strong kick with strength $h = 2J$ so that the effect of the drive is clearly visible in the dynamics. Initialize the system in a Haar-random state and simulate the evolution for $10$ Floquet periods. Plot a heatmap of $\langle Z_i(t)\rangle$, with qubit index on the x-axis, Floquet step (time) on the y-axis, and color representing the local magnetization.
}

\quantumScientist{

\textbf{Action:} I'll help you simulate Floquet dynamics of a periodically kicked Ising chain using CUDA-Q. This is an interesting quantum dynamics problem. Let me first search the CUDA-Q manual to understand the best approach for implementing this simulation.

\textbf{Tool used:} cudaq deepsearch

\underline{query:} state vector simulation time evolution expectation value measurement spin operators

\underline{query:} cudaq get\_state evolve simulation circuit state vector apply gate
}

\quantumScientist{

\textbf{Action:} I now have sufficient information from the CUDA-Q manual. Based on my research, I can formulate a high-level plan for this Floquet dynamics simulation.

\textbf{High-Level Plan: Floquet Dynamics of Kicked Ising Chain in CUDA-Q
} \\
\textbf{Physics Summary}
\begin{itemize}
    \item Hamiltonian during evolution: $H_{ZZ} = J \sum_{i=0}^{N-2} Z_i Z_{i+1}$ with $J=1$
    \item Kick operator: $U_{\text{kick}} = \exp(-i h \sum_i X_i)$ with $h = 2J = 2$
    \item Floquet unitary: $U_F = U_{\text{kick}} \cdot U_{ZZ}$ where $U_{ZZ} = \exp(-i H_{ZZ} T)$
\end{itemize}

\textbf{Implementation Strategy.} CUDA-Q's state-in/state-out pattern is well-suited:
\begin{itemize}
    \item Generate Haar-random initial state using NumPy
    \item Build circuit kernels for ZZ evolution and X-kick
    \item Iteratively apply Floquet unitary using cudaq.get\_state()
    \item Measure $\langle Z_i \rangle$ after each period
\end{itemize}

\textbf{Proposed Workflow (2 Steps)}
\begin{itemize}
    \item \textbf{Step 1: Simulation}
    \begin{itemize}
        \item Generate a Haar-random initial state vector (10 qubits = 1024 amplitudes)
        \item Construct the Floquet unitary as a CUDA-Q kernel:
        \begin{itemize}
            \item ZZ evolution: Trotterized with CNOT-RZ-CNOT decomposition for each bond
            \item X-kick: Apply RX(2h) = RX(4) gates globally (since kick = exp(-ihX))
        \end{itemize}
        \item Simulate 10 Floquet periods, recording $\langle Z_i \rangle$ after each period
        \item Store the 10×10 matrix of local magnetizations
    \end{itemize}
    \item \textbf{Step 2: Visualization}
    \begin{itemize}
        \item Create publication-quality heatmap:
        \begin{itemize}
            \item x-axis: qubit index (0-9)
            \item y-axis: Floquet step (0-10, including initial state)
            \item colormap: local magnetization $\langle Z_i(t) \rangle \in [-1, 1]$
        \end{itemize}
    \end{itemize}
\end{itemize}
\textbf{Do you want me to proceed with this plan?} (yes / no / modify)
If you'd like, I can also:
\begin{itemize}
    \item Modify the number of Floquet periods
    \item Add entanglement entropy tracking
    \item Include uncertainty estimation by averaging over multiple random initial states
\end{itemize}
}

\user{}

\quantumScientist{

\textbf{Action:} Proceeding with the plan. I'll delegate the simulation task to the CUDA-Q expert.

\textbf{Tool used:}  cudaq expert

\textbf{message2agent}  \textit{Omitting for brevity}

}

\cudaexpert{

\textbf{Action:} I'll first search the CUDA-Q manual to understand the correct API for state initialization and state evolution patterns.

\textbf{Tool used:} cudaq deepsearch

\underline{query:} cudaq.State.from\_data from\_state initialize state numpy array get\_state state-in state-out

\underline{query:} cudaq.observe spin.z single qubit expectation value measurement
}

\cudaexpert{

\textbf{Action:} Now I have the necessary information about CUDA-Q APIs. Let me implement a minimal test version first (4 qubits, 2 periods) to verify the circuit works correctly.

\textbf{Tool used:}  submit\_python\_slurm

\textbf{Code:} \textit{Omitting for brevity}
}

\cudaexpert{

\textbf{Action:} The minimal test works correctly. Now let me implement the full simulation with 10 qubits, 10 Floquet periods, including the heatmap visualization.

\textbf{Tool used:}  submit\_python\_slurm

\textbf{Code:} \textit{Omitting for brevity}
}

\cudaexpert{

\textbf{Action:} The Floquet dynamics simulation completed successfully. Delegating to the quantum scientist for analysis.

\textbf{Message2agent:} \textit{Omitting for brevity}

}

\quantumScientist{

\textbf{Action:} The simulation executed successfully. We observe irregular red--blue patches fluctuating around white, with all magnetization values remaining very small ($\lVert \langle Z \rangle \rVert < 0.09$). 
There is no discernible spatial pattern or temporal ordering.

\textbf{Physical interpretation.} 
\begin{itemize}
    \item \textbf{Thermalization:} The near-zero magnetization across the entire space--time grid indicates that the system has effectively thermalized to an infinite-temperature state, as expected for chaotic Floquet dynamics.
    \item \textbf{Scrambling:} The rapid, sign-changing fluctuations reflect strong information scrambling. The large kicks ($h = 2J$) drive efficient mixing of quantum states and suppress any coherent structure.
\end{itemize}

The featureless, noise-like fluctuations around zero confirm that the kicked Ising chain  is deep in the chaotic regime, acting as a quantum thermalizer rather than supporting stable or long-lived coherent dynamics.

}

\subsection{Time-evolving block decimation (TEBD)}
\prompt{

Simulate Floquet dynamics with TD-DMRG for a 1D disordered spin-1/2 chain. Use a periodically driven Hamiltonian with period $T=\tau_1+\tau_2$. During $0<t<\tau_1$, evolve under a strong global $x$-field plus interactions and disorder:
$H = \sum_i \Omega_x S_i^x + \Delta_i S_i^z + J(S_i^xS_{i+1}^x + S_i^yS_{i+1}^y - S_i^zS_{i+1}^z)$.
During $\tau_1<t<T$, switch to a global $y$-field,
$H = \sum_i \Omega_y(1-\varepsilon) S_i^y + \Delta_i S_i^z + H_{i,i+1}$.
Choose $\Omega_x\tau_1 \approx \pi$ so each period approximately flips the $x$-polarization. Use the following parameters: system size $L=10$ spins, maximum bond dimension $\chi_{\max}=100$, time step $\delta t=0.03$, evolve for $N_{\text{periods}}=50$ Floquet periods, average over $N_{\text{disorder}}=5$ disorder realizations, and scan $\varepsilon \in \{0, 0.15, 0.3\}$. Keep the Hamiltonian parameters as before: $J=1$, $\Omega_x = \Omega_y = 4J$, $\tau_1 = \pi/(2\Omega_x) \approx 0.393$, $\tau_2 = 0.5$, and disorder strength $W=2J$. 

Initialize the system in a product state polarized along $+x$. Evolve stroboscopically for many periods using TD-DMRG (TEBD or TDVP), keeping truncation errors under control. Use disorder $\Delta_i\in(-W,W)$ and average over several realizations. After each Floquet period, measure the bulk $x$-magnetization and plot it versus the number of periods. Vary $\varepsilon$ to identify regimes with long-lived period-doubling (subharmonic $2T$ oscillations) versus rapid decay due to heating. Track entanglement growth and discuss how it limits the reachable simulation time.
}

\quantumScientist{

\textbf{Action:}  This is a fascinating Floquet Many-Body Localization / Discrete Time Crystal (DTC) simulation! The physics here involves:
\begin{itemize}
    \item Period-doubling oscillations: When $\varepsilon\approx0$, the system shows robust 2T-periodic response (DTC phase)
    \item Heating to infinite temperature: For large $\varepsilon$, the drive heats the system and destroys coherent oscillations
    \item Disorder-induced stabilization: Strong disorder (MBL) protects the DTC from thermalization
\end{itemize}
This task is well-suited for TenPy's TEBD implementation. Let me propose a structured plan:

\textbf{Proposed High-Level Plan}
\begin{itemize}
    \item \textbf{Step 1: Code Generation \& Validation}
    \begin{itemize}
        \item Generate TenPy code implementing the two-phase Floquet protocol
        \item Use TEBD for time evolution with the specified Hamiltonians
        \item Initialize $\ket{+x}$product state
        \item Implement disorder sampling and stroboscopic measurement
    \end{itemize}
    \item \textbf{Step 2: Production Simulations (all parallelizable)}
    \begin{itemize}
        \item Run simulations for $\varepsilon$ $\in \{0, 0.15, 0.3\}$
        \item Average over 5 disorder realizations per $\varepsilon$
        \item Track x-magnetization and entanglement entropy at each period
        \item Monitor truncation errors
    \end{itemize}
    \item \textbf{Step 3: Analysis \& Visualization}
    \begin{itemize}
        \item Plot disorder-averaged magnetization vs. Floquet period for each $\varepsilon$
        \item Plot entanglement entropy growth to assess simulation reliability
        \item Identify DTC regime (persistent oscillations) vs. heating regime (decay)
        \item Discuss bond dimension limitations on simulation time
    \end{itemize}
\end{itemize}
\textbf{Would you like me to proceed with this plan?}

Also, would you like me to perform a brief literature search on Floquet MBL/DTC simulations to ensure we follow best practices for this type of calculation?
}

\user

\quantumScientist{

\textbf{Action:} I'll proceed with the plan. Let me delegate the code generation to the TenPy expert.

\textbf{Tool used:} tenpy expert

\textbf{Message2agent:} \textit{Omitting for brevity}
}

\tenpyexpert{

\textbf{Action:} I'll help you create a TenPy simulation for Floquet dynamics of a disordered spin-1/2 chain (Discrete Time Crystal / Floquet MBL). Let me first search the TenPy manual to ensure I use the correct API for TEBD time evolution.

\textbf{Tool used:} tenpy deepsearch

\underline{query:} TEBD time evolution engine spin chain MPS time-dependent Hamiltonian

\underline{query:} SpinHalfChain custom model NearestNeighborModel add\_onsite add\_coupling disorder random field TEBD trunc\_params chi\_max

\underline{query:} MPS from\_product\_state x direction Sx expectation value entanglement\_entropy measurement

\underline{query:} TEBDEngine run real time evolution run\_evolution dt N\_steps real time not imaginary
}

\tenpyexpert{

\textbf{Action:} Now I have all the information needed. Let me create a complete TenPy simulation for the Floquet DTC dynamics. I'll first run a minimal test to verify the code works, then run the full production simulation.

\textbf{Tool used:} submit\_python\_slurm

\textbf{code:} \textit{Omitting for brevity}
}

\tenpyexpert{

\textbf{Action:} The minimal test works correctly. The code shows expected DTC behavior with oscillating magnetization. Now let me run the full production simulation with all parameters and generate publication-quality plots.

\textbf{Tool used:} submit\_python\_slurm

\textbf{code:} \textit{Omitting for brevity}
}

\tenpyexpert{

\textbf{Action:} The simulation completed successfully. Now let me save the Python code file as requested:

\textbf{Tool used:} submit\_python\_slurm

\textbf{code:} \textit{Omitting for brevity}
}

\tenpyexpert{

\textbf{Action:} I have successfully created and executed a complete TenPy simulation for Floquet dynamics of a 1D disordered spin-1/2 chain (Discrete Time Crystal / MBL). 

}

\quantumScientist{

\textbf{Action:} Excellent! The simulation has completed successfully. Let me view the main results to analyze the DTC dynamics.

\textbf{Figure Observations:}
\begin{itemize}
    \item \textbf{Panel (a) - Magnetization:} Stroboscopic $\langle S^{x}\rangle$ vs Floquet period with clear period-doubling oscillations. The alternating pattern demonstrates discrete-time crystal behavior, with the decay depending on the imperfection $\varepsilon$.
    \item \textbf{Panel (b) - Entanglement:} Slow logarithmic growth of entanglement entropy $S$, confirming MBL protection ($S \approx 2.0 << 3.47$ thermal limit). Three curves for $\varepsilon \in {0, 0.15, 0.3}$ with shaded error bands ($\pm 1 \sigma$ from disorder averaging).
\end{itemize}

\textbf{Key Physics Observations}
This simulation provides direct numerical evidence for the MBL-protected discrete time crystal phase: a driven quantum system that spontaneously oscillates at twice the driving period, protected from thermalization by many-body localization.
}

\subsection{Phase diagrams of quantum systems}

\prompt{Use DMRG to prepare the ground state of the one-dimensional transverse field cluster Hamiltonian over a ring of $N$ qubits (interpolate with cos and sin between these cluster Hamiltonian and the magnetic field). Compute the string-order parameter $ZXIX\cdots IXZ$ for multiple strengths of the cluster Hamiltonian over 11, 21, and 41 qubits. Plot the string-order parameter as a function of the cluster Hamiltonian strength for the different system sizes and estimate the critical point.}

\quantumScientist{

\textbf{Action:} This is a tensor network simulation task for studying a symmetry-protected topological (SPT) phase transition. Let me outline my understanding and propose a plan.

\textbf{Problem Understanding}
\begin{itemize}
    \item \textbf{Hamiltonian:} The transverse field cluster model on a ring: $$H(\theta) = \cos(\theta) \sum_i X_i + \sin(\theta) \sum_i Z_{i-1} X_i Z_{i+1}$$
    \begin{itemize}
        \item At $\theta = 0$: paramagnetic phase (trivial)
        \item At $\theta = \pi/2$: cluster Hamiltonian (SPT phase)
    \end{itemize}
    \item \textbf{String Order Parameter:} $\langle Z_1 X_2 I_3 X_4 I_5 \cdots X_{N-1} Z_N \rangle$ — this detects the hidden topological order characteristic of the cluster state phase.
    \item \textbf{Goal:} Map out the phase transition by computing string order vs. $\theta$ for N = 11, 21, 41.
\end{itemize}

\textbf{Proposed High-Level Plan
}
\begin{itemize}
    \item \textbf{Step 1: DMRG Ground State Calculations + String Order Computation}\\
    For each system size (N = 11, 21, 41) and for multiple values of $\theta \in [0, \pi/2]$ (e.g., 20-30 points):
    \begin{itemize}
        \item Construct the transverse field cluster Hamiltonian on a ring
        \item Run DMRG to obtain the ground state
        \item Compute the string order parameter $\langle ZXIX\cdots IXZ \rangle$
    \end{itemize}
    \item \textbf{Step 2: Analysis and Visualization}
    \begin{itemize}
        \item Plot string order parameter vs. $\theta$ for all three system sizes on the same figure
        \item Estimate the critical point from the crossing/transition behavior
        \item Apply finite-size scaling if needed
    \end{itemize}
\end{itemize}

\textbf{Key Considerations}
\begin{itemize}
    \item Periodic boundary conditions (ring geometry) require careful handling in DMRG — may need to use infinite DMRG or handle the long-range bond appropriately
    \item Bond dimension must be sufficient to capture entanglement near criticality
    \item String order computation requires measuring a product of non-local operators
\end{itemize}
\textbf{Would you like me to proceed with this plan?
}

Also, would you like me to perform a literature search first to:
\begin{itemize}
    \item Confirm the known critical point for benchmarking
    \item Verify the exact form of the string order parameter used in the literature
\end{itemize}
}

\user{}

\quantumScientist{

\textbf{Action:} I'll proceed with the plan. Let me first do a quick literature/documentation search to ensure we implement this correctly, then delegate to the TenPy expert.

\textbf{Tool used:} tenpy deepsearch

\underline{query:} string order parameter correlation function DMRG cluster state SPT

\underline{query:} three-body interaction ZXZ cluster Hamiltonian custom model periodic boundary conditions ring
}

\quantumScientist{

\textbf{Action:} Excellent! I have the necessary information. Let me summarize the key findings:

\textbf{Literature/Documentation Findings:}
\begin{itemize}
    \item Three-body cluster term: Use add\_multi\_coupling() with ('Sigmaz', 0, 0), ('Sigmax', 1, 0), ('Sigmaz', 2, 0)
    \item Periodic boundary conditions: Set bc\_x='periodic' (note: MPS stays open but couplings wrap around)
    \item String order parameter: Use psi.correlation\_function() with opstr argument for inserting operators between endpoints
    \item Conservation: Must use conserve='None' since X terms break Sz conservation
\end{itemize}
Now I'll delegate to the TenPy expert to implement this calculation.

\textbf{Tool used:} tenpy expert

\textbf{Message2agent:} \textit{Omitting for brevity}

}

\tenpyexpert{

\textbf{Action:} I'll help you implement the DMRG study of the transverse field cluster model. Let me first search the tenpy manual to ensure I use the correct API for tenpy 1.11.2.

\textbf{Tool used:} tenpy deepsearch

\underline{query:} CouplingMPOModel add\_multi\_coupling three-body term periodic boundary conditions

\underline{query:} expectation\_value\_multi\_sites string order parameter MPS

\underline{query:}  SpinHalfSite Sigmax Sigmaz Pauli operators conserve None
}

\tenpyexpert{

\textbf{Action:} Now I have all the information I need. Let me first create a minimal debug version to test the implementation:

\textbf{Tool used:} submit\_python\_slurm 

\textbf{Code:} \textit{Omitting for brevity}
}

\tenpyexpert{

\textbf{Action:} The debug test works correctly. The string order parameter shows the expected behavior:
\begin{itemize}
    \item $\theta=0$ (trivial phase): string order $\approx 0$
    \item $\theta=\pi/2$ (SPT cluster phase): string order $= -1.0$ (fully ordered)
\end{itemize}

\textbf{Tool used:} submit\_python\_slurm 

\textbf{Code:} \textit{Omitting for brevity}
}

\tenpyexpert{

\textbf{Action:} The simulation completed successfully. The data shows a clear phase transition in the string order parameter. Now let me delegate the visualization and critical point analysis to the quantum scientist.

\textbf{Message2agent:} \textit{Omitting for brevity} 

}

\quantumScientist{

\textbf{Action:} Analysis Complete. I have created publication-quality plots and analyzed the DMRG results for the transverse field cluster model.

I implemented DMRG simulations (tolerance $10^{-5}$) for the transverse field cluster Hamiltonian:
\begin{itemize}
    \item \textbf{Model:} $H(\theta) = -\cos(\theta) \sum_i Z_{i-1} X_i Z_{i+1} - \sin(\theta) \sum_i X_i$ where $\theta$ interpolates between the cluster symmetry-protected topological (SPT) phase ($\theta=0$) and trivial paramagnetic phase ($\theta=\pi/2$). 
\end{itemize}

\textbf{Physical Interpretation:} 
\begin{itemize}
    \item SPT Phase ($\theta < \theta_c$): String order $O \approx 1$, indicating symmetry-protected topological order
    \item Trivial Phase ($\theta > \theta_c$): String order $\theta \rightarrow 0$, no topological order 
    \item Critical Point: $\theta_c / \pi \approx 0.236$ (extrapolated), close to theoretical $\pi/4 = 0.25$
    \item The transition sharpens with increasing system size, consistent with a quantum phase transition.
\end{itemize}  
}

\subsection{Bell-state correlations under depolarizing noise}

\prompt{ Read the \texttt{Qiskit} manual and prepare the Bell state $(|00\rangle + |11\rangle)/\sqrt{2}$. Start in $|00\rangle$, apply a Hadamard gate on qubit 0, and then a CNOT with control qubit 0 and target qubit 1. Add a depolarizing noise channel with probability $p$ to both qubits. Simulate the circuit for $p \in \{0, 0.05, 0.1, 0.2, 0.3\}$. For each value of $p$, run 4096 shots in the $Z$ basis, return the measurement counts, and compute $\langle Z \otimes Z \rangle$. Then insert Hadamard gates on both qubits to measure in the $X$ basis, run 4096 shots, return the measurement counts, and compute $\langle X \otimes X \rangle$. Plot both correlators as functions of $p$ and comment on how the noise degrades the Bell-state correlations.}

\quantumScientist{

\textbf{Action:}  I'll help you prepare a Bell state with depolarizing noise in Qiskit. Let me first search the Qiskit manual to understand how to implement noise channels and measurements.

\textbf{Tool used:} qiskit deepsearch

\underline{query:} depolarizing noise channel circuit simulation shots measurement

\underline{query:} NoiseModel depolarizing channel probability create add noise gates

\underline{query:} ExpectationValue Paulistring Z measurement counts bitstring sampling

\underline{query:} Qiskit noise BitFlip DepolarizingError NoiseModel constructor example

}

\quantumScientist{

\textbf{Action:} Based on my search of the qiskit manual, I now have a good understanding of how to implement this task. 

\textbf{Proposed High-Level Plan}

\begin{itemize}
    \item \textbf{Step 1: Build and simulate Bell state circuit with depolarizing noise}
    \begin{itemize}
        \item Create Bell state circuit: H(0) + CNOT(0,1)
        \item For each $p \in \{0, 0.05, 0.1, 0.2, 0.3\}$:
        \begin{itemize}
            \item Apply depolarizing noise (level=2 for after CNOT) to both qubits
            \item Z-basis: Simulate with 4096 shots, collect counts, compute $\langle Z \otimes Z\rangle$
            \item X-basis: Add H gates on both qubits, simulate with 4096 shots, collect counts, compute $\langle X \otimes X\rangle$
        \end{itemize}
    \end{itemize}
    \item \textbf{Step 2: Plot and analyze results}
    \begin{itemize}
        \item Plot $\langle Z \otimes Z\rangle$ and $\langle X \otimes X\rangle$ vs $p$.
        \item Comment on how noise degrades Bell-state correlations
    \end{itemize}
\end{itemize}
\textbf{Should I proceed with this plan?} (yes / no / modify)
}

\user{}

\quantumScientist{

\textbf{Action:} I'll delegate to the qiskit expert to implement the noisy Bell state simulation.

\textbf{Tool used:} qiskit expert.

\textbf{Message2agent:} \textit{Omitting for brevity}
}

\tequilaexpert{

\textbf{Action:}  Let me first search the Qiskit manual to understand the correct syntax for noise models and measurements.

\textbf{Tool used:} qiskit deepsearch

\underline{query:} depolarizing noise channel simulation measurement samples shots

\underline{query:} NoiseModel depolarizing add\_noise circuit gates qiskit create noise model

\underline{query:} qiskit circuit noise BitFlip DepolarizingError create noise probability qubit

\underline{query:} qiskit gates H Hadamard CNOT X controlled gate circuit construction

}

\tequilaexpert{

\textbf{Action:} Now I have enough information to create the simulation. Let me write a comprehensive Python script that handles all the requirements

\textbf{Tool used:} submit\_python\_slurm

\textbf{Code:} \textit{Omitting for brevity}
}

\tequilaexpert{

\textbf{Action:} I see the issue - there's a qiskit version compatibility problem (qiskit 2.2.3 has breaking changes). Let me implement a solution using Qiskit Aer directly for proper noisy simulation:

\textbf{Tool used:} submit\_python\_slurm

\textbf{Code:} \textit{Omitting for brevity}
}

\tequilaexpert{

\textbf{Action:} Excellent! The simulation worked perfectly! The results show clear noise degradation. Now let me delegate the plotting to the quantum scientist:

\textbf{Tool used:} submit\_python\_slurm

\textbf{Code:} \textit{Omitting for brevity}
}

\quantumScientist{

\textbf{Action:} The simulation executed successfully. I prepared a Bell state simulation with depolarizing noise and analyzed the degradation of quantum correlations. Here are the key results:
\begin{itemize}
    \item Ideal Bell state produces only correlated outcomes ($\ket{00}$, $\ket{00}$)
    \item  Depolarizing noise introduces anti-correlated outcomes ($\ket{01}$, $\ket{10}$) 
    \item High noise limit: Distribution approaches uniform (maximally mixed state)
    \item X-basis degradation is 2-3x faster due to additional Hadamard gates also experiencing noise. 
\end{itemize}
This demonstrates why the choice of measurement basis matters in noisy quantum systems. More gates equal more accumulated errors.
}

\quantumScientist{

\textbf{Action:} I'll proceed with Step 1 - constructing the Hamiltonian and preparing for resource estimation. Let me delegate this to the PennyLane expert.

\textbf{Tool used:} pennylane expert.

\textbf{Message2agent:} \textit{Omitting for brevity}
}

\subsection{Estimating qubit requirements for quantum simulation}

\prompt{ Read the \texttt{PennyLane} manual and construct the second-quantized electronic Hamiltonian of water in the STO-3G basis. Use the Jordan-Wigner transformation to map the Hamiltonian to a qubit representation. Then build a logical (non–fault-tolerant) quantum phase estimation (QPE) circuit that would estimate the ground-state energy of water within chemical accuracy, and use \texttt{PennyLane}'s resource-estimation tools to approximate: (1) the number of logical qubits required and (2) the number of non-Clifford (e.g., $T$) gates arising from the Hamiltonian-simulation segments of the QPE routine. Report and discuss these logical-level resource estimates.
}

\quantumScientist{

\textbf{Action:} I'll analyze this task and develop a high-level plan. This involves quantum chemistry Hamiltonian construction and quantum resource estimation using PennyLane. Let me first search the PennyLane manual to understand the available tools for molecular Hamiltonians, QPE, and resource estimation.

\textbf{Tool used:} pennylane deepsearch

\underline{query:} molecular Hamiltonian second quantized Jordan Wigner transformation water H2O STO-3G basis

\underline{query:} quantum phase estimation QPE resource estimation T gates non-Clifford gates logical qubits

\underline{query:} DoubleFactorization error chemical accuracy precision target\_error one\_body two\_body electron integrals
}

\quantumScientist{

\textbf{Action:} Excellent! I now have comprehensive information to formulate a plan. Based on my search of the PennyLane documentation, here's my analysis:

\textbf{Task Analysis}
\begin{itemize}
    \item \textbf{Objective:} Estimate logical-level QPE resources for the water molecule (H$_{2}$O) in STO-3G basis using PennyLane.
    \item \textbf{Key technical aspects:}
    \begin{itemize}
        \item \textbf{H$_{2}$O in STO-3G:} 10 electrons, 7 spatial orbitals $\to$ 14 spin-orbitals $\to$ 14 qubits under Jordan-Wigner
        \item \textbf{Chemical accuracy:} 0.0016 Hartree ($\sim$1 kcal/mol), which is PennyLane's default for resource estimation
        \item \textbf{Resource estimation approach:} PennyLane provides qml.resource.DoubleFactorization which estimates Toffoli gates and logical qubits for QPE using the double-factorized Hamiltonian representation (state-of-the-art for second-quantized molecular simulation)
        \item \textbf{T-gate conversion:} Each Toffoli gate decomposes into $\sim$7 T gates (standard Clifford+T decomposition)
    \end{itemize}
\end{itemize}
\textbf{Proposed High-Level Plan}
\begin{itemize}
    \item \textbf{Step 1}
    \begin{itemize}
        \item Construct H$_{2}$O molecule in STO-3G basis
        \item Generate second-quantized fermionic Hamiltonian
        \item Apply Jordan-Wigner transformation $\to$ qubit Hamiltonian
        \item Extract one- and two-electron integrals for resource estimation
    \end{itemize}
    \item \textbf{Step 2}
    \begin{itemize}
        \item Use qml.resource.DoubleFactorization with error=0.0016 (chemical accuracy) to estimate logical qubits and Toffoli gates
        \item Convert Toffoli count to approximate T-gate count
        \item Report and discuss results (qubit count, gate cost, implications for near-term/fault-tolerant hardware)
    \end{itemize}
\end{itemize}
\textbf{Would you like me to proceed with this plan?} Also, do you want me to perform a brief literature search on QPE resource estimation methodologies before execution, or shall I proceed directly with the PennyLane implementation?
}

\pennylaneexpert{

\textbf{Action:} Let me first search the PennyLane manual to understand the correct API for molecular Hamiltonians, QPE, and resource estimation tools.

\textbf{Tool used:} pennylane deepsearch

\underline{query:} molecular Hamiltonian water qchem Jordan-Wigner second quantization STO-3G basis

\underline{query:} quantum phase estimation QPE circuit resource estimation logical qubits T gates non-Clifford

\underline{query:} DoubleFactorization electron\_integrals QPE gate cost qubit cost Hamiltonian simulation Toffoli
}

\pennylaneexpert{

\textbf{Action:} Now I have comprehensive information. Let me search for more details about inspecting the Hamiltonian and any additional resource estimation details.

\textbf{Tool used:} pennylane deepsearch

\underline{query:}  Hamiltonian coefficients terms pauli words print inspect number terms Jordan Wigner fermionic
}

\pennylaneexpert{

\textbf{Action:} Excellent! I now have all the information needed. Let me create a comprehensive script that:
\begin{itemize}
    \item Constructs water's molecular Hamiltonian in STO-3G basis with Jordan-Wigner mapping
    \item Analyzes the Hamiltonian structure
    \item Uses DoubleFactorization for QPE resource estimation (logical qubits and non-Clifford gates)
\end{itemize}

\textbf{Tool used:} submit\_python\_slurm

\textbf{Code:} \textit{Omitting for brevity}
}

\pennylaneexpert{

\textbf{Action:} Let me fix the electron integrals handling and re-run the script:

\textbf{Tool used:} submit\_python\_slurm

\textbf{Code:} \textit{Omitting for brevity}
}

\pennylaneexpert{

\textbf{Action:} Excellent! The simulation completed successfully. Let me now delegate the visualization and detailed interpretation to the quantum scientist.

\textbf{Message2agent} \textit{Omitting for brevity}
}

\quantumScientist{

\textbf{Action:} Excellent! Task Complete: Water QPE Resource Estimation

\textbf{Summary Report}\\
I have successfully constructed the second-quantized electronic Hamiltonian of water (H$_{2}$O) in STO-3G basis, applied the Jordan-Wigner transformation, and estimated the logical-level resources for quantum phase estimation (QPE).

\textbf{Figure Content:}
\begin{itemize}
    \item \textbf{(a) Left panel, qubit breakdown.} 14 system (4.8$\%$) vs 276 ancilla (95.2$\%$) = 290 total. 
    \item \textbf{(b) Right panel, required Toffoli gates.} Toffoli gates vs error (log-scale), showing $\mathcal{O}(1/\epsilon)$ scaling with chemical accuracy marked in green. 
\end{itemize}
The resource estimates demonstrate that even for a small molecule like water, fault-tolerant QPE requires significant quantum resources, highlighting the importance of algorithmic optimizations (double factorization) and the path toward utility-scale quantum chemistry.

\textbf{Key takeaway:} For water QPE at chemical accuracy $\to$ 290 logical qubits and $\sim$112M Toffoli gates ($\sim$447M T-gates).

}

\subsection{Quantum Error Correction}

\prompt{ Implement and run a surface-code logical memory experiment in the Z basis with the \texttt{Stim}. Use the following parameters:

\textbf{Distances:} d = 3, 5, 7. \textbf{Rounds:} R = d stabilizer-measurement rounds. \textbf{Shots:} N = 3000000 per distance \textbf{State:} prepare logical $\ket{0_{L}}$ (Z-basis memory). \textbf{Noise:} apply two-qubit depolarizing noise with probability p = 0.003 on every CX (CNOT) gate only. \textbf{Sampling:} for each shot, sample \texttt{Stim} detector events and final data-qubit Z measurements. \textbf{Logical observable:} compute measured $Z_{L}$ as the parity of the final data-qubit Z outcomes along the chosen logical $Z_{L}$ string. \textbf{Decoding:} build the \texttt{Stim}-generated detector error model (DEM) and decode using \texttt{PyMatching} (MWPM). \textbf{Plots:} report a single plot of d vs logic error gate for the data with and without the decoder, and add theoretical estimation.
}

\quantumScientist{

\textbf{Action:}  This is a quantum error correction (QEC) task involving surface code simulation with CUDA-Q QEC and the Stim backend. Let me first search the CUDA-Q documentation to understand the QEC capabilities and Stim integration before proposing a plan.

\textbf{Tool used:} cudaq deepsearch

\underline{query:}  CUDA-Q QEC surface code Stim backend quantum error correction logical memory experiment

\underline{query:} PyMatching decoder MWPM Stim detector error model two-qubit depolarizing noise CNOT gate
}

\quantumScientist{

\textbf{Action:} Based on my searches, I have a comprehensive understanding of the task.

\textbf{Proposed High-Level Plan
} 
\begin{itemize}
    \item \textbf{Step 1}
    \begin{itemize}
        \item builds rotated surface code memory circuits for d=3,5,7 with R=d rounds
        \item applies DEPOLARIZE p=0.003 only on CX gates
        \item samples 3M shots per distance
        \item decodes with PyMatching
        \item computes raw vs decoded logical error rates
    \end{itemize}
    \item \textbf{Step 2:}
    \begin{itemize}
        \item Execute code
        \item Collect results
        \item produce publication-quality plot: d vs logical error rate (raw, decoded, theoretical)
    \end{itemize}
\end{itemize}
\textbf{Should I proceed with this plan?}
}

\user{}

\quantumScientist{

\textbf{Action:} I'll proceed with the plan. Let me delegate to the CUDA-Q expert to implement the surface code memory experiment using Stim + PyMatching.

\textbf{Tool used:} cudaq expert

\textbf{Message2agent:} \textit{Omitting for brevity}

}

\cudaexpert{

\textbf{Action:} Let me implement the surface-code logical memory experiment with:
\begin{itemize}
    \item Stim for circuit generation and sampling
    \item PyMatching for MWPM decoding
    \item Two-qubit depolarizing noise on CNOT gates only
\end{itemize}

\textbf{Tool used:} submit\_python\_slurm

\textbf{Code:} \textit{Omitting for brevity}
}

\cudaexpert{

\textbf{Action:} The simulation completed successfully with excellent results. Now let me delegate the visualization to the quantum scientist expert to create the requested plot.

\textbf{Message2agent:} \textit{Omitting for brevity}
}

\quantumScientist{

\textbf{Action:} Task completed. Publication-quality plot created successfully for the surface code logical memory experiment.

\textbf{Plot Overview}\\
The plot displays logical error rate vs. code distance for a Z-basis surface code memory experiment with three curves:
\begin{itemize}
    \item Red(raw): Error rate without decoder. Increases with distance
    \item Blue (decoder): Error rate with MWPM decoder. Decreases with distance
    \item Green (Theory): Threshold model: $p_L \sim A(p/p_{th})^{(d+1)/2}$. 	Decreases with distance
\end{itemize}

\textbf{Physical Interpretation
}
\begin{itemize}
    \item \textbf{Why Raw Error INCREASES with Distance}
    \begin{itemize}
        \item More physical qubits $\to$ more CNOT gates $\to$ more noise opportunities
        \item Without correction, errors simply accumulate: $p_{raw} \propto (\text{number of CNOTs}) \times p$
        \item Scales roughly as $\sim d^2 \cdot d = d^3$ (qubits × rounds)
    \end{itemize}
    \item \textbf{Why Decoded Error DECREASES with Distance}
    \begin{itemize}
        \item Surface code can correct up to $\lfloor(d-1)/2\rfloor$ errors
        \item Below threshold, probability of $\geq d/2$ correlated errors is exponentially small
        \item \textbf{Exponential suppression:} $p_L \sim (p/p_{th})^{(d+1)/2}$
    \end{itemize}
    \item \textbf{Below-Threshold Regime}
    \begin{itemize}
        \item Physical error: $p = 0.003$
        \item Threshold: $p_{th} \approx 0.01$ (for depolarizing noise on surface code)
        \item Ratio: $p/p_{th} = 0.3$ — \textbf{well below threshold}
        \item This is the regime where QEC provides exponential benefit
    \end{itemize}
\end{itemize}

The plot demonstrates the central promise of quantum error correction: below the threshold, increasing code distance exponentially suppresses logical errors despite linearly increasing physical resources. At $p=0.003$, the surface code is firmly in the fault-tolerant regime.
}

\section{Evaluation Rubrics}\label{sec:eval_rubrics}

The benchmark task was graded on a 100-point scale using three fixed criteria. The benchmark covers five representative quantum-simulation problems: variational quantum eigensolver (VQE), Bell-state preparation, transverse-field Ising dynamics, open-system Lindblad dynamics, and hierarchical equations of motion (HEOM). Criterion A (30 points) evaluates the correctness of the technical implementation, including proper construction of the model and correct use of the target software framework. Criterion B (30 points) assesses the correctness and validation of the computed results, including appropriate observables, numerical consistency, and comparisons where required. In turn, criterion C (40 points) evaluates the clarity and quality of the final outputs, including figures, analysis, and physical interpretation. For each criterion, Tables \ref{tab:rubric_vqe}--\ref{tab:rubric_heom} provide explicit point assignments that define full, partial, and zero credit.

\begin{table}[ht!]
\centering
\caption{Grading rubric for the VQE benchmark (H$_2$).}
\label{tab:rubric_vqe}

\renewcommand{\arraystretch}{1.35}
\setlength{\tabcolsep}{6pt}

\begin{tabularx}{\linewidth}
{c c >{\raggedright\arraybackslash}X}
\toprule
\textbf{Criterion} & \textbf{Points} & \textbf{Description} \\
\midrule

\rowcolor{gray!10}
A & 30 & Correct Hamiltonian, ansatz, optimizer, and bond-distance sweep. \\
\rowcolor{gray!10}
  & 20 & Minor setup issues but VQE runs correctly. \\
\rowcolor{gray!10}
  & 10 & Incomplete or poorly configured VQE. \\
\rowcolor{gray!10}
  & 0  & No working VQE. \\
\midrule

B & 30 & Exact reference included and correctly compared. \\
  & 20 & Comparison shown but weakly justified. \\
  & 10 & Partial or unclear comparison. \\
  & 0  & No comparison to exact result. \\
\midrule

\rowcolor{gray!10}
C & 40 & Clear dissociation curve with correct physical interpretation. \\
\rowcolor{gray!10}
  & 30 & Correct plot with limited physical insight. \\
\rowcolor{gray!10}
  & 20 & Plot present but interpretation weak. \\
\rowcolor{gray!10}
  & 0--10 & Missing or incorrect plot. \\
\bottomrule
\end{tabularx}
\end{table}

\begin{table}[ht!]
\centering
\caption{Grading rubric for the Bell-state preparation benchmark.}
\label{tab:rubric_bell}

\renewcommand{\arraystretch}{1.35}
\setlength{\tabcolsep}{6pt}

\begin{tabularx}{\linewidth}
{c c >{\raggedright\arraybackslash}X}
\toprule
\textbf{Criterion} & \textbf{Points} & \textbf{Description} \\
\midrule

\rowcolor{gray!10}
A & 30 & Correct Bell-state circuit, measurement bases, and shot count. \\
\rowcolor{gray!10}
  & 20 & Minor circuit or basis issues with correct intent. \\
\rowcolor{gray!10}
  & 10 & Circuit mostly correct but measurements flawed. \\
\rowcolor{gray!10}
  & 0  & Incorrect circuit or missing measurements. \\
\midrule

B & 30 & Correct computation of $\langle Z\!\otimes\!Z\rangle$ and $\langle X\!\otimes\!X\rangle$. \\
  & 20 & Correct method with numerical or normalization errors. \\
  & 10 & Partial or unclear computation. \\
  & 0  & Observables not computed. \\
\midrule

\rowcolor{gray!10}
C & 40 & Clear plots with correct interpretation. \\
\rowcolor{gray!10}
  & 30 & Correct plots with superficial analysis. \\
\rowcolor{gray!10}
  & 20 & Poorly explained or hard-to-read plots. \\
\rowcolor{gray!10}
  & 0--10 & No meaningful plot or analysis. \\
\bottomrule
\end{tabularx}
\end{table}

\begin{table}[ht!]
\centering
\caption{Grading rubric for the transverse-field Ising dynamics benchmark.}
\label{tab:rubric_ising}

\renewcommand{\arraystretch}{1.35}
\setlength{\tabcolsep}{6pt}

\begin{tabularx}{\linewidth}
{c c >{\raggedright\arraybackslash}X}
\toprule
\textbf{Criterion} & \textbf{Points} & \textbf{Description} \\
\midrule

\rowcolor{gray!10}
A & 30 & Correct Hamiltonian, evolution method, parameters, and initialization. \\
\rowcolor{gray!10}
  & 20 & Minor errors with sensible dynamics. \\
\rowcolor{gray!10}
  & 10 & Significant issues in implementation. \\
\rowcolor{gray!10}
  & 0  & Incorrect time evolution. \\
\midrule

B & 30 & Correct extraction of $\langle Z\rangle$ for all qubits. \\
  & 20 & Partial or noisy measurements. \\
  & 10 & Incorrect observable handling. \\
  & 0  & No meaningful measurement. \\
\midrule

\rowcolor{gray!10}
C & 40 & Clear heatmaps with insightful regime comparison. \\
\rowcolor{gray!10}
  & 30 & Correct heatmaps with brief analysis. \\
\rowcolor{gray!10}
  & 20 & Poor visualization. \\
\rowcolor{gray!10}
  & 0--10 & Missing or unreadable plots. \\
\bottomrule
\end{tabularx}
\end{table}

\begin{table}[ht!]
\centering
\caption{Grading rubric for the Lindblad dynamics benchmark.}
\label{tab:rubric_lindblad}

\renewcommand{\arraystretch}{1.35}
\setlength{\tabcolsep}{6pt}

\begin{tabularx}{\linewidth}
{c c >{\raggedright\arraybackslash}X}
\toprule
\textbf{Criterion} & \textbf{Points} & \textbf{Description} \\
\midrule

\rowcolor{gray!10}
A & 30 & Correct Hamiltonian, collapse operators, and solver configuration. \\
\rowcolor{gray!10}
  & 20 & Minor implementation issues. \\
\rowcolor{gray!10}
  & 10 & Incomplete setup. \\
\rowcolor{gray!10}
  & 0  & Incorrect dynamics. \\
\midrule

B & 30 & Correct $\langle Z(t)\rangle$ for both initial states. \\
  & 20 & Partial or noisy results. \\
  & 10 & Incorrect observables. \\
  & 0  & Missing data. \\
\midrule

\rowcolor{gray!10}
C & 40 & Clear comparison with correct physical explanation. \\
\rowcolor{gray!10}
  & 30 & Mostly correct explanation. \\
\rowcolor{gray!10}
  & 20 & Limited insight. \\
\rowcolor{gray!10}
  & 0--10 & No meaningful discussion. \\
\bottomrule
\end{tabularx}
\end{table}

\begin{table}[ht!]
\centering
\caption{Grading rubric for the HEOM benchmark.}
\label{tab:rubric_heom}

\renewcommand{\arraystretch}{1.35}
\setlength{\tabcolsep}{6pt}

\begin{tabularx}{\linewidth}
{c c >{\raggedright\arraybackslash}X}
\toprule
\textbf{Criterion} & \textbf{Points} & \textbf{Description} \\
\midrule

\rowcolor{gray!10}
A & 30 & Correct Hamiltonian, bath parameters, hierarchy, and time grid. \\
\rowcolor{gray!10}
  & 20 & Minor setup or parameter issues. \\
\rowcolor{gray!10}
  & 10 & Incomplete setup. \\
\rowcolor{gray!10}
  & 0  & Incorrect formulation. \\
\midrule

B & 30 & Stable propagation with physically consistent populations. \\
  & 20 & Partial or weak validation. \\
  & 10 & Unclear or poorly validated dynamics. \\
  & 0  & No validation or unstable evolution. \\
\midrule

\rowcolor{gray!10}
C & 40 & Clear population plots at both temperatures with correct physical interpretation. \\
\rowcolor{gray!10}
  & 30 & Correct results with limited clarity or depth. \\
\rowcolor{gray!10}
  & 20 & Poor visualization or weak interpretation. \\
\rowcolor{gray!10}
  & 0--10 & Missing or incorrect outputs. \\
\bottomrule

\end{tabularx}
\end{table}


%% file: preprint.bbl
\begin{thebibliography}{156}
\providecommand{\natexlab}[1]{#1}
\providecommand{\url}[1]{\texttt{#1}}
\expandafter\ifx\csname urlstyle\endcsname\relax
  \providecommand{\doi}[1]{doi: #1}\else
  \providecommand{\doi}{doi: \begingroup \urlstyle{rm}\Url}\fi

\bibitem[Levine(2014)]{levine2009quantum}
Ira~N. Levine.
\newblock \emph{Quantum Chemistry}.
\newblock Pearson, seventh edition, 2014.
\newblock ISBN 978-0-321-80345-0.

\bibitem[Cao et~al.(2019)Cao, Romero, Olson, Degroote, Johnson, Kieferov{\'a}, Kivlichan, Menke, Peropadre, Sawaya, et~al.]{cao2019quantum}
Yudong Cao, Jonathan Romero, Jonathan~P. Olson, Matthias Degroote, Peter~D. Johnson, M{\'a}ria Kieferov{\'a}, Ian~D. Kivlichan, Tim Menke, Borja Peropadre, Nicolas~P.D. Sawaya, et~al.
\newblock Quantum chemistry in the age of quantum computing.
\newblock \emph{Chem. Rev.}, 119\penalty0 (19):\penalty0 10856--10915, 2019.
\newblock \doi{10.1021/acs.chemrev.8b00803}.

\bibitem[Marder(2010)]{marder2010condensed}
Michael~P. Marder.
\newblock \emph{Condensed matter physics}.
\newblock John Wiley \& Sons, 2010.
\newblock ISBN 9780470617984.

\bibitem[Anderson(2018)]{anderson2018basic}
Philip~W. Anderson.
\newblock \emph{Basic notions of condensed matter physics}.
\newblock CRC press, 2018.
\newblock ISBN 0805302204.

\bibitem[Chaikin et~al.(1995)Chaikin, Lubensky, and Witten]{chaikin1995principles}
Paul~M. Chaikin, Tom~C. Lubensky, and Thomas~A. Witten.
\newblock \emph{Principles of condensed matter physics}, volume~10.
\newblock Cambridge university press, 1995.
\newblock ISBN 9781139643054.

\bibitem[Scully and Zubairy(1997)]{scully1997quantum}
Marlan~O. Scully and M.~Suhail Zubairy.
\newblock \emph{Quantum optics}.
\newblock Cambridge university press, 1997.
\newblock ISBN 9780511813993.

\bibitem[Gerry and Knight(2023)]{gerry2023introductory}
Christopher~C. Gerry and Peter~L. Knight.
\newblock \emph{Introductory quantum optics}.
\newblock Cambridge university press, 2023.
\newblock ISBN 9780511791239.

\bibitem[Longair(2011)]{longair2011high}
Malcolm~S. Longair.
\newblock \emph{High energy astrophysics}.
\newblock Cambridge university press, 2011.
\newblock ISBN 9780511778346.

\bibitem[Perkins(2000)]{perkins2000introduction}
Donald~H. Perkins.
\newblock \emph{Introduction to high energy physics}.
\newblock Cambridge university press, 2000.
\newblock ISBN 9780511809040.

\bibitem[Chawla(2012)]{chawla2012composite}
Krishan~K. Chawla.
\newblock \emph{Composite materials: science and engineering}.
\newblock Springer Science \& Business Media, 2012.
\newblock ISBN 9780387743646.

\bibitem[Callister and Rethwisch(2000)]{callister2000fundamentals}
William~D. Callister and David~G. Rethwisch.
\newblock \emph{Fundamentals of materials science and engineering}.
\newblock Wiley London, 2000.
\newblock ISBN 9781118319222.

\bibitem[Georgescu et~al.(2014)Georgescu, Ashhab, and Nori]{georgescu2014quantum}
Iulia~M. Georgescu, Sahel Ashhab, and Franco Nori.
\newblock Quantum simulation.
\newblock \emph{Rev. Mod. Phys.}, 86\penalty0 (1):\penalty0 153--185, 2014.
\newblock \doi{10.1103/revmodphys.86.153}.

\bibitem[Troyer and Wiese(2005)]{troyer2005computational}
Matthias Troyer and Uwe-Jens Wiese.
\newblock Computational complexity and fundamental limitations to fermionic quantum monte carlo simulations.
\newblock \emph{Phys. Rev. Lett.}, 94\penalty0 (17):\penalty0 170201, 2005.
\newblock \doi{10.1103/physrevlett.94.170201}.

\bibitem[Aspuru-Guzik et~al.(2005)Aspuru-Guzik, Dutoi, Love, and Head-Gordon]{aspuru2005simulated}
Al{\'a}n Aspuru-Guzik, Anthony~D. Dutoi, Peter~J. Love, and Martin Head-Gordon.
\newblock Simulated quantum computation of molecular energies.
\newblock \emph{Science}, 309\penalty0 (5741):\penalty0 1704--1707, 2005.
\newblock \doi{10.1126/science.1113479}.

\bibitem[Schollw{\"o}ck(2011)]{schollwock2011density}
Ulrich Schollw{\"o}ck.
\newblock The density-matrix renormalization group in the age of matrix product states.
\newblock \emph{Ann. Phys.}, 326\penalty0 (1):\penalty0 96--192, 2011.
\newblock \doi{10.1016/j.aop.2010.09.012}.

\bibitem[Eisert et~al.(2010)Eisert, Cramer, and Plenio]{eisert2010colloquium}
Jens Eisert, Marcus Cramer, and Martin~B. Plenio.
\newblock Colloquium: Area laws for the entanglement entropy.
\newblock \emph{Rev. Mod. Phys.}, 82\penalty0 (1):\penalty0 277--306, 2010.
\newblock \doi{10.1103/revmodphys.82.277}.

\bibitem[Or{\'u}s(2014)]{orus2014practical}
Rom{\'a}n Or{\'u}s.
\newblock A practical introduction to tensor networks: Matrix product states and projected entangled pair states.
\newblock \emph{Ann. Phys.}, 349:\penalty0 117--158, 2014.
\newblock \doi{10.1016/j.aop.2014.06.013}.

\bibitem[De~Vega and Alonso(2017)]{de2017dynamics}
In{\'e}s De~Vega and Daniel Alonso.
\newblock Dynamics of non-markovian open quantum systems.
\newblock \emph{Rev. Mod. Phys.}, 89\penalty0 (1):\penalty0 015001, 2017.
\newblock \doi{10.1103/revmodphys.89.015001}.

\bibitem[Beck et~al.(2000)Beck, J{\"a}ckle, Worth, and Meyer]{beck2000multiconfiguration}
Michael~H. Beck, Andreas J{\"a}ckle, Graham~A. Worth, and H-D Meyer.
\newblock The multiconfiguration time-dependent hartree (mctdh) method: a highly efficient algorithm for propagating wavepackets.
\newblock \emph{Phys. Rep.}, 324\penalty0 (1):\penalty0 1--105, 2000.
\newblock \doi{10.1016/s0370-1573(99)00047-2}.

\bibitem[Chen and Franco(2025)]{Chen2025}
Xinxian Chen and Ignacio Franco.
\newblock Tree tensor network hierarchical equations of motion based on time-dependent variational principle for efficient open quantum dynamics in structured thermal environments.
\newblock \emph{J. Chem. Phys.}, 163:\penalty0 104109, 9 2025.
\newblock \doi{10.1063/5.0278591}.

\bibitem[Peruzzo et~al.(2014)Peruzzo, McClean, Shadbolt, Yung, Zhou, Love, Aspuru-Guzik, and O’brien]{peruzzo2014variational}
Alberto Peruzzo, Jarrod McClean, Peter Shadbolt, Man-Hong Yung, Xiao-Qi Zhou, Peter~J. Love, Al{\'a}n Aspuru-Guzik, and Jeremy~L. O’brien.
\newblock A variational eigenvalue solver on a photonic quantum processor.
\newblock \emph{Nat. Commun.}, 5\penalty0 (1):\penalty0 4213, 2014.
\newblock \doi{10.1038/ncomms5213}.

\bibitem[McClean et~al.(2016)McClean, Romero, Babbush, and Aspuru-Guzik]{mcclean2016theory}
Jarrod~R. McClean, Jonathan Romero, Ryan Babbush, and Al{\'a}n Aspuru-Guzik.
\newblock The theory of variational hybrid quantum-classical algorithms.
\newblock \emph{New J. Phys.}, 18\penalty0 (2):\penalty0 023023, 2016.
\newblock \doi{10.1088/1367-2630/18/2/023023}.

\bibitem[Cerezo et~al.(2021)Cerezo, Arrasmith, Babbush, Benjamin, Endo, Fujii, McClean, Mitarai, Yuan, Cincio, et~al.]{cerezo2021variational}
Marco Cerezo, Andrew Arrasmith, Ryan Babbush, Simon~C. Benjamin, Suguru Endo, Keisuke Fujii, Jarrod~R. McClean, Kosuke Mitarai, Xiao Yuan, Lukasz Cincio, et~al.
\newblock Variational quantum algorithms.
\newblock \emph{Nat. Rev. Phys.}, 3\penalty0 (9):\penalty0 625--644, 2021.
\newblock \doi{10.1038/s42254-021-00348-9}.

\bibitem[Calderón et~al.(2024)Calderón, Raussendorf, Feldmann, and Bondarenko]{mantilla2025mbqml}
Luis~Mantilla Calderón, Robert Raussendorf, Polina Feldmann, and Dmytro Bondarenko.
\newblock Measurement-based quantum machine learning.
\newblock \emph{ArXiv preprint arXiv:2405.08319}, 2024.
\newblock \doi{10.48550/arXiv.2405.08319}.

\bibitem[Bharti et~al.(2022)Bharti, Cervera-Lierta, Kyaw, Haug, Alperin-Lea, Anand, Degroote, Heimonen, Kottmann, Menke, et~al.]{bharti2022noisy}
Kishor Bharti, Alba Cervera-Lierta, Thi~Ha Kyaw, Tobias Haug, Sumner Alperin-Lea, Abhinav Anand, Matthias Degroote, Hermanni Heimonen, Jakob~S Kottmann, Tim Menke, et~al.
\newblock Noisy intermediate-scale quantum algorithms.
\newblock \emph{Reviews of Modern Physics}, 94\penalty0 (1):\penalty0 015004, 2022.

\bibitem[Heim et~al.(2020)Heim, Soeken, Marshall, Granade, Roetteler, Geller, Troyer, and Svore]{heim2020quantum}
Bettina Heim, Mathias Soeken, Sarah Marshall, Chris Granade, Martin Roetteler, Alan Geller, Matthias Troyer, and Krysta Svore.
\newblock Quantum programming languages.
\newblock \emph{Nat. Rev. Phys.}, 2\penalty0 (12):\penalty0 709--722, 2020.
\newblock \doi{10.1038/s42254-020-00245-7}.

\bibitem[Fingerhuth et~al.(2018)Fingerhuth, Babej, and Wittek]{fingerhuth2018open}
Mark Fingerhuth, Tom{\'a}{\v{s}} Babej, and Peter Wittek.
\newblock Open source software in quantum computing.
\newblock \emph{PLoS ONE}, 13\penalty0 (12):\penalty0 e0208561, 2018.
\newblock \doi{10.1371/journal.pone.0208561}.

\bibitem[LaRose(2019)]{larose2019overview}
Ryan LaRose.
\newblock Overview and comparison of gate level quantum software platforms.
\newblock \emph{Quantum}, 3:\penalty0 130, 2019.
\newblock \doi{10.22331/q-2019-03-25-130}.

\bibitem[Ramos et~al.(2025)Ramos, Collison, and White]{ramos2025review}
Mayk~Caldas Ramos, Christopher~J Collison, and Andrew~D White.
\newblock A review of large language models and autonomous agents in chemistry.
\newblock \emph{Chem. Sci.}, 2025.
\newblock \doi{10.1039/d4sc03921a}.

\bibitem[Wu et~al.(2024)Wu, Zheng, Qiu, Wang, Gu, Shen, Qin, Zhu, Zhu, Liu, et~al.]{wu2024survey}
Likang Wu, Zhi Zheng, Zhaopeng Qiu, Hao Wang, Hongchao Gu, Tingjia Shen, Chuan Qin, Chen Zhu, Hengshu Zhu, Qi~Liu, et~al.
\newblock A survey on large language models for recommendation.
\newblock \emph{World Wide Web}, 27\penalty0 (5):\penalty0 60, 2024.
\newblock \doi{10.1007/s11280-024-01291-2}.

\bibitem[Brown et~al.(2020)Brown, Mann, Ryder, Subbiah, Kaplan, Dhariwal, Neelakantan, Shyam, Sastry, Askell, Agarwal, Herbert-Voss, Krueger, Henighan, Child, Ramesh, Ziegler, Wu, Winter, Hesse, Chen, Sigler, Litwin, Gray, Chess, Clark, Berner, McCandlish, Radford, Sutskever, and Amodei]{brown2020language}
Tom~B. Brown, Benjamin Mann, Nick Ryder, Melanie Subbiah, Jared Kaplan, Prafulla Dhariwal, Arvind Neelakantan, Pranav Shyam, Girish Sastry, Amanda Askell, Sandhini Agarwal, Ariel Herbert-Voss, Gretchen Krueger, Tom Henighan, Rewon Child, Aditya Ramesh, Daniel~M. Ziegler, Jeffrey Wu, Clemens Winter, Christopher Hesse, Mark Chen, Eric Sigler, Mateusz Litwin, Scott Gray, Benjamin Chess, Jack Clark, Christopher Berner, Sam McCandlish, Alec Radford, Ilya Sutskever, and Dario Amodei.
\newblock Language models are few-shot learners.
\newblock In \emph{Adv. Neural Inf. Process. Syst. (NeurIPS)}, NIPS '20, Red Hook, NY, USA, 2020. Curran Associates Inc.
\newblock ISBN 9781713829546.

\bibitem[Ouyang et~al.(2022)Ouyang, Wu, Jiang, Almeida, Wainwright, Mishkin, Zhang, Agarwal, Slama, Ray, Schulman, Hilton, Kelton, Miller, Simens, Askell, Welinder, Christiano, Leike, and Lowe]{ouyang2022training}
Long Ouyang, Jeffrey Wu, Xu~Jiang, Diogo Almeida, Carroll Wainwright, Pamela Mishkin, Chong Zhang, Sandhini Agarwal, Katarina Slama, Alex Ray, John Schulman, Jacob Hilton, Fraser Kelton, Luke Miller, Maddie Simens, Amanda Askell, Peter Welinder, Paul~F. Christiano, Jan Leike, and Ryan Lowe.
\newblock Training language models to follow instructions with human feedback.
\newblock In S.~Koyejo, S.~Mohamed, A.~Agarwal, D.~Belgrave, K.~Cho, and A.~Oh, editors, \emph{Adv. Neural Inf. Process. Syst.}, volume~35, pages 27730--27744. Curran Associates, Inc., 2022.
\newblock ISBN 9781713871088.

\bibitem[Christiano et~al.(2017)Christiano, Leike, Brown, Martic, Legg, and Amodei]{christiano2017deep}
Paul~F. Christiano, Jan Leike, Tom~B. Brown, Miljan Martic, Shane Legg, and Dario Amodei.
\newblock Deep reinforcement learning from human preferences.
\newblock In \emph{Adv. Neural Inf. Process. Syst.}, NIPS'17, page 4302–4310, Red Hook, NY, USA, 2017. Curran Associates Inc.
\newblock ISBN 9781510860964.

\bibitem[Chen(2021)]{chen2021evaluating}
Mark Chen.
\newblock Evaluating large language models trained on code.
\newblock \emph{arXiv preprint arXiv:2107.03374}, 2021.
\newblock \doi{10.48550/arXiv.2107.03374}.

\bibitem[Yao et~al.(2023{\natexlab{a}})Yao, Zhao, Yu, Du, Shafran, Narasimhan, and Cao]{yao2022react}
Shunyu Yao, Jeffrey Zhao, Dian Yu, Nan Du, Izhak Shafran, Karthik~R Narasimhan, and Yuan Cao.
\newblock React: Synergizing reasoning and acting in language models.
\newblock In \emph{Int. Conf. Learn. Represent. (ICLR)}, 2023{\natexlab{a}}.
\newblock URL \url{https://openreview.net/forum?id=WE_vluYUL-X}.

\bibitem[Madaan et~al.(2023)Madaan, Tandon, Gupta, Hallinan, Gao, Wiegreffe, Alon, Dziri, Prabhumoye, Yang, Gupta, Majumder, Hermann, Welleck, Yazdanbakhsh, and Clark]{madaan2023self}
Aman Madaan, Niket Tandon, Prakhar Gupta, Skyler Hallinan, Luyu Gao, Sarah Wiegreffe, Uri Alon, Nouha Dziri, Shrimai Prabhumoye, Yiming Yang, Shashank Gupta, Bodhisattwa~Prasad Majumder, Katherine Hermann, Sean Welleck, Amir Yazdanbakhsh, and Peter Clark.
\newblock Self-refine: Iterative refinement with self-feedback.
\newblock In \emph{Adv. Neural Inf. Process. Syst. (NeurIPS)}, 2023.
\newblock URL \url{https://openreview.net/forum?id=S37hOerQLB}.

\bibitem[Wei et~al.(2022)Wei, Wang, Schuurmans, Bosma, ichter, Xia, Chi, Le, and Zhou]{wei2022chain}
Jason Wei, Xuezhi Wang, Dale Schuurmans, Maarten Bosma, brian ichter, Fei Xia, Ed~Chi, Quoc~V Le, and Denny Zhou.
\newblock Chain-of-thought prompting elicits reasoning in large language models.
\newblock In S.~Koyejo, S.~Mohamed, A.~Agarwal, D.~Belgrave, K.~Cho, and A.~Oh, editors, \emph{Adv. Neural Inf. Process. Syst.}, volume~35, pages 24824--24837. Curran Associates, Inc., 2022.
\newblock URL \url{https://proceedings.neurips.cc/paper_files/paper/2022/file/9d5609613524ecf4f15af0f7b31abca4-Paper-Conference.pdf}.

\bibitem[OpenAI(2024)]{openai2024}
OpenAI.
\newblock Openai o1 system card.
\newblock \emph{arXiv preprint arXiv:2412.16720}, 2024.
\newblock \doi{10.48550/arXiv.2412.16720}.

\bibitem[DeepSeek-AI(2025)]{deepseek-ai2025}
DeepSeek-AI.
\newblock Deepseek-r1: Incentivizing reasoning capability in llms via reinforcement learning.
\newblock \emph{arXiv preprint arXiv:2501.12948}, 2025.
\newblock \doi{10.48550/arXiv.2501.12948}.
\newblock URL \url{https://arxiv.org/abs/2501.12948}.

\bibitem[Gemini~Team(2025)]{comanici2025}
Google Gemini~Team.
\newblock Gemini 2.5: Pushing the frontier with advanced reasoning, multimodality, long context, and next generation agentic capabilities.
\newblock \emph{arXiv preprint arXiv:2507.06261}, 2025.
\newblock \doi{10.48550/arXiv.2507.06261}.
\newblock URL \url{https://arxiv.org/abs/2507.06261}.

\bibitem[Nvidia et~al.(2024)Nvidia, :, Adler, Agarwal, Aithal, Anh, Bhattacharya, Brundyn, Casper, Catanzaro, Clay, Cohen, Das, Dattagupta, Delalleau, Derczynski, Dong, Egert, Evans, Ficek, Fridman, Ghosh, Ginsburg, Gitman, Grzegorzek, Hero, Huang, Jawa, Jennings, Jhunjhunwala, Kamalu, Khan, Kuchaiev, LeGresley, Li, Liu, Liu, Long, Mahabaleshwarkar, Majumdar, Maki, Martinez, de~Melo, Moshkov, Narayanan, Narenthiran, Navarro, Nguyen, Nitski, Noroozi, Nutheti, Parisien, Parmar, Patwary, Pawelec, Ping, Prabhumoye, Roy, Saar, Sabavat, Satheesh, Scowcroft, Sewall, Shamis, Shen, Shoeybi, Sizer, Smelyanskiy, Soares, Sreedhar, Su, Subramanian, Sun, Toshniwal, Wang, Wang, You, Zeng, Zhang, Zhang, Zhang, Zhang, and Zhu]{nvidia2024}
Nvidia, :, Bo~Adler, Niket Agarwal, Ashwath Aithal, Dong~H. Anh, Pallab Bhattacharya, Annika Brundyn, Jared Casper, Bryan Catanzaro, Sharon Clay, Jonathan Cohen, Sirshak Das, Ayush Dattagupta, Olivier Delalleau, Leon Derczynski, Yi~Dong, Daniel Egert, Ellie Evans, Aleksander Ficek, Denys Fridman, Shaona Ghosh, Boris Ginsburg, Igor Gitman, Tomasz Grzegorzek, Robert Hero, Jining Huang, Vibhu Jawa, Joseph Jennings, Aastha Jhunjhunwala, John Kamalu, Sadaf Khan, Oleksii Kuchaiev, Patrick LeGresley, Hui Li, Jiwei Liu, Zihan Liu, Eileen Long, Ameya~Sunil Mahabaleshwarkar, Somshubra Majumdar, James Maki, Miguel Martinez, Maer~Rodrigues de~Melo, Ivan Moshkov, Deepak Narayanan, Sean Narenthiran, Jesus Navarro, Phong Nguyen, Osvald Nitski, Vahid Noroozi, Guruprasad Nutheti, Christopher Parisien, Jupinder Parmar, Mostofa Patwary, Krzysztof Pawelec, Wei Ping, Shrimai Prabhumoye, Rajarshi Roy, Trisha Saar, Vasanth Rao~Naik Sabavat, Sanjeev Satheesh, Jane~Polak Scowcroft, Jason Sewall, Pavel Shamis, Gerald Shen, Mohammad
  Shoeybi, Dave Sizer, Misha Smelyanskiy, Felipe Soares, Makesh~Narsimhan Sreedhar, Dan Su, Sandeep Subramanian, Shengyang Sun, Shubham Toshniwal, Hao Wang, Zhilin Wang, Jiaxuan You, Jiaqi Zeng, Jimmy Zhang, Jing Zhang, Vivienne Zhang, Yian Zhang, and Chen Zhu.
\newblock Nemotron-4 340b technical report.
\newblock \emph{arXiv preprint arXiv:2406.11704}, 2024.
\newblock \doi{10.48550/arXiv.2406.11704}.
\newblock URL \url{https://arxiv.org/abs/2406.11704}.

\bibitem[{OpenAI}(2025)]{openai_gpt5_card}
{OpenAI}.
\newblock Gpt-5.2 system card.
\newblock System card, OpenAI, 2025.
\newblock URL \url{https://cdn.openai.com/pdf/3a4153c8-c748-4b71-8e31-aecbde944f8d/oai_5_2_system-card.pdf}.

\bibitem[{Google DeepMind}(2024)]{deepmind_gemini_card}
{Google DeepMind}.
\newblock Gemini 3 pro model card.
\newblock \url{https://storage.googleapis.com/deepmind-media/Model-Cards/Gemini-3-Pro-Model-Card.pdf}, 2024.

\bibitem[{Anthropic}(2025)]{anthropic_claude}
{Anthropic}.
\newblock Claude opus 4.5 system card.
\newblock System card, Anthropic, 2025.
\newblock URL \url{https://assets.anthropic.com/m/64823ba7485345a7/Claude-Opus-4-5-System-Card.pdf}.

\bibitem[Wei et~al.(2025)Wei, Yang, Zhang, Chen, Zhuang, Gao, Zhou, Wang, Gao, Cao, et~al.]{wei2025ai}
Jiaqi Wei, Yuejin Yang, Xiang Zhang, Yuhan Chen, Xiang Zhuang, Zhangyang Gao, Dongzhan Zhou, Guangshuai Wang, Zhiqiang Gao, Juntai Cao, et~al.
\newblock From ai for science to agentic science: A survey on autonomous scientific discovery.
\newblock \emph{arXiv preprint arXiv:2508.14111}, 2025.
\newblock \doi{10.48550/arXiv.2508.14111}.

\bibitem[Huang et~al.(2025)Huang, Zhang, Wang, Qu, Lu, Roohani, Li, Qiu, Li, Zhang, et~al.]{huang2025biomni}
Kexin Huang, Serena Zhang, Hanchen Wang, Yuanhao Qu, Yingzhou Lu, Yusuf Roohani, Ryan Li, Lin Qiu, Gavin Li, Junze Zhang, et~al.
\newblock Biomni: A general-purpose biomedical ai agent.
\newblock \emph{bioRxiv}, 2025.
\newblock \doi{10.1101/2025.05.30.656746}.

\bibitem[Team et~al.(2025)Team, Zhang, Feng, Yan, Yuan, Yu, He, Huang, Hou, Nie, et~al.]{team2025novelseek}
NovelSeek Team, Bo~Zhang, Shiyang Feng, Xiangchao Yan, Jiakang Yuan, Zhiyin Yu, Xiaohan He, Songtao Huang, Shaowei Hou, Zheng Nie, et~al.
\newblock Novelseek: When agent becomes the scientist--building closed-loop system from hypothesis to verification.
\newblock \emph{arXiv preprint arXiv:2505.16938}, 2025.
\newblock \doi{10.48550/arXiv.2505.16938}.

\bibitem[Lála et~al.(2023)Lála, O'Donoghue, Shtedritski, Cox, Rodriques, and White]{lala2023paperqa}
Jakub Lála, Odhran O'Donoghue, Aleksandar Shtedritski, Sam Cox, Samuel~G. Rodriques, and Andrew~D. White.
\newblock Paperqa: Retrieval-augmented generative agent for scientific research.
\newblock \emph{arXiv preprint arXiv:2312.07559}, 2023.
\newblock \doi{10.48550/arXiv.2312.07559}.

\bibitem[Mitchener et~al.(2025)Mitchener, Yiu, Chang, Bourdenx, Nadolski, Sulovari, Landsness, Barabasi, Narayanan, Evans, et~al.]{mitchener2025kosmos}
Ludovico Mitchener, Angela Yiu, Benjamin Chang, Mathieu Bourdenx, Tyler Nadolski, Arvis Sulovari, Eric~C Landsness, Daniel~L Barabasi, Siddharth Narayanan, Nicky Evans, et~al.
\newblock Kosmos: An ai scientist for autonomous discovery.
\newblock \emph{arXiv preprint arXiv:2511.02824}, 2025.
\newblock \doi{10.48550/arXiv.2511.02824}.

\bibitem[Swanson et~al.(2025)Swanson, Wu, Bulaong, Pak, and Zou]{swanson2025virtuallab}
Kyle Swanson, Wesley Wu, Nash~L. Bulaong, John~E. Pak, and James~Z. Zou.
\newblock The virtual lab of ai agents designs new sars-cov-2 nanobodies.
\newblock \emph{Nature}, 2025.
\newblock \doi{10.1038/s41586-025-09442-9}.

\bibitem[Lu et~al.(2024)Lu, Lu, Lange, Foerster, Clune, and Ha]{lu2024ai}
Chris Lu, Cong Lu, Robert~Tjarko Lange, Jakob Foerster, Jeff Clune, and David Ha.
\newblock The ai scientist: Towards fully automated open-ended scientific discovery.
\newblock \emph{arXiv preprint arXiv:2408.06292}, 2024.
\newblock \doi{10.48550/arXiv.2408.06292}.

\bibitem[Yamada et~al.(2025)Yamada, Lange, Lu, Hu, Lu, Foerster, Clune, and Ha]{yamada2025ai}
Yutaro Yamada, Robert~Tjarko Lange, Cong Lu, Shengran Hu, Chris Lu, Jakob Foerster, Jeff Clune, and David Ha.
\newblock The ai scientist-v2: Workshop-level automated scientific discovery via agentic tree search.
\newblock \emph{arXiv preprint arXiv:2504.08066}, 2025.
\newblock \doi{10.48550/arXiv.2504.08066}.

\bibitem[Darvish et~al.(2024)Darvish, Skreta, Zhao, Yoshikawa, Som, Bogdanovic, and …]{darvish2024organa}
Kourosh Darvish, Marta Skreta, Yuchi Zhao, Naruki Yoshikawa, Sagnik Som, Miroslav Bogdanovic, and ….
\newblock Organa: A robotic assistant for automated chemistry experimentation and characterization.
\newblock \emph{arXiv preprint arXiv:2401.06949}, 2024.
\newblock \doi{10.48550/arXiv.2401.06949}.

\bibitem[Bran et~al.(2023)Bran, Cox, Schilter, Baldassari, White, and Schwaller]{bran2023chemcrow}
Andres~M Bran, Sam Cox, Oliver Schilter, Carlo Baldassari, Andrew~D White, and Philippe Schwaller.
\newblock Chemcrow: Augmenting large-language models with chemistry tools.
\newblock \emph{arXiv preprint arXiv:2304.05376}, 2023.
\newblock \doi{10.48550/arXiv.2304.05376}.

\bibitem[Boiko et~al.(2023)Boiko, MacKnight, Kline, and Gomes]{boiko2023autonomous}
Daniil~A Boiko, Robert MacKnight, Ben Kline, and Gabe Gomes.
\newblock Autonomous chemical research with large language models.
\newblock \emph{Nature}, 624\penalty0 (7992):\penalty0 570--578, 2023.
\newblock \doi{10.1038/s41586-023-06792-0}.

\bibitem[Tang et~al.(2025)Tang, Hu, Ye, Shao, Yin, Ouyang, Zhou, Lu, Zhang, Zhao, et~al.]{tang2025chemagent}
Xiangru Tang, Tianyu Hu, Muyang Ye, Yanjun Shao, Xunjian Yin, Siru Ouyang, Wangchunshu Zhou, Pan Lu, Zhuosheng Zhang, Yilun Zhao, et~al.
\newblock Chemagent: Self-updating library in large language models improves chemical reasoning.
\newblock \emph{arXiv preprint arXiv:2501.06590}, 2025.
\newblock \doi{10.48550/arXiv.2501.06590}.

\bibitem[Zou et~al.(2025)Zou, Cheng, Aldossary, Bai, Leong, Campos-Gonzalez-Angulo, Choi, Ser, Tom, Wang, Zhang, Yakavets, Hao, Crebolder, Bernales, and Aspuru-Guzik]{elagenteOG}
Yunheng Zou, Austin~H. Cheng, Abdulrahman Aldossary, Jiaru Bai, Shi~Xuan Leong, Jorge~Arturo Campos-Gonzalez-Angulo, Changhyeok Choi, Cher~Tian Ser, Gary Tom, Andrew Wang, Zijian Zhang, Ilya Yakavets, Han Hao, Chris Crebolder, Varinia Bernales, and Alán Aspuru-Guzik.
\newblock El agente: An autonomous agent for quantum chemistry.
\newblock \emph{Matter}, 8\penalty0 (7):\penalty0 102263, 2025.
\newblock ISSN 2590-2385.
\newblock \doi{https://doi.org/10.1016/j.matt.2025.102263}.

\bibitem[N{\"a}gele and Marquardt(2025)]{nagele2025agentic}
Maximilian N{\"a}gele and Florian Marquardt.
\newblock Agentic exploration of physics models.
\newblock \emph{arXiv preprint arXiv:2509.24978}, 2025.

\bibitem[Breen et~al.(2025)Breen, Del~Tredici, McCarran, Mijares, Yin, Sulimany, Taylor, Koppens, and Englund]{breen2025ax}
Benjamin Breen, Marco Del~Tredici, Jacob McCarran, Javier~Aspuru Mijares, Weichen~Winston Yin, Kfir Sulimany, Jacob~M Taylor, Frank~HL Koppens, and Dirk Englund.
\newblock Ax-prover: A deep reasoning agentic framework for theorem proving in mathematics and quantum physics.
\newblock \emph{arXiv preprint arXiv:2510.12787}, 2025.
\newblock \doi{10.48550/arXiv.2510.12787}.

\bibitem[Wang et~al.(2025)Wang, Zhang, Li, Hu, Mao, Chen, Du, Wang, Wu, Yang, et~al.]{wang2025starwhisper}
Cunshi Wang, Yu~Zhang, Yuyang Li, Xinjie Hu, Yiming Mao, Xunhao Chen, Pengliang Du, Rui Wang, Ying Wu, Hang Yang, et~al.
\newblock Starwhisper telescope: an ai framework for automating end-to-end astronomical observations.
\newblock \emph{Commun. Eng.}, 4\penalty0 (1):\penalty0 184, 2025.
\newblock \doi{10.1038/s44172-025-00520-4}.

\bibitem[Naumov et~al.(2025)Naumov, Zagirova, Lin, Xie, Gou, Urban, Tikhonova, Alawi, Durymanov, Galkin, et~al.]{naumov2025dora}
Vladimir Naumov, Diana Zagirova, Sha Lin, Yupeng Xie, Wenhao Gou, Anatoly Urban, Nina Tikhonova, Khadija Alawi, Mike Durymanov, Fedor Galkin, et~al.
\newblock Dora ai scientist: Multi-agent virtual research team for scientific exploration discovery and automated report generation.
\newblock \emph{bioRxiv}, 2025.
\newblock \doi{10.1101/2025.03.06.641840}.

\bibitem[Alexeev et~al.(2025)Alexeev, Farag, Patti, Wolf, Ares, Aspuru-Guzik, Benjamin, Cai, Cao, Chamberland, et~al.]{alexeev2025artificial}
Yuri Alexeev, Marwa~H. Farag, Taylor~L. Patti, Mark~E. Wolf, Natalia Ares, Al{\'a}n Aspuru-Guzik, Simon~C. Benjamin, Zhenyu Cai, Shuxiang Cao, Christopher Chamberland, et~al.
\newblock Artificial intelligence for quantum computing.
\newblock \emph{Nat. Commun.}, 16\penalty0 (1):\penalty0 10829, 2025.

\bibitem[Arlt et~al.(2025{\natexlab{a}})Arlt, Gu, and Krenn]{arlt2025towards}
S{\"o}ren Arlt, Xuemei Gu, and Mario Krenn.
\newblock Towards autonomous quantum physics research using llm agents with access to intelligent tools.
\newblock \emph{arXiv preprint arXiv:2511.11752}, 2025{\natexlab{a}}.

\bibitem[Cao et~al.(2025)Cao, Zhang, Alghadeer, Fasciati, Piscitelli, Bakr, Leek, and Aspuru-Guzik]{cao2025automating}
Shuxiang Cao, Zijian Zhang, Mohammed Alghadeer, Simone~D Fasciati, Michele Piscitelli, Mustafa Bakr, Peter Leek, and Al{\'a}n Aspuru-Guzik.
\newblock Automating quantum computing laboratory experiments with an agent-based ai framework.
\newblock \emph{Patterns}, 6\penalty0 (10), 2025.
\newblock \doi{10.1016/j.patter.2025.101372}.

\bibitem[Flores-Garrigos et~al.(2025)Flores-Garrigos, Dev, Falkenthal, Cadavid, Simen, Kumar, Solano, and Hegade]{flores2025quantum}
Carlos Flores-Garrigos, Gaurav Dev, Michael Falkenthal, Alejandro~Gomez Cadavid, Anton Simen, Shubham Kumar, Enrique Solano, and Narendra~N Hegade.
\newblock Quantum combinatorial reasoning for large language models.
\newblock \emph{arXiv preprint arXiv:2510.24509}, 2025.
\newblock \doi{10.48550/arXiv.2510.24509}.

\bibitem[Sharma et~al.(2025)Sharma, Ansari, Fu, Iyer, Matres, Tamas, Akdeniz, Englund, and Poon]{sharma2025towards}
Ankita Sharma, Vahid Ansari, Yuqi Fu, Rishabh Iyer, Joaquin Matres, Troy Tamas, Onur Akdeniz, Dirk~R Englund, and Joyce~KS Poon.
\newblock Towards ai agents for photonic integrated circuit design automation.
\newblock In \emph{CLEO: Science and Innovations}, page SS186\_1. Optica Publishing Group, 2025.
\newblock \doi{10.1364/cleo\_si.2025.ss186\_1}.

\bibitem[Saggio et~al.(2021)Saggio, Asenbeck, Hamann, Str{\"o}mberg, Schiansky, Dunjko, Friis, Harris, Hochberg, Englund, et~al.]{saggio2021experimental}
Valeria Saggio, Beate~E. Asenbeck, Arne Hamann, Teodor Str{\"o}mberg, Peter Schiansky, Vedran Dunjko, Nicolai Friis, Nicholas~C. Harris, Michael Hochberg, Dirk Englund, et~al.
\newblock Experimental quantum speed-up in reinforcement learning agents.
\newblock \emph{Nature}, 591\penalty0 (7849):\penalty0 229--233, 2021.

\bibitem[Elliott et~al.(2022)Elliott, Gu, Garner, and Thompson]{elliott2022quantum}
Thomas~J Elliott, Mile Gu, Andrew~JP Garner, and Jayne Thompson.
\newblock Quantum adaptive agents with efficient long-term memories.
\newblock \emph{Phys. Rev. X}, 12\penalty0 (1):\penalty0 011007, 2022.

\bibitem[Thompson et~al.(2025)Thompson, Riechers, Garner, Elliott, and Gu]{thompson2025energetic}
Jayne Thompson, Paul~M. Riechers, Andrew J.~P. Garner, Thomas~J. Elliott, and Mile Gu.
\newblock Energetic advantages for quantum agents in online execution of complex strategies.
\newblock \emph{Phys. Rev. Lett.}, 135\penalty0 (16):\penalty0 160402, 2025.

\bibitem[Yun et~al.(2022)Yun, Kwak, Kim, Cho, Jung, Park, and Kim]{yun2022quantum}
Won~Joon Yun, Yunseok Kwak, Jae~Pyoung Kim, Hyunhee Cho, Soyi Jung, Jihong Park, and Joongheon Kim.
\newblock Quantum multi-agent reinforcement learning via variational quantum circuit design.
\newblock In \emph{Proc. IEEE 42nd Int. Conf. Distrib. Comput. Syst. (ICDCS)}, pages 1332--1335. IEEE, 2022.

\bibitem[Sultanow et~al.(2025)Sultanow, Tehrani, Dutta, Buchanan, and Khan]{sultanow2025quantum}
Eldar Sultanow, Madjid Tehrani, Siddhant Dutta, William~J Buchanan, and Muhammad~Shahbaz Khan.
\newblock Quantum agents.
\newblock \emph{arXiv preprint arXiv:2506.01536}, 2025.

\bibitem[AI()]{lemma_website}
Axiomatic AI.
\newblock Lemma closed beta release.
\newblock URL \url{https://axiomatic-ai.com/products/lemma/}.
\newblock Accessed: 2025-12-15.

\bibitem[{Kipu Quantum}(2025)]{kipu_linkedin_roadmap}
{Kipu Quantum}.
\newblock Roadmap toward qc and qc+ai products (linkedin post).
\newblock LinkedIn, 2025.
\newblock URL \url{https://www.linkedin.com/posts/kipu-quantum_kipu-quantum-announces-the-dawn-of-industrial-activity-7401969816447819776-IE6w?utm_source=social_share_send&utm_medium=member_desktop_web&rcm=ACoAAD89xJkBnOvIyxxqAC0rLGy8nmks-NVSUGU}.

\bibitem[Team(2025)]{cuda-q_2025_15407754}
The CUDA-Q~Development Team.
\newblock Cuda-q, 2025.
\newblock URL \url{https://doi.org/10.5281/zenodo.15407754}.

\bibitem[Bergholm et~al.(2018)Bergholm, Izaac, Schuld, Gogolin, Ahmed, Ajith, Alam, Alonso-Linaje, AkashNarayanan, Asadi, et~al.]{bergholm2018pennylane}
Ville Bergholm, Josh Izaac, Maria Schuld, Christian Gogolin, Shahnawaz Ahmed, Vishnu Ajith, M~Sohaib Alam, Guillermo Alonso-Linaje, B~AkashNarayanan, Ali Asadi, et~al.
\newblock Pennylane: Automatic differentiation of hybrid quantum-classical computations.
\newblock \emph{arXiv preprint arXiv:1811.04968}, 2018.
\newblock \doi{10.48550/arXiv.1811.04968}.

\bibitem[Javadi-Abhari et~al.(2024)Javadi-Abhari, Treinish, Krsulich, Wood, Lishman, Gacon, Martiel, Nation, Bishop, Cross, et~al.]{javadiabhari2024quantumcomputingqiskit}
Ali Javadi-Abhari, Matthew Treinish, Kevin Krsulich, Christopher~J Wood, Jake Lishman, Julien Gacon, Simon Martiel, Paul~D Nation, Lev~S Bishop, Andrew~W Cross, et~al.
\newblock Quantum computing with qiskit.
\newblock \emph{arXiv preprint arXiv:2405.08810}, 2024.
\newblock \doi{10.48550/arXiv.2405.08810}.

\bibitem[Lambert et~al.(2026)Lambert, Gigu{\`e}re, Menczel, Li, Hopf, Su{\'a}rez, Gali, Lishman, Gadhvi, Agarwal, et~al.]{QuTiP2025}
Neill Lambert, Eric Gigu{\`e}re, Paul Menczel, Boxi Li, Patrick Hopf, Gerardo Su{\'a}rez, Marc Gali, Jake Lishman, Rushiraj Gadhvi, Rochisha Agarwal, et~al.
\newblock Qutip 5: The quantum toolbox in python.
\newblock \emph{Phys. Rep.}, 1153:\penalty0 1--62, 2026.
\newblock \doi{10.1016/j.physrep.2025.10.001}.

\bibitem[Hauschild et~al.(2024)Hauschild, Unfried, Anand, Andrews, Bintz, Borla, Divic, Drescher, Geiger, Hefel, Hémery, Kadow, Kemp, Kirchner, Liu, Möller, Parker, Rader, Romen, Scalet, Schoonderwoerd, Schulz, Soejima, Thoma, Wu, Zechmann, Zweng, Mong, Zaletel, and Pollmann]{tenpy2024}
Johannes Hauschild, Jakob Unfried, Sajant Anand, Bartholomew Andrews, Marcus Bintz, Umberto Borla, Stefan Divic, Markus Drescher, Jan Geiger, Martin Hefel, Kévin Hémery, Wilhelm Kadow, Jack Kemp, Nico Kirchner, Vincent~S. Liu, Gunnar Möller, Daniel Parker, Michael Rader, Anton Romen, Samuel Scalet, Leon Schoonderwoerd, Maximilian Schulz, Tomohiro Soejima, Philipp Thoma, Yantao Wu, Philip Zechmann, Ludwig Zweng, Roger S.~K. Mong, Michael~P. Zaletel, and Frank Pollmann.
\newblock {Tensor network Python (TeNPy) version 1}.
\newblock \emph{SciPost Phys. Codebases}, page~41, 2024.
\newblock \doi{10.21468/SciPostPhysCodeb.41}.
\newblock URL \url{https://scipost.org/10.21468/SciPostPhysCodeb.41}.

\bibitem[Kottmann et~al.(2020)Kottmann, Alperin-Lea, Tamayo-Mendoza, Cervera-Lierta, Lavigne, Tzu-Ching, Verteletsky, Schleich, Degroote, Chaney, Kesibo, Curnow, Solo, Tsilimigkounakis, Zendejas-Morales, Izmaylov, Aspuru-Guzik, Scala, and Saxena]{Kottmann_Tequila_A_platform_2020}
Jakob~S. Kottmann, Sumner Alperin-Lea, Teresa Tamayo-Mendoza, Alba Cervera-Lierta, Cyrille Lavigne, Yen Tzu-Ching, Vladislav Verteletsky, Philipp Schleich, Matthias Degroote, Skylar Chaney, Maha Kesibo, Naomi~G. Curnow, Brandon Solo, Georgios Tsilimigkounakis, Claudia Zendejas-Morales, Artur Izmaylov, Alan Aspuru-Guzik, Francesco Scala, and Gaurav Saxena.
\newblock {Tequila: A platform for rapid development of quantum algorithms}, November 2020.
\newblock URL \url{https://github.com/tequilahub/tequila}.

\bibitem[Wang et~al.(2023)Wang, Xie, Jiang, Mandlekar, Xiao, Zhu, Fan, and Anandkumar]{wang2023voyager}
Guanzhi Wang, Yuqi Xie, Yunfan Jiang, Ajay Mandlekar, Chaowei Xiao, Yuke Zhu, Linxi Fan, and Anima Anandkumar.
\newblock Voyager: An open-ended embodied agent with large language models.
\newblock \emph{arXiv preprint arXiv:2305.16291}, 2023.
\newblock \doi{10.48550/arXiv.2305.16291}.

\bibitem[{Significant Gravitas} et~al.(2023)]{AutoGPT}
{Significant Gravitas} et~al.
\newblock {Auto-GPT: An Autonomous GPT-4 Experiment}.
\newblock \url{https://github.com/Significant-Gravitas/Auto-GPT}, Apr 2023.
\newblock Accessed: 2025-12-16.

\bibitem[Yao et~al.(2023{\natexlab{b}})Yao, Zhao, Yu, Du, Shafran, Narasimhan, and Cao]{Yao2023react}
Shunyu Yao, Jeffrey Zhao, Dian Yu, Nan Du, Izhak Shafran, Karthik Narasimhan, and Yuan Cao.
\newblock React: Synergizing reasoning and acting in language models, 2023{\natexlab{b}}.
\newblock URL \url{https://arxiv.org/abs/2210.03629}.

\bibitem[Jordan and Wigner(1928)]{jordan1928paulische}
Pascual Jordan and Eugene Wigner.
\newblock {\"U}ber das paulische {\"a}quivalenzverbot.
\newblock \emph{Zeitschrift f{\"u}r Physik}, 47\penalty0 (9):\penalty0 631--651, 1928.
\newblock \doi{10.1007/BF01331938}.

\bibitem[Somma et~al.(2002)Somma, Ortiz, Gubernatis, Knill, and Laflamme]{somma2002simulating}
Rolando Somma, Gerardo Ortiz, James~E. Gubernatis, Emanuel Knill, and Raymond Laflamme.
\newblock Simulating physical phenomena by quantum networks.
\newblock \emph{Phys. Rev. A}, 65\penalty0 (4):\penalty0 042323, 2002.

\bibitem[Seeley et~al.(2012)Seeley, Richard, and Love]{seeley2012bravyi}
Jacob~T Seeley, Martin~J Richard, and Peter~J Love.
\newblock The bravyi-kitaev transformation for quantum computation of electronic structure.
\newblock \emph{J. Chem. Phys.}, 137\penalty0 (22), 2012.
\newblock \doi{10.1063/1.4768229}.

\bibitem[Bartlett and Musia{\l}(2007)]{bartlett2007coupled}
Rodney~J Bartlett and Monika Musia{\l}.
\newblock Coupled-cluster theory in quantum chemistry.
\newblock \emph{Rev. Mod. Phys.}, 79\penalty0 (1):\penalty0 291--352, 2007.
\newblock \doi{10.1103/revmodphys.79.291}.

\bibitem[Romero et~al.(2018)Romero, Babbush, McClean, Hempel, Love, and Aspuru-Guzik]{romero2018strategies}
Jonathan Romero, Ryan Babbush, Jarrod~R McClean, Cornelius Hempel, Peter~J Love, and Al{\'a}n Aspuru-Guzik.
\newblock Strategies for quantum computing molecular energies using the unitary coupled cluster ansatz.
\newblock \emph{Quantum Sci. Technol.}, 4\penalty0 (1):\penalty0 014008, 2018.
\newblock \doi{10.1088/2058-9565/aad3e4}.

\bibitem[Anand et~al.(2022)Anand, Schleich, Alperin-Lea, Jensen, Sim, D{\'\i}az-Tinoco, Kottmann, Degroote, Izmaylov, and Aspuru-Guzik]{anand2022quantum}
Abhinav Anand, Philipp Schleich, Sumner Alperin-Lea, Phillip~W.K. Jensen, Sukin Sim, Manuel D{\'\i}az-Tinoco, Jakob~S. Kottmann, Matthias Degroote, Artur~F. Izmaylov, and Al{\'a}n Aspuru-Guzik.
\newblock A quantum computing view on unitary coupled cluster theory.
\newblock \emph{Chem. Soc. Revi.}, 51\penalty0 (5):\penalty0 1659--1684, 2022.

\bibitem[Nielsen and Chuang(2010)]{nielsen_chuang_2010}
Michael~A. Nielsen and Isaac~L. Chuang.
\newblock \emph{Quantum Computation and Quantum Information}.
\newblock Cambridge University Press, 10th anniversary edition edition, 2010.
\newblock \doi{10.1017/cbo9780511976667}.

\bibitem[Bell(1964)]{bell1964einstein}
John~S. Bell.
\newblock On the einstein podolsky rosen paradox.
\newblock \emph{Physics}, 1\penalty0 (3):\penalty0 195--200, 1964.
\newblock \doi{10.1142/9789812386540\_0002}.

\bibitem[Bennett et~al.(1993)Bennett, Brassard, Cr{\'e}peau, Jozsa, Peres, and Wootters]{bennett1993teleporting}
Charles~H. Bennett, Gilles Brassard, Claude Cr{\'e}peau, Richard Jozsa, Asher Peres, and William~K. Wootters.
\newblock Teleporting an unknown quantum state via dual classical and {E}instein--{P}odolsky--{R}osen channels.
\newblock \emph{Phys. Rev. Lett.}, 70\penalty0 (13):\penalty0 1895--1899, 1993.
\newblock \doi{10.1103/physrevlett.70.1895}.

\bibitem[Bennett and Wiesner(1992)]{bennett1992communication}
Charles~H. Bennett and Stephen~J. Wiesner.
\newblock Communication via one- and two-particle operators on {E}instein--{P}odolsky--{R}osen states.
\newblock \emph{Phys. Rev. Lett.}, 69\penalty0 (20):\penalty0 2881--2884, 1992.
\newblock \doi{10.1103/physrevlett.69.2881}.

\bibitem[Horodecki et~al.(2009)Horodecki, Horodecki, Horodecki, and Horodecki]{horodecki2009quantum}
Ryszard Horodecki, Pawe{\l} Horodecki, Micha{\l} Horodecki, and Karol Horodecki.
\newblock Quantum entanglement.
\newblock \emph{Rev. Mod. Phys.}, 81\penalty0 (2):\penalty0 865--942, 2009.
\newblock \doi{10.1103/revmodphys.81.865}.

\bibitem[Essler et~al.(2005)Essler, Frahm, G{\"o}hmann, Kl{\"u}mper, and Korepin]{essler2005one}
Fabian~HL Essler, Holger Frahm, Frank G{\"o}hmann, Andreas Kl{\"u}mper, and Vladimir~E Korepin.
\newblock \emph{The one-dimensional Hubbard model}.
\newblock Cambridge University Press, 2005.
\newblock \doi{10.1017/cbo9780511534843}.

\bibitem[Feiguin and White(2005)]{feiguin2005finite}
Adrian~E Feiguin and Steven~R White.
\newblock Finite-temperature density matrix renormalization using an enlarged hilbert space.
\newblock \emph{Phys. Rev. B}, 72\penalty0 (22):\penalty0 220401, 2005.
\newblock \doi{10.1103/physrevb.72.220401}.

\bibitem[Verstraete et~al.(2004)Verstraete, Garcia-Ripoll, and Cirac]{verstraete2004matrix}
Frank Verstraete, Juan~J Garcia-Ripoll, and Juan~Ignacio Cirac.
\newblock Matrix product density operators: Simulation of finite-temperature and dissipative systems.
\newblock \emph{Phys. Rev. Lett.}, 93\penalty0 (20):\penalty0 207204, 2004.
\newblock \doi{10.1103/physrevlett.93.207204}.

\bibitem[Suzuki(1990)]{suzuki1990fractal}
Masuo Suzuki.
\newblock Fractal decomposition of exponential operators with applications to many-body theories and monte carlo simulations.
\newblock \emph{Phys. Lett. A}, 146\penalty0 (6):\penalty0 319--323, 1990.
\newblock \doi{10.1016/0375-9601(90)90962-n}.

\bibitem[Childs et~al.(2021)Childs, Su, Tran, Wiebe, and Zhu]{childs2021theory}
Andrew~M Childs, Yuan Su, Minh~C Tran, Nathan Wiebe, and Shuchen Zhu.
\newblock Theory of trotter error with commutator scaling.
\newblock \emph{Phys. Rev. X}, 11\penalty0 (1):\penalty0 011020, 2021.
\newblock \doi{10.1103/physrevx.11.011020}.

\bibitem[Bagherimehrab et~al.(2024)Bagherimehrab, Mantilla~Calderon, Berry, Schleich, Vakili, Aldossary, Angulo, Gorgulla, and Aspuru-Guzik]{bagherimehrab2025faster}
Mohsen Bagherimehrab, Luis Mantilla~Calderon, Dominic~W Berry, Philipp Schleich, Mohammad~Ghazi Vakili, Abdulrahman Aldossary, Jorge~A Angulo, Christoph Gorgulla, and Alan Aspuru-Guzik.
\newblock Faster algorithmic quantum and classical simulations by corrected product formulas.
\newblock \emph{arXiv preprint arXiv:2409.08265}, 2024.
\newblock \doi{10.48550/arXiv.2409.08265}.

\bibitem[Gustin et~al.(2023)Gustin, Kim, McCamant, and Franco]{gustin2023mapping}
Ignacio Gustin, Chang~Woo Kim, David~W McCamant, and Ignacio Franco.
\newblock Mapping electronic decoherence pathways in molecules.
\newblock \emph{Proc. Natl. Acad. Sci.}, 120\penalty0 (49):\penalty0 e2309987120, 2023.
\newblock \doi{10.26434/chemrxiv-2023-ld0d4}.

\bibitem[Joos et~al.(2013)Joos, Zeh, Kiefer, Giulini, Kupsch, and Stamatescu]{joos2013decoherence}
Erich Joos, H~Dieter Zeh, Claus Kiefer, Domenico~JW Giulini, Joachim Kupsch, and Ion-Olimpiu Stamatescu.
\newblock \emph{Decoherence and the appearance of a classical world in quantum theory}.
\newblock Springer Science \& Business Media, 2013.
\newblock \doi{10.1007/978-3-662-03263-3}.

\bibitem[Gustin et~al.(2025{\natexlab{a}})Gustin, Chen, and Franco]{gustin2025decoherence}
Ignacio Gustin, Xinxian Chen, and Ignacio Franco.
\newblock Decoherence dynamics in molecular qubits: Exponential, gaussian and beyond.
\newblock \emph{J. Chem. Phys.}, 162\penalty0 (6), 2025{\natexlab{a}}.
\newblock \doi{10.1063/5.0246970}.

\bibitem[Schlosshauer(2007)]{schlosshauer2007decoherence}
Maximilian~A Schlosshauer.
\newblock \emph{Decoherence: and the quantum-to-classical transition}.
\newblock Springer Science \& Business Media, 2007.
\newblock ISBN 978-3-540-35773-5.

\bibitem[Hu et~al.(2022)Hu, Gustin, Krauss, and Franco]{hu2022tuning}
Wenxiang Hu, Ignacio Gustin, Todd~D Krauss, and Ignacio Franco.
\newblock Tuning and enhancing quantum coherence time scales in molecules via light-matter hybridization.
\newblock \emph{J. Phys. Chem. Lett.}, 13\penalty0 (49):\penalty0 11503--11511, 2022.
\newblock \doi{10.1021/acs.jpclett.2c02877}.

\bibitem[Gustin et~al.(2025{\natexlab{b}})Gustin, Kim, and Franco]{gustin2025dissipation}
Ignacio Gustin, Chang~Woo Kim, and Ignacio Franco.
\newblock Dissipation pathways in a photosynthetic complex.
\newblock \emph{J. Phys. Chem. Lett.}, 16:\penalty0 13093--13101, 2025{\natexlab{b}}.
\newblock \doi{10.1021/acs.jpclett.5c02945}.

\bibitem[Weiss(2012)]{weiss2012quantum}
Ulrich Weiss.
\newblock \emph{Quantum dissipative systems}.
\newblock World Scientific, 2012.
\newblock \doi{10.1142/1476}.

\bibitem[Gustin et~al.(2025{\natexlab{c}})Gustin, Kim, and Franco]{gustin2025general}
Ignacio Gustin, Chang~Woo Kim, and Ignacio Franco.
\newblock General framework for quantifying dissipation pathways in open quantum systems. iii. off-diagonal system-bath couplings.
\newblock \emph{arXiv preprint arXiv:2510.04372}, 2025{\natexlab{c}}.
\newblock \doi{10.48550/arXiv.2510.04372}.

\bibitem[Whitfield et~al.(2010)Whitfield, Rodr{\'\i}guez-Rosario, and Aspuru-Guzik]{whitfield2010quantum}
James~D Whitfield, C{\'e}sar~A Rodr{\'\i}guez-Rosario, and Al{\'a}n Aspuru-Guzik.
\newblock Quantum stochastic walks: A generalization of classical random walks and quantum walks.
\newblock \emph{Phys. Rev. A}, 81\penalty0 (2):\penalty0 022323, 2010.

\bibitem[Rebentrost et~al.(2009)Rebentrost, Mohseni, Kassal, Lloyd, and Aspuru-Guzik]{rebentrost2009environment}
Patrick Rebentrost, Masoud Mohseni, Ivan Kassal, Seth Lloyd, and Al{\'a}n Aspuru-Guzik.
\newblock Environment-assisted quantum transport.
\newblock \emph{New Journal of Physics}, 11\penalty0 (3):\penalty0 033003, 2009.

\bibitem[Gorini et~al.(1976)Gorini, Kossakowski, and Sudarshan]{Gorini1976}
V.~Gorini, A.~Kossakowski, and E.~C.~G. Sudarshan.
\newblock Completely positive dynamical semigroups of $n$-level systems.
\newblock \emph{J. Math. Phys.}, 17:\penalty0 821, 1976.
\newblock \doi{10.1063/1.522979}.

\bibitem[Lindblad(1976)]{Lindblad1976}
G.~Lindblad.
\newblock On the generators of quantum dynamical semigroups.
\newblock \emph{Commun. Math. Phys.}, 48:\penalty0 119, 1976.
\newblock \doi{10.1007/bf01608499}.

\bibitem[Breuer and Petruccione(2002)]{BreuerPetruccione2002}
H.~P. Breuer and F.~Petruccione.
\newblock \emph{The Theory of Open Quantum Systems}.
\newblock Oxford University Press, 2002.
\newblock \doi{10.1093/acprof:oso/9780199213900.001.0001}.

\bibitem[Tanimura(2020)]{tanimura2020numerically}
Yoshitaka Tanimura.
\newblock Numerically ``exact'' approach to open quantum dynamics: The hierarchical equations of motion (heom).
\newblock \emph{J. Chem. Phys.}, 153\penalty0 (2):\penalty0 020901, 2020.
\newblock \doi{10.1063/5.0011599}.

\bibitem[Ikeda and Scholes(2020)]{Ikeda2020}
Tatsushi Ikeda and Gregory~D. Scholes.
\newblock Generalization of the hierarchical equations of motion theory for efficient calculations with arbitrary correlation functions.
\newblock \emph{J. Chem. Phys.}, 152\penalty0 (20):\penalty0 204101, 2020.
\newblock \doi{10.1063/5.0007327}.

\bibitem[Adolphs and Renger(2006)]{adolphs2006proteins}
Julia Adolphs and Thomas Renger.
\newblock How proteins trigger excitation energy transfer in the fmo complex of green sulfur bacteria.
\newblock \emph{Biophys. J.}, 91\penalty0 (8):\penalty0 2778--2797, 2006.

\bibitem[Kreisbeck et~al.(2011)Kreisbeck, Kramer, Rodriguez, and Hein]{kreisbeck2011high}
Christoph Kreisbeck, Tobias Kramer, Mirta Rodriguez, and Birgit Hein.
\newblock High-performance solution of hierarchical equations of motion for studying energy transfer in light-harvesting complexes.
\newblock \emph{J. Chem. Theory Comput.}, 7\penalty0 (7):\penalty0 2166--2174, 2011.

\bibitem[Kreisbeck et~al.(2014)Kreisbeck, Kramer, and Aspuru-Guzik]{kreisbeck2014scalable}
Christoph Kreisbeck, Tobias Kramer, and Al{\'a}n Aspuru-Guzik.
\newblock Scalable high-performance algorithm for the simulation of exciton dynamics. application to the light-harvesting complex ii in the presence of resonant vibrational modes.
\newblock \emph{J. Chem. Theory Comput.}, 10\penalty0 (9):\penalty0 4045--4054, 2014.

\bibitem[Khaneja et~al.(2005)Khaneja, Reiss, Kehlet, Schulte-Herbr{\"u}ggen, and Glaser]{khaneja2005optimal}
Navin Khaneja, Tilman Reiss, Conny Kehlet, Thomas Schulte-Herbr{\"u}ggen, and Steffen~J. Glaser.
\newblock Optimal control of coupled spin dynamics in {NMR}.
\newblock \emph{J. Magn. Reson.}, 172\penalty0 (2):\penalty0 296--305, 2005.
\newblock \doi{10.1016/j.jmr.2004.11.004}.

\bibitem[Vitanov et~al.(2017)Vitanov, Rangelov, Shore, and Bergmann]{vitanov2017stimulated}
Nikolay~V. Vitanov, Andon~A. Rangelov, Bruce~W. Shore, and Klaus Bergmann.
\newblock Stimulated raman adiabatic passage in physics chemistry and beyond.
\newblock \emph{Rev. Mod. Phys.}, 89\penalty0 (1):\penalty0 015006, 2017.
\newblock \doi{10.1103/revmodphys.89.015006}.

\bibitem[Wiebe et~al.(2011)Wiebe, Berry, H{\o}yer, and Sanders]{wiebe2011simulating}
Nathan Wiebe, Dominic~W Berry, Peter H{\o}yer, and Barry~C Sanders.
\newblock Simulating quantum dynamics on a quantum computer.
\newblock \emph{J. Phys. A Math. Theor.}, 44\penalty0 (44):\penalty0 445308, 2011.
\newblock \doi{10.1088/1751-8113/44/44/445308}.

\bibitem[Prosen(2002)]{prosen2002general}
Toma{\v{z}} Prosen.
\newblock General relation between quantum ergodicity and fidelity of quantum dynamics.
\newblock \emph{Phys. Rev. E.}, 65\penalty0 (3):\penalty0 036208, 2002.
\newblock \doi{10.1103/physreve.65.036208}.

\bibitem[Else et~al.(2020)Else, Monroe, Nayak, and Yao]{else2020discrete}
Dominic~V Else, Christopher Monroe, Chetan Nayak, and Norman~Y Yao.
\newblock Discrete time crystals.
\newblock \emph{Annu. Rev. Condens. Matter Phys.}, 11\penalty0 (1):\penalty0 467--499, 2020.
\newblock \doi{10.1146/annurev-conmatphys-031119-050658}.

\bibitem[Wilczek(2012)]{wilczek2012quantum}
Frank Wilczek.
\newblock Quantum time crystals.
\newblock \emph{Phys. Rev. Lett.}, 109\penalty0 (16):\penalty0 160401, 2012.

\bibitem[Else et~al.(2017)Else, Bauer, and Nayak]{else2017prethermal}
Dominic~V Else, Bela Bauer, and Chetan Nayak.
\newblock Prethermal phases of matter protected by time-translation symmetry.
\newblock \emph{Phys. Rev. X}, 7\penalty0 (1):\penalty0 011026, 2017.
\newblock \doi{10.1103/physrevx.7.011026}.

\bibitem[Vidal(2003)]{vidal2003efficient}
Guifr{\'e} Vidal.
\newblock Efficient classical simulation of slightly entangled quantum computations.
\newblock \emph{Phys. Rev. Lett.}, 91\penalty0 (14):\penalty0 147902, 2003.
\newblock \doi{10.1103/physrevlett.91.147902}.

\bibitem[Zeng and Sheng(2017)]{zeng2017prethermal}
Tian-Sheng Zeng and DN~Sheng.
\newblock Prethermal time crystals in a one-dimensional periodically driven floquet system.
\newblock \emph{Phys. Rev. B}, 96\penalty0 (9):\penalty0 094202, 2017.
\newblock \doi{10.1103/physrevb.96.094202}.

\bibitem[Doherty and Bartlett(2009)]{doherty2009identifying}
Andrew~C Doherty and Stephen~D Bartlett.
\newblock Identifying phases of quantum many-body systems that are universal<? format?> for quantum computation.
\newblock \emph{Phys. Rev. Lett.}, 103\penalty0 (2):\penalty0 020506, 2009.
\newblock \doi{10.1103/physrevlett.103.020506}.

\bibitem[Miyake(2010)]{miyake2010quantum}
Akimasa Miyake.
\newblock Quantum computation on the edge of a symmetry-protected topological order.
\newblock \emph{Phys. Rev. Lett.}, 105\penalty0 (4):\penalty0 040501, 2010.
\newblock \doi{10.1103/physrevlett.105.040501}.

\bibitem[Raussendorf et~al.(2023)Raussendorf, Yang, and Adhikary]{raussendorf2023measurement}
Robert Raussendorf, Wang Yang, and Arnab Adhikary.
\newblock Measurement-based quantum computation in finite one-dimensional systems: string order implies computational power.
\newblock \emph{Quantum}, 7:\penalty0 1215, 2023.
\newblock \doi{10.22331/q-2023-12-28-1215}.

\bibitem[Kitaev(1995)]{kitaev1995quantum}
A~Yu Kitaev.
\newblock Quantum measurements and the abelian stabilizer problem.
\newblock \emph{arXiv preprint quant-ph/9511026}, 1995.
\newblock \doi{10.48550/arXiv.quant-ph/9511026}.

\bibitem[Shor(1995)]{shor1995scheme}
Peter~W Shor.
\newblock Scheme for reducing decoherence in quantum computer memory.
\newblock \emph{Phys. Rev. A}, 52\penalty0 (4):\penalty0 R2493, 1995.
\newblock \doi{10.1103/physreva.52.r2493}.

\bibitem[Steane(1996)]{steane1996error}
Andrew~M Steane.
\newblock Error correcting codes in quantum theory.
\newblock \emph{Phys. Rev. Lett.}, 77\penalty0 (5):\penalty0 793, 1996.
\newblock \doi{10.1103/physrevlett.77.793}.

\bibitem[Knill et~al.(1998)Knill, Laflamme, and Zurek]{knill1998resilient}
Emanuel Knill, Raymond Laflamme, and Wojciech~H Zurek.
\newblock Resilient quantum computation.
\newblock \emph{Science}, 279\penalty0 (5349):\penalty0 342--345, 1998.
\newblock \doi{10.1126/science.279.5349.342}.

\bibitem[Devitt et~al.(2013)Devitt, Munro, and Nemoto]{devitt2013quantum}
Simon~J Devitt, William~J Munro, and Kae Nemoto.
\newblock Quantum error correction for beginners.
\newblock \emph{Rep. Prog. Phys.}, 76\penalty0 (7):\penalty0 076001, 2013.
\newblock \doi{10.1088/0034-4885/76/7/076001}.

\bibitem[Higgott and Gidney(2025)]{Higgott2025sparseblossom}
Oscar Higgott and Craig Gidney.
\newblock Sparse {B}lossom: correcting a million errors per core second with minimum-weight matching.
\newblock \emph{{Quantum}}, 9:\penalty0 1600, January 2025.
\newblock ISSN 2521-327X.
\newblock \doi{10.22331/q-2025-01-20-1600}.

\bibitem[Delfosse and Nickerson(2021)]{delfosse2021almost}
Nicolas Delfosse and Naomi~H Nickerson.
\newblock Almost-linear time decoding algorithm for topological codes.
\newblock \emph{Quantum}, 5:\penalty0 595, 2021.
\newblock \doi{10.22331/q-2021-12-02-595}.

\bibitem[Kitaev(2003)]{kitaev2003fault}
A~Yu Kitaev.
\newblock Fault-tolerant quantum computation by anyons.
\newblock \emph{Ann. Phys.}, 303\penalty0 (1):\penalty0 2--30, 2003.
\newblock \doi{10.1016/s0003-4916(02)00018-0}.

\bibitem[Fowler et~al.(2012)Fowler, Mariantoni, Martinis, and Cleland]{fowler2012surface}
Austin~G Fowler, Matteo Mariantoni, John~M Martinis, and Andrew~N Cleland.
\newblock Surface codes: Towards practical large-scale quantum computation.
\newblock \emph{Phys. Rev. A}, 86\penalty0 (3):\penalty0 032324, 2012.
\newblock \doi{10.1103/physreva.86.032324}.

\bibitem[AI(2023)]{google2023suppressing}
Google~Quantum AI.
\newblock Suppressing quantum errors by scaling a surface code logical qubit.
\newblock \emph{Nature}, 614\penalty0 (7949):\penalty0 676--681, 2023.
\newblock \doi{10.1038/s41586-022-05434-1}.

\bibitem[Bluvstein et~al.(2024)Bluvstein, Evered, Geim, Li, Zhou, Manovitz, Ebadi, Cain, Kalinowski, Hangleiter, et~al.]{bluvstein2024logical}
Dolev Bluvstein, Simon~J Evered, Alexandra~A Geim, Sophie~H Li, Hengyun Zhou, Tom Manovitz, Sepehr Ebadi, Madelyn Cain, Marcin Kalinowski, Dominik Hangleiter, et~al.
\newblock Logical quantum processor based on reconfigurable atom arrays.
\newblock \emph{Nature}, 626\penalty0 (7997):\penalty0 58--65, 2024.
\newblock \doi{10.1038/s41586-023-06927-3}.

\bibitem[goo(2025)]{google2025quantum}
Quantum error correction below the surface code threshold.
\newblock \emph{Nature}, 638\penalty0 (8052):\penalty0 920--926, 2025.
\newblock \doi{10.1038/s41586-024-08449-y}.

\bibitem[Aghaee~Rad et~al.(2025)Aghaee~Rad, Ainsworth, Alexander, Altieri, Askarani, Baby, Banchi, Baragiola, Bourassa, Chadwick, et~al.]{aghaee2025scaling}
H~Aghaee~Rad, T~Ainsworth, RN~Alexander, B~Altieri, MF~Askarani, R~Baby, L~Banchi, BQ~Baragiola, JE~Bourassa, RS~Chadwick, et~al.
\newblock Scaling and networking a modular photonic quantum computer.
\newblock \emph{Nature}, 638\penalty0 (8052):\penalty0 912--919, 2025.
\newblock \doi{10.1038/s41586-024-08406-9}.

\bibitem[Zhao et~al.(2022)Zhao, Ye, Huang, Zhang, Wu, Guan, Zhu, Wei, He, Cao, et~al.]{zhao2022realization}
Youwei Zhao, Yangsen Ye, He-Liang Huang, Yiming Zhang, Dachao Wu, Huijie Guan, Qingling Zhu, Zuolin Wei, Tan He, Sirui Cao, et~al.
\newblock Realization of an error-correcting surface code with superconducting qubits.
\newblock \emph{Phys. Rev. Lett.}, 129\penalty0 (3):\penalty0 030501, 2022.
\newblock \doi{10.1103/physrevlett.129.030501}.

\bibitem[Gidney(2021)]{gidney2021stim}
Craig Gidney.
\newblock Stim: a fast stabilizer circuit simulator.
\newblock \emph{Quantum}, 5:\penalty0 497, 2021.
\newblock \doi{10.22331/q-2021-07-06-497}.

\bibitem[Arlt et~al.(2025{\natexlab{b}})]{Arlt2025AIMandel}
J.~Arlt et~al.
\newblock Ai-mandel: Autonomous idea generation and experimental design in quantum physics.
\newblock \emph{arXiv preprint arXiv:2511.11752}, 2025{\natexlab{b}}.
\newblock \doi{10.48550/arXiv.2511.11752}.

\bibitem[Murali et~al.(2019)]{Murali2019NoiseAdaptive}
Prakash Murali et~al.
\newblock Noise-adaptive compiler mappings for noisy intermediate-scale quantum computers.
\newblock In \emph{ASPLOS}, 2019.
\newblock \doi{10.1145/3297858.3304075}.

\bibitem[Sivarajah et~al.(2020)]{Sivarajah2020TKet}
Simon Sivarajah et~al.
\newblock t|ket⟩: A retargetable compiler for nisq devices.
\newblock \emph{Quantum Sci. Technol.}, 6:\penalty0 014003, 2020.
\newblock \doi{10.1088/2058-9565/ab8e92}.

\bibitem[Bayraktar et~al.(2023)Bayraktar, Charara, Clark, Cohen, Costa, Fang, Gao, Guan, Gunnels, Haidar, et~al.]{bayraktar2023cuquantum}
Harun Bayraktar, Ali Charara, David Clark, Saul Cohen, Timothy Costa, Yao-Lung~L Fang, Yang Gao, Jack Guan, John Gunnels, Azzam Haidar, et~al.
\newblock cuquantum sdk: A high-performance library for accelerating quantum science.
\newblock In \emph{Proc. IEEE Int. Conf. Quantum Comput. Eng. (QCE)}, volume~1, pages 1050--1061. IEEE, 2023.
\newblock \doi{10.1109/qce57702.2023.00119}.

\bibitem[Brunner et~al.()Brunner, Clark, Finger, Greene-Diniz, Kalidindi, Koziell-Pipe, Manrique, Meichanetzidis, Rapp, Fawzi, Fawzi, He, Romera-Paredes, and Yin]{algorithm_discovery_quantinuum}
Eric Brunner, Steve Clark, Fabian Finger, Gabriel Greene-Diniz, Pranav Kalidindi, Alexander Koziell-Pipe, David~Zsolt Manrique, Konstantinos Meichanetzidis, Frederic Rapp, Alhussein Fawzi, Hamza Fawzi, Kerry He, Bernardino Romera-Paredes, and Kante Yin.
\newblock Automated quantum algorithm discovery for quantum chemistry.
\newblock https://www.quantinuum.com/blog/automated-quantum-algorithm-discovery-for-quantum-chemistry.
\newblock Accessed: 2025-12-15.

\bibitem[Tom et~al.(2024)Tom, Schmid, Baird, Cao, Darvish, Hao, Lo, Pablo-Garc{\'\i}a, Rajaonson, Skreta, et~al.]{tom2024self}
Gary Tom, Stefan~P Schmid, Sterling~G Baird, Yang Cao, Kourosh Darvish, Han Hao, Stanley Lo, Sergio Pablo-Garc{\'\i}a, Ella~M Rajaonson, Marta Skreta, et~al.
\newblock Self-driving laboratories for chemistry and materials science.
\newblock \emph{Chem. Rev.}, 124\penalty0 (16):\penalty0 9633--9732, 2024.
\newblock \doi{10.1021/acs.chemrev.4c00055}.

\bibitem[Bai et~al.(2022)Bai, Cao, Mosbach, Akroyd, Lapkin, and Kraft]{bai2022platform}
Jiaru Bai, Liwei Cao, Sebastian Mosbach, Jethro Akroyd, Alexei~A Lapkin, and Markus Kraft.
\newblock From platform to knowledge graph: evolution of laboratory automation.
\newblock \emph{JACS Au}, 2\penalty0 (2):\penalty0 292--309, 2022.
\newblock \doi{10.1021/jacsau.1c00438}.

\bibitem[Vogler et~al.(2024)Vogler, Steensen, Ram{\'\i}rez, Merker, Busk, Carlsson, Rieger, Zhang, Liot, Pizzi, et~al.]{vogler2024autonomous}
Monika Vogler, Simon~Krarup Steensen, Francisco~Fernando Ram{\'\i}rez, Leon Merker, Jonas Busk, Johan~Martin Carlsson, Laura~Hannemose Rieger, Bojing Zhang, Fran{\c{c}}ois Liot, Giovanni Pizzi, et~al.
\newblock Autonomous battery optimization by deploying distributed experiments and simulations.
\newblock \emph{Advanced Energy Materials}, 14\penalty0 (46):\penalty0 2403263, 2024.
\newblock \doi{10.1002/aenm.202403263}.

\bibitem[Bai et~al.(2024)Bai, Mosbach, Taylor, Karan, Lee, Rihm, Akroyd, Lapkin, and Kraft]{bai2024dynamic}
Jiaru Bai, Sebastian Mosbach, Connor~J Taylor, Dogancan Karan, Kok~Foong Lee, Simon~D Rihm, Jethro Akroyd, Alexei~A Lapkin, and Markus Kraft.
\newblock A dynamic knowledge graph approach to distributed self-driving laboratories.
\newblock \emph{Nat. Commun.}, 15\penalty0 (1):\penalty0 462, 2024.
\newblock \doi{10.1038/s41467-023-44599-9}.

\bibitem[Rauschen et~al.(2024)Rauschen, Guy, Hein, and Cronin]{rauschen2024universal}
Robert Rauschen, Mason Guy, Jason~E Hein, and Leroy Cronin.
\newblock Universal chemical programming language for robotic synthesis repeatability.
\newblock \emph{Nature Synthesis}, 3\penalty0 (4):\penalty0 488--496, 2024.
\newblock \doi{10.1038/s44160-023-00473-6}.

\bibitem[Caramelli et~al.(2018)Caramelli, Salley, Henson, Camarasa, Sharabi, Keenan, and Cronin]{caramelli2018networking}
Dario Caramelli, Daniel Salley, Alon Henson, Gerardo~Aragon Camarasa, Salah Sharabi, Graham Keenan, and Leroy Cronin.
\newblock Networking chemical robots for reaction multitasking.
\newblock \emph{Nat. Commun.}, 9\penalty0 (1):\penalty0 3406, 2018.
\newblock \doi{10.1038/s41467-018-05828-8}.

\bibitem[Krenn et~al.(2022)Krenn, Pollice, Guo, Aldeghi, Cervera-Lierta, Friederich, dos Passos~Gomes, H{\"a}se, Jinich, Nigam, et~al.]{krenn2022scientific}
Mario Krenn, Robert Pollice, Si~Yue Guo, Matteo Aldeghi, Alba Cervera-Lierta, Pascal Friederich, Gabriel dos Passos~Gomes, Florian H{\"a}se, Adrian Jinich, AkshatKumar Nigam, et~al.
\newblock On scientific understanding with artificial intelligence.
\newblock \emph{Nat. Rev. Phys.}, 4\penalty0 (12):\penalty0 761--769, 2022.

\end{thebibliography}
